\documentclass[aps,10pt,prb,twocolumn,superscriptaddress,longbibliography,floatfix]{revtex4-2}
\usepackage{diagbox} 
\usepackage{graphicx}
\usepackage{pstricks}
\usepackage{pst-node}
\usepackage{pstricks-add}
\usepackage{tikz}
\usepackage{amsfonts,amssymb,amsmath}
\usepackage[colorlinks=true,citecolor=green,linkcolor=red,urlcolor=blue]{hyperref}
\usepackage{subfigure}
\usepackage{calligra}

\usepackage{dsfont}

\newcommand{\maC}{\mathcal{C}}
\newcommand{\maG}{\mathcal{G}}

\newcommand{\maO}{\mathcal{O}}
\newcommand{\beq}{\begin{equation}}
\newcommand{\be}{\begin{equation}}
\newcommand{\beqn}{\begin{eqnarray}}
\newcommand{\eeq}{\end{equation}}
\newcommand{\ee}{\end{equation}}
\newcommand{\eeqn}{\end{eqnarray}}
\newcommand{\Inv}{\mathrm{Inv}}
\newcommand{\Stab}{\mathrm{Stab}}
\newcommand{\vecg}{\mathrm{Vec}_G}
\newcommand{\repg}{\mathrm{Rep}_G}
\newcommand{\HomZvecg}{\mathrm{Hom}_{\mathcal{Z}(\mathrm{Vec}_G)}}
\newcommand{\cl}{\mathrm{cl}}
\newcommand{\id}{\mathrm{id}}
\newcommand{\1}{\mathbf{1}}
\newcommand{\nn}{\nonumber}
\newcommand{\R}{\Gamma}
\newcommand{\genus}{\mathfrak{g}}

\begin{document}
\title{Partition function of the Kitaev quantum double model}
\author{Anna Ritz-Zwilling}
\email{anna.ritz\_zwilling@sorbonne-universite.fr }
\affiliation{Sorbonne Universit\'e, CNRS, Laboratoire de Physique Th\'eorique de la Mati\`ere Condens\'ee, LPTMC, F-75005 Paris, France}

\author{Beno\^it Dou\c{c}ot}
\email{benoit.doucot@sorbonne-universite.fr}
\affiliation{Sorbonne Universit\'e, CNRS, Laboratoire de Physique Th\'eorique des Hautes \'Energies, LPTHE, F-75005 Paris, France}

\author{Steven H. Simon}
\email{steven.simon@physics.ox.ac.uk}
\affiliation{Rudolf Peierls Centre for Theoretical Physics, Clarendon Laboratory, Oxford, OX1 3PU, United Kingdom}

\author{Julien Vidal}
\email{julien.vidal@sorbonne-universite.fr}
\affiliation{Sorbonne Universit\'e, CNRS, Laboratoire de Physique Th\'eorique de la Mati\`ere Condens\'ee, LPTMC, F-75005 Paris, France}

\author{Jean-No\"el Fuchs}
\email{jean-noel.fuchs@sorbonne-universite.fr}
\affiliation{Sorbonne Universit\'e, CNRS, Laboratoire de Physique Th\'eorique de la Mati\`ere Condens\'ee, LPTMC, F-75005 Paris, France}

\begin{abstract}

We compute the degeneracy of energy levels in the Kitaev quantum double model for any discrete group $G$ on any planar graph forming the skeleton of a closed orientable surface of arbitrary genus. The derivation is based on the fusion rules of the properly identified vertex and plaquette excitations, which are selected among the anyons, i.e., the simple objects of the Drinfeld center $\mathcal{Z}(\vecg)$. These degeneracies are given in terms of the quantum dimensions of the anyons and allow one to  obtain the exact finite-temperature partition function of the model, valid for any finite-size system.  

\end{abstract}
\date{\today}

\maketitle

\section{Introduction}
Anyons~\cite{Leinaas77,Wilczek82_1,Wilczek82_2,Goldin85} are the building blocks of topological quantum computation~\cite{Wang_book,Freedman03,Nayak08}. First observed in the fractional quantum Hall effect~\cite{Tsui82,Stormer99}, these exotic quasiparticles that obey fractional statistics~\cite{Greiter24} have since triggered much interest due to their potential use in performing nontrivial braiding operations that could act as logic gates for a quantum computer. Additionally, the nonlocal nature of the information that can be encoded in anyons makes them promising candidates for topological quantum memories~\cite{Dennis02}. As early realized by Kitaev in his seminal paper on fault-tolerant quantum computation~\cite{Kitaev03}, this nonlocality ensures a topological protection against external perturbations~\cite{Bravyi10} (see also Ref.~\cite{Doucot12} for a review on physical implementation of topologically protected qubits). Recent experiments have leveraged this protection to implement and stabilize the ground-state wavefunction of the paradigmatic toric code~\cite{Kitaev03} in a quantum processor~\cite{Satzinger21,Zhao22}, as well as in Rydberg atom array~\cite{Semeghini21} (see also Refs.~\cite{Xu23,Iqbal24} for the realization of non-Abelian topological order). Although directly implementing large-scale surface code Hamiltonians (not just the ground state) remains challenging, understanding the role of thermal fluctuations in these systems is important. 

For two decades, this problem has been the subject of numerous studies (see, for example, Refs.~\cite{Castelnovo07_2, Nussinov08,Alicki09, Iblisdir09, Iblisdir10,Landon13, Brown16}).  In two dimensions, it has been shown that topological order is destroyed at any finite temperature in the thermodynamical limit~\cite{Hastings11} for local commuting projector Hamiltonians (such as the  Kitaev quantum double (KQD) Hamiltonian~\cite{Kitaev03} or the Levin-Wen (string-net) Hamiltonian~\cite{Levin05}), which host achiral topological phases. However, a competition between the temperature $T$ and the system size $L$ allows one to preserve the main characteristics of the topological order at temperatures smaller than a  size-dependent crossover temperature $T^*\propto 1/\ln L$. To gain a deeper understanding of these thermal fluctuations, access to the partition function of the problem  at hand is essential.

In the present work, we focus on the KQD model, which is a lattice realization of a topological quantum field theory (TQFT). The excitations of this TQFT are anyons labeled by the simple objects of a unitary modular tensor category (UMTC) (see, e.g., Ref.~\cite{Simon_book} for a review). For a KQD model based on a group $G$, this UMTC is the Drinfeld center $\mathcal{Z}(\vecg)$ of the tensor category $\vecg$ built from the group $G$. 
The partition function for the toric code model, the simplest of the KQD models, was computed by several authors~\cite{Castelnovo07_2, Nussinov08, Iblisdir10, Gregor11}. A formal expression was also given for the partition function of the KQD model in the case of a non-Abelian group~\cite{Iblisdir09,Iblisdir10}. 
Here, we provide a simple and compact expression for the partition function of the KQD model for an arbitrary finite group $G$ and for any planar graph forming the skeleton of any closed orientable surface of arbitrary genus, which we believe to be a new result. 

A non-trivial part of deriving the partition function for this model is to obtain its spectral degeneracies, which depend on the topological nature of the anyons present in the system. Our derivation closely follows the approach recently developed for string-net models~\cite{Vidal22,Ritz24_1,Ritz24_2,Ritz_thesis,Soares25}. The main idea for computing the spectral degeneracies is to ignore the microscopic degrees of freedom, and to focus on the anyonic excitations and their fusion rules. 
In short, ``only fusion matters", in most cases. However, in addition to this (topological) degeneracy of fusion channels, vertex excitations of the KQD model feature an internal multiplicity when the group $G$ is non-Abelian, which leads to additional (nontopological) degeneracies.

This article is structured as follows: in Sec.~\ref{sec:KQD}, we introduce the KQD model and identify the elementary excitations that correspond to vertex and plaquette constraint violations. We also discuss the notion of subtype introduced in Ref.~\cite{Kitaev03}, which is crucial for computing degeneracies when $G$ is a non-Abelian group. In Sec.~\ref{sec:degen}, we explain how to compute the degeneracy of any energy levels of the KQD model on a closed orientable surface of genus $\genus$ in terms of fusion trees (see Fig.~\ref{fig:fusiontreekqd} for an illustration). Using the Moore-Seiberg-Banks formula~\cite{Moore89}, we obtain a simple, compact expression for these degeneracies [see Eq.~\eqref{eq:deg_final}], which only involves the quantum dimensions of the anyons of the Drinfeld center $\mathcal{Z}(\vecg)$. 
Finally, the finite-temperature partition function is derived in Sec.~\ref{sec:pf} for any finite-size system [see Eq.~\eqref{eq:kqdPF}]. In Appendix~\ref{app:refined}, we consider a refined version of the KQD model introduced in Ref.~\cite{Komar17}, which distinguishes between different excitations that may exist in the KQD model. Using the same approach as for the KQD model, we compute the exact partition function of this refined model. Appendix~\ref{app:examples} contains two examples of groups for which we discuss the relation between energy eigenspaces and anyons. Appendix~\ref{app:benoit} provides a (technical) proof of the main results, based on very general arguments and valid for a general lattice gauge theory. Appendix~\ref{app:steve} gives an alternative group-theoretical proof of the degeneracy formula using lattice manipulations. Eventually, in Appendix~\ref{app:sn}, we compare the KQD model with string-net models.

\section{Kitaev quantum double model}
\label{sec:KQD}

\subsection{Definition}
The KQD model~\cite{Kitaev03} is defined on any planar graph forming the skeleton of a two-dimensional closed manifold (meaning the graph goes around each noncontractable cycle and plaquettes are filled in accordingly~\footnote{~\label{CW}Strictly speaking, we are defining a CW-complex.}). It assigns every oriented edge of this graph a value out of the elements of a discrete finite group $G$. The Hilbert space $\mathcal{H}$ is spanned by all possible labelings. Since there are $|G|$ possible choices for every link (with $|G|$ the number of elements in the group), one has
%
%
\be
\textnormal{dim }\mathcal{H}=|G|^{N_{\rm l}},
\label{eq:dimHKQD}
\ee
%
%
where $N_{\rm l}$ is the number of links (or edges) of the lattice. For notational convenience, we index the vertices of the lattice with roman letters $a,b,c...$, so that any edge label can be written $g_{ab} \in G$, with the subscript $ab$ indicating that the edge is pointing from the vertex $a$ towards the vertex $b$. Reversing the orientation of an edge changes the label from $g_{ab}$ to $g_{ba}= g_{ab}^{-1}$. 

At each vertex $v$, one defines a set of local operators $A_v^h$, with $h \in G$. These operators preserve the group operation 
%
%
\be A_v^h A_v^{h'} = A_v^{hh'},
\label{eq:preserve}
\ee
%
%
and thus form a representation of $G$ (see, e.g., Ref.~\cite{Preskill_notes,Komar17}). The action of an operator $A_v^h$ is to pre-multiply all (outward oriented) edges $g_{av}$ around vertex $v$ by the group element $h$. For example, on a square lattice, one has 
%
%
\be
A_v^h \, \, \, \,  \raisebox{-8ex}{\includegraphics[scale=0.6]{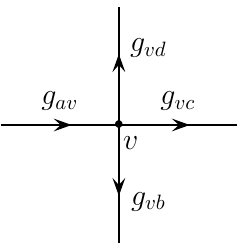}} = \raisebox{-8ex}{\includegraphics[scale=0.6]{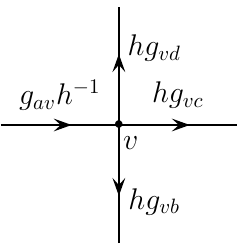}}.
\ee
%
%
These operators implement local gauge transformations on the lattice. In fact, if two operators $A_{v_1}^{h}$ and $A_{v_2}^h$ act on two vertices connected by an edge labeled $g_{v_2v_1}$, this edge label transforms as $h g_{v_2v_1} h^{-1}$. Next, one defines the vertex operator $A_v$ as
%
%
\be
A_v = \frac{1}{|G|} \sum_{h \in G} A_v^{h}.
\label{eq:trivirrep}
\ee
%
%
Using Eq.~\eqref{eq:preserve}, one can easily show that $A_v$ is a projector ($A_v^2=A_v$), and that it is invariant under the action of all $A^h_v$. Therefore, $A_v$ projects on gauge-invariant states at the vertex $v$ (states without ``electric charge''). In other words, it is the projector on the trivial representation of $G$, $\Gamma_1^G$,  at $v$~\cite{Komar17}.
Using the operators $A_v^h$, one can also construct projectors onto any other irreducible representation (irrep) of $G$ at vertex $v$, [see Eq.~\eqref{eq:anyirrep}]. 

For every plaquette, one can define projectors $B_p^h$, which ensure that the product of all labels around the plaquette $p$ equals the group element $h$. For example, on a square lattice,
%
%
\be
B_p^h\, \,   \raisebox{-6ex}{\includegraphics[scale=0.6]{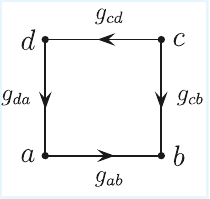}} = \delta_{h, g_{ab}g_{cb}^{-1}g_{cd}g_{da}} \, \raisebox{-6ex}{\includegraphics[scale=0.6]{p.pdf}}.
\label{eq:Bp}
\ee
%
%
By convention, the product is always taken counterclockwise around a plaquette [see Eq.~\eqref{eq:Bp}]. A $B_p^h$ projects onto a ``flux'' $h$ in plaquette $p$ - however, the gauge-invariant quantity is not the group element $h$ but its conjugacy class $C_h$ [see also Eq.~\eqref{eq:cc}]. In particular, the plaquette operator $B_p \equiv B_p^e$ projects onto states for which the oriented product of labels around a plaquette is equal to the identity $e$ of the group $G$ (i.e., it projects onto the conjugacy class $C_e = \lbrace e \rbrace$). This corresponds to a trivial ``magnetic flux'' in the plaquette $p$ or to a flat connection. 

The Hamiltonian of the KQD model~\cite{Kitaev03} is the sum of the local commuting projectors $A_v$ acting on the vertices $v$  and $B_p$ acting on the plaquettes $p$ of the lattice:
%
%
\be
H = -J_{\rm v} \sum_v A_v -J_{\rm p} \sum_p B_p, 
\label{eq:KQDham}
\ee
%
%
where $J_{\rm v}\geqslant 0$ and $J_{\rm p}\geqslant 0$.  A more general version of this model, introduced by 
Ref.~\cite{Komar17}, is discussed in Appendix~\ref{app:refined}.
For the model defined by Eq.~\eqref{eq:KQDham}, all plaquette and vertex operators commute with each other. A ground state of the KQD Hamiltonian is invariant under the action of $A_v$ and $B_p$, i.e., it corresponds to a state $|\psi\rangle$ where $A_v|\psi\rangle= B_p|\psi\rangle =|\psi\rangle,  \,\,\forall \,\, p,v$. A vertex excitation (or ``electric charge'') at a given vertex $v$ corresponds to $A_v=0$, whereas a plaquette excitation at a given plaquette $p$ (or ``magnetic flux'') corresponds to $B_p=0$.

The ground-state energy is thus 
%
%
\be 
E_0 = -J_{\rm v} N_{\rm v} -J_{\rm p} N_{\rm p}, 
\label{eq:gse2}
\ee
%
%
where $N_{\rm v}$ is the total number of vertices $N_{\rm p}$ is the total number of plaquettes. On a closed orientable surface of genus $\genus$, the Euler-Poincar\'e characteristics is given by
%
%
\be 
N_{\rm v}-N_{\rm l}+N_{\rm p}=2-2g.
\label{eq:Euler}
\ee
%
%

Excited energy levels of the KQD model are easily obtained. A state with $n$ excited vertices and $m$ excited plaquettes has energy:
%
%
\be 
E_{n,m}=E_0+J_{\rm v} n + J_{\rm p} m \leqslant 0.
\label{eq:kqdenergy}
\ee
%
%
%

\subsection{Energy eigenspaces versus anyons}
\label{subsec_energy_eigenspaces}
On the one hand, as can be seen from Eq.~\eqref{eq:kqdenergy}, the energy levels of the KQD model are determined by the number of vertex and plaquette excitations, which we will call elementary excitations. 
On the other hand, in the low-energy limit, the lattice model should be described by a TQFT, whose topological sectors are labeled by the anyons of the Drinfeld center $\mathcal{Z}(\vecg)$~\cite{Kitaev03}. These anyons can be split into three categories: \textit{chargeons} carry non-trivial charge, \textit{fluxons} carry non-trivial flux, and \textit{dyons} carry both non-trivial charge and flux. However, these anyons generically live on a \textit{site} of the lattice, i.e., the combination of a vertex and an adjacent plaquette \cite{Kitaev03}. As a result, the relation between elementary excitations on the lattice and anyons is not one-to-one. In short, vertex excitations are chargeons but plaquette excitations are not exactly fluxons. Additionnally, there may be different ways to produce a dyon on a site as a combination of a vertex and a plaquette excitation. This distinction was already made by K\'om\'ar and Landon-Cardinal~\cite{Komar17}, who introduced the notion of ``energy eigenspaces'' (closely related to lattice excitations), that we will adopt. Below, we discuss the difference between energy eigenspaces and anyons. Energy eigenspaces are best understood by using a refined Hamiltonian $H_R$ introduced in Ref.~\cite{Komar17} (see Appendix~\ref{app:refined}). We give two examples of the relation between energy eigenspaces and anyons (for the groups $\mathbb{Z}_N$ and $S_3$) in Appendix~\ref{app:examples}.

\subsubsection{Energy eigenspaces and excitations}
In the KQD model, the operator $A_v$ projects onto the trivial irrep $\Gamma_1^G$ and $B_p$ projects onto a plaquette having a trivial flux $e \in C_e$. Therefore, energy eigenspaces are naturally labeled by $(C_g,\Gamma^G)$, where $C_g$ is a conjugacy class and $\Gamma^G$ an irrep of the group $G$~\cite{Komar17}. The absence of a lattice excitation (corresponding to $A_v=1$ and $B_p=1$) is called the \textit{vacuum} and is labeled by
%
%
\be
\mathbf{1}=(C_e, \Gamma_1^G).
\ee
%
%
Apart from the vacuum, other energy eigenspaces correspond to various lattice excitations, which we now detail. 

\textit{Vertex excitations} correspond to $A_v=0$ and $B_p=1$, and are labeled by
%
%
\be
(C_e, \Gamma^G) \in \text{Ve},
\ee
%
%
where $\Gamma^G \neq \Gamma_1^G$ is a nontrivial irrep. The set of vertex excitations is denoted by Ve. Its cardinality is $|\text{Ve}|=M$, where we define $M+1$ to be the number of irreps of $G$, which is also the number of its conjugacy classes. 

The (gauge-invariant) \textit{plaquette excitations} correspond to $B_p=0$ and $A_v=1$, and are labeled by 
%
%
\be
(C_g,\Gamma^G_1) \in \text{Pl},
\ee
%
%
where $C_g$ with $g\neq e$ is a nontrivial conjugacy class. The set of plaquette excitations is denoted by Pl and one has $|\text{Pl}|=M$. 

Finally, we introduce the set of \textit{site excitations} that violate both the plaquette and the vertex constraints ($A_v=0$ and $B_p=0$). They are labeled by
%
%
\be
(C_g,\Gamma^G)  \in \text{Si},
\ee
%
%
where $C_g$ and $\Gamma^G$ are both nontrivial. The set of site excitations is called Si and has a cardinality $|\text{Si}|=M^2$. 

The total number of energy eigenspaces is $(M+1)^2$, which splits in a unique  vacuum, $M$ vertex, $M$ plaquette, and $M^2$ site excitations. We will also need the notations
%
%
\be
\underline{\text{Ve}} = \text{Ve} \cup \{\mathbf{1}\}, \text{ and } \underline{\text{Pl}} = \text{Pl} \cup \{\mathbf{1}\},
\ee
%
%
for the union of vertex (or plaquette) excitations with the vacuum.

\subsubsection{Anyons}
\label{subsub:anyons}

A simple object of the Drinfeld center $\mathcal{Z}(\vecg)$, called an anyon, is labeled by $(C_g, \Gamma^{\mathcal{N}_g})$, where $C_g$ is the conjugacy class of the element $g \in G$ and $\Gamma^{\mathcal{N}_g}$ is an irreducible representation of the centralizer (or normalizer) $\mathcal{N}_g$ of $g$. Special kinds of anyons can be distinguished. As for energy eigenspaces, the vacuum is 
%
%
\be
\mathbf{1}= (C_e, \Gamma_1^{G}).
\ee
%
%

\textit{Chargeons} correspond to the conjugacy class $C_e$ of the identity and are indexed by the nontrivial irreps $\Gamma^G$ of $G$ (as the centralizer of $e$ is the whole group $G$). They are written
%
%
\be 
(C_e, \Gamma^{G}) \in \text{Ch},
\ee
%
%
where Ch denotes the set of chargeons. Actually, \mbox{Ch = Ve}, and $|\text{Ch}|=M$. 

\textit{Fluxons} are indexed by the nontrivial conjugacy classes $C_g$, and correspond to the trivial irreducible representation $\Gamma_1^{\mathcal{N}_g}$ of the respective centralizers
%
%
\be  
(C_g, \Gamma_1^{\mathcal{N}_g}) \in \text{Fl},
\ee
%
%
where Fl denotes the set of fluxons. The cardinal of Fl is $|\text{Fl}|=M$, but Fl $\neq$ Pl (see below).

Eventually, the other anyons (corresponding to \mbox{$C_g \neq C_e$} and $\Gamma^{\mathcal{N}_g} \neq \Gamma_1^{\mathcal{N}_g}$) are called \textit{dyons} and denoted
%
%
\be 
(C_g, \Gamma^{\mathcal{N}_g}) \in \text{Dy}.
\ee
%
%
The set of dyon excitations is denoted by Dy and one has $|\text{Dy}|\leqslant M^2$. Importantly, Dy and Si are generally different.

The total number of anyon types, i.e., the number of simple objects of $\mathcal{Z}(\vecg)$, decomposes into one vacuum, $M$ chargeons, $M$ fluxons, and $|\text{Dy}|$ dyons, and is always smaller or equal to the number of energy eigenspaces  $(M+1)^2$. In the following, we will also use the notations:
%
%
\be 
\underline{\text{Ch}} = \text{Ch} \cup \{\mathbf{1}\} \text{ and } \underline{\text{Fl}} = \text{Fl} \cup \{\mathbf{1}\}.
\label{eq:underline}
\ee
%
%

\subsubsection{Plaquette excitations are not exactly fluxons}

\label{subsub:plaquettes}

Vertex excitations $(C_e,\Gamma^G)$ are the same as chargeons $(C_e,\Gamma^{\mathcal{N}_e})$ as $\mathcal{N}_e=G$, i.e., Ve = Ch. One can therefore label vertex excitations with the corresponding anyon label in the Drinfeld center $\mathcal{Z}(\vecg)$. However, plaquette excitations $(C_g,\Gamma_1^G)$ are not exactly fluxons $(C_g,\Gamma_1^{\mathcal{N}_g})$. This distinction is only relevant when $G$ is a non-Abelian group: in this case, $\mathcal{N}_g$ is different from $G$ (except for $g=e$). Thus, for non-Abelian $G$, Pl $\neq$ Fl. There is nevertheless an injective mapping between plaquette excitations and fluxons: 
%
%
\be 
(C_g,\Gamma_1^G)\in \text{Pl} \to (C_g,\Gamma_1^{\mathcal{N}_g})\in \text{Fl}\subset \mathcal{Z}(\vecg).
\label{eq:map}
\ee
%
%
A convenient way to label a plaquette excitation $(C_g,\Gamma_1^G)$ is therefore to use the corresponding fluxon label \mbox{$J=(C_g,\Gamma_1^{\mathcal{N}_g}) \in \mathcal{Z}(\vecg)$}. However, as we will see in the following, the dimension of the subspace associated with fluxons is larger or equal to the dimension of the subspace associated with plaquette excitations.

\subsubsection{Quantum dimensions and fusion of anyons}
In a TQFT, it is expected that an anyon $J$ corresponds to a subspace of states of dimension $d_J$. This quantum dimension is topological and results from the channel degeneracy appearing through multiple fusion processes. In Ref.~\cite{Komar17}, it is called a global or topological dimension $d_\text{global}=d_J$. 
The quantum dimension of an anyon \mbox{$J=(C_g,\Gamma^{\mathcal{N}_g})\in \mathcal{Z}(\vecg)$} is given by
%
%
\be 
 d_J = |\Gamma^{\mathcal{N}_g}||C_g|,
\label{eq:quantum_dimension_anyon}
\ee
%
%
 where $|{\Gamma^{\mathcal{N}_g}}|$ is the dimension of the irreducible representation and $|C_g|$ is the cardinal of the conjugacy class. For chargeons, $d_J = |\Gamma^{\mathcal{N}_g}|$, and for fluxons, $d_J=|C_g|$. The total quantum dimension of the Drinfeld center $\mathcal{Z}(\vecg)$ is
%
%
\be  
\mathcal{D}=\sqrt{\sum_{J \in \mathcal{Z}(\vecg)} d_J^2}=|G|.
\label{eq:totqudim}
\ee
%
%

Fusion rules of the anyons are described by fusion matrices $N_A$: 
%
%
\be  
A \times B = \sum_{D \in \mathcal{Z}(\vecg)} N_{A B}^D\, D,  \label{eq:NABC}
\ee
%
%
where the matrix element $[N_A]_{B,D}=N_{A B}^D$ and $A,B,D\in \mathcal{Z}(\vecg)$ (we avoid the letter $C$ to denote anyons, as we use it extensively for conjugacy classes). 
Mutual exchange statistics of the anyons are encoded in the $S$-matrix, which is unitary and symmetric. In particular, its first row has a simple expression in terms of quantum dimensions: $S_{\mathbf{1}, A} = d_A/\mathcal{D}$. Additionally, this matrix diagonalizes the fusion matrices, which leads to the Verlinde equation~\cite{Verlinde88}: 
%
%
\be  N_{A B}^D = \sum_{J \in \mathcal{Z}(\vecg)} \frac{S_{A,J}S_{B,J}S_{D,J}^*}{S_{\mathbf{1},J}}. 
\label{Verlinde}
\ee
%
%

\subsubsection{Local degrees of freedom: Subtypes and internal multiplicities}
In the KQD model, which is a lattice realization of a TQFT, on top of the topological dimension discussed in the previous section, the subspace associated with an anyon $J$ also has a local dimension $d_\text{local}$ equal to its internal multiplicity (or number of subtypes) $n_J$ \cite{Kitaev03}, which we explain in the following. \\
An excitation $(C_g,\Gamma^G)$ such that $|\Gamma^G|>1$ has an internal multiplicity, i.e., it can exist in several internal states called subtypes and corresponding to local degrees of freedom. 
For the elementary excitations such as vertex and plaquette, the number of subtypes is directly given by the dimension of the irrep $\Gamma^G$~\cite{Komar17}: vertex excitations $(C_e,\Gamma^G)$ have $|\Gamma^G|$ subtypes, whereas plaquette excitations $(C_g,\Gamma_1^G)$ do not have subtypes, as $|\Gamma_1^G|=1$. It is convenient to gather these internal multiplicities for vertex and plaquette excitations  into two vectors ($\vec{n}^\text{Ve}$ and $\vec{n}^\text{Pl}$) with components
\beqn 
n_J^\text{Ve}&=&d_J \, \delta_{J\in \underline{\text{Ch}}},
\label{eq:nJVe} \\
n_J^\text{Pl}&=& \delta_{J\in \underline{\text{Fl}}},
\label{eq:nJPl}
\eeqn
%
%
labeled by $J\in \mathcal{Z}(\vecg)$, following Eq.~\eqref{eq:map}. Note that the above definitions also include the vacuum as we use the notations introduced in Eq.~\eqref{eq:underline}. These vectors, whose components are integers, satisfy the following identities (see, e.g., Ref.~\cite{Ritz24_1}):
%
%
\beqn  
S\:\vec{n}^{\text{Ve}}&=& \vec{n}^{\text{Ve}}, \nonumber \\
S\:\vec{n}^{\text{Pl}}&=& \vec{n}^{\text{Pl}}.
\label{eq:Smatrixrelations}
\eeqn
%
%

The notion of internal multiplicity (or subtypes) can be extended from excitations to anyons. The internal multiplicity $n_J$ of an anyon $J \in \mathcal{Z}(\vecg)$ is obtained through the fusion of a vertex excitation $A$ and a plaquette excitation $B$ as
\beqn
n_J &=& \sum_{A, B \in \mathcal{Z}(\vecg)} N_{A B}^J \:n_{A}^{\textnormal{Ve}}\:n_{B}^{\textnormal{Pl}},
 \label{eq:dimJ}
 \eeqn
where, in the fusion coefficient $N_{A B}^J$, $A$ and $B$ are simple objects of the Drinfeld center, following Eq.~\eqref{eq:map}. 
 Using the Verlinde formula~\eqref{Verlinde} together with Eq.~\eqref{eq:Smatrixrelations}, one can show that Eq.~\eqref{eq:dimJ} becomes
%
%
\be  
n_J = d_J. 
\label{eq:NJdJ}
\ee
%
%
The internal multiplicity $n_J$ of an anyon $J$ therefore equals its quantum dimension $d_J$ (this is different in string-net models, see Refs.~\cite{Ritz24_1,Soares25} and Appendix \ref{app:sn}). Another derivation of this equality is given in Appendix~\ref{subsubsec_tensor_products}. As a consequence, the total dimension of the subspace associated with an anyon $J$ is the product of the global quantum dimension $d_{\text{global}} = d_J$ and the local dimension $d_{\text{local}}=n_J$~\cite{Komar17}:
%
%
\be 
d_{\text{global}} \times d_{\text{local}} = d_J^2.
\ee
%
%
In particular, when $G$ is a non-Abelian group, a fluxon may have a quantum dimension larger than $1$. Then, a fluxon has several subtypes and the plaquette excitation corresponds to only one of its subtypes. The precise relation between a plaquette excitation and a fluxon is actually a relation between the irreps of the group and the irreps of its centralizer subgroups. This issue is discussed in detail in Ref.~\cite{Komar17} and relies on the notions of induced irreps and restricted irreps. Finally, note that the elements of Si correspond to other subtypes of fluxons or subtypes of dyons (for an illustration, see Appendix~\ref{app:examples}).

\section{Degeneracies}
\label{sec:degen}

We now turn to compute the degeneracies of the excited states of the KQD model. We obtain those by counting the number of configurations of plaquette and vertex excitations. Through the mapping between elementary excitations and anyons [see Eq.~\eqref{eq:map}], our formulas involve only a subset of the anyons of the Drinfeld center, namely chargeons and fluxons. Dyons never appear explicitely in the formula, as they can be obtained through combinations of vertex and plaquette excitations. The main physical insight of this section is contained in Eqs.~\eqref{eq:kqdMS} and~\eqref{eq:kqdMSinternaldeg}, the proof of which is provided in Appendixes~\ref{app:benoit} and \ref{app:steve}. The final expression for the energy-level degeneracies is given in Eq.~\eqref{eq:deg_final}. 

\subsection{Fusion channel degeneracy}

Let us consider a two-dimensional closed orientable manifold of genus $\genus$ with $n$ vertex excitations, $A_1,..., A_n$ $\in \textnormal{Ch}$, and $m$ plaquette excitations, $B_1,...,B_m \in \textnormal{Fl}$. We start by discussing two simple cases.

First, if $\genus=0$ and if there are non-Abelian anyons among these excitations, this information is insufficient to uniquely specify a quantum state in the Hilbert space, and there is a degeneracy associated with the possibility of having multiple fusion channels, with the constraint that ``everything fuses to the vacuum". 

Second, if there are no excitations, it is a well-known feature of TQFTs that such systems have a ground-state degeneracy which depends on the genus $\genus$ and is given by~\cite{Verlinde88} 
\begin{equation}\label{eq:gsd0}
    \textnormal{dim}(\genus;0,0)= \sum_{A\in \mathcal{Z}(\vecg)}S_{\mathbf{1},A}^{2-2\genus}.
\end{equation}
This is due to the fact that there can be non-trivial anyonic fluxes going through the $\genus$ handles of the surface. 

\begin{figure}[t]
  \centering 
  \includegraphics[scale=0.8]{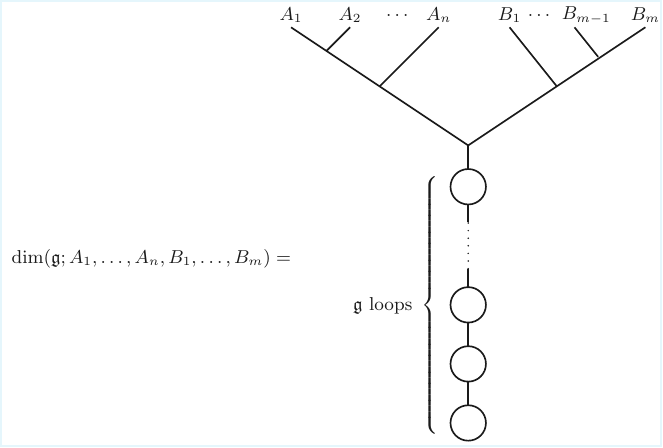}
\caption{Fusion tree for degeneracies for a eigenenergy level of the KQD model defined on a surface of genus $\genus$, with $n$ vertex excitations $A_1, A_2, \ldots, A_n\in$ Ch and $m$ plaquette excitations $B_1, B_2, \ldots, B_m \in$ Fl. Each vertex excitation $A$ can exist in $n_{A}^\text{Ve} = d_A$ subtypes.}
\label{fig:fusiontreekqd}
\end{figure}

Therefore, combining the two previous situations and considering the general case of excitations on a manifold with $\genus\geqslant 1$, one thus has to count the possible fusion channels arising from fusing together the excitations and the anyons in the $\genus$ handles. This is equivalent to counting the number of ways to label a fusion diagram such as the one shown in Fig.~\ref{fig:fusiontreekqd}, where the fusion rules of $\mathcal{Z}(\vecg)$ are respected at every vertex (see also Ref.~\cite{Ritz24_1} for a similar discussion). Such a diagram can be restructured as long as the number of fusion vertices is conserved. On Fig.~\ref{fig:fusiontreekqd}, for convenience, we have represented all vertex excitations on one side and all plaquette excitations on the other, even though this does not reflect their positions on the lattice. Using the Moore-Seiberg-Banks formula~\cite{Moore89}, one can then obtain the fusion channel degeneracy:
%
%
\begin{align}
&\textnormal{dim}(\genus; A_1, ..., A_n, B_1, ..., B_m)\nonumber \\ &= \sum_{A\in \mathcal{Z}(\vecg)}\left [ \prod_{j=1}^n S_{A_j, A} \prod_{k=1}^m S_{B_k, A}\right ] S_{\mathbf{1},A}^{2-2\genus-(m+n)}.
\label{eq:kqdMS}
\end{align} 
%
%
A local Hamiltonian cannot lift such a degeneracy, which is usually called a topological degeneracy. 
%
%
%
%
\subsection{Internal multiplicities}
The KQD Hamiltonian $H$ does not distinguish between the different subtypes of vertex or plaquette excitations. Therefore, to obtain the full degeneracy for a state with $n$ excited vertices and $m$ excited plaquettes, the internal degeneracy $n_{A}^\text{Ve}=d_A$ of the vertex excitations (the plaquette excitations have only one subtype $n_{A}^\text{Pl}=1$), must be taken into account. Therefore, the number of states with vertex excitations $A_1,...,A_n$ and plaquette excitations $B_1,...,B_m$ is
%
%
\beqn
&&\textnormal{dim}(\genus; A_1, ..., A_n, B_1, ..., B_m) \prod_{j=1}^n n_{A_j}^\text{Ve} \prod_{k=1}^m n_{B_k}^\text{Pl}\nonumber \\
&&= \textnormal{dim}(\genus; A_1, ..., A_n, B_1, ..., B_m) \prod_{j=1}^n d_{A_j} .
\label{eq:kqdMSinternaldeg}
\eeqn
%
%
This additional degeneracy, which stems from the internal multiplicities of vertex excitations, can be lifted by a local perturbation of the Hamiltonian. It is therefore non-topological.

Equations~\eqref{eq:kqdMS} and \eqref{eq:kqdMSinternaldeg} constitute the main result of this section. We provide two different proofs of this result in the Appendixes~\ref{app:benoit} and \ref{app:steve}. We also performed numerical exact diagonalization of the Hamiltonian for the $\mathbb{Z}_2$ and $S_3$ groups on small systems of different topologies to check the predicted degeneracies. 
The main features of this formula are:\\ 
(i) it corresponds to a fusion process in which ``everything fuses to the vacuum,'' and which is represented by a fusion tree (see Fig.~\ref{fig:fusiontreekqd}); \\
(ii) extremities of the tree are either vertex excitations ($A_1,...,A_n$), plaquette excitations ($B_1,...,B_m$) or noncontractible cycles ($\genus$); \\
(iii) there are internal multiplicities for the vertex excitations only.

In the following, we use this formula to compute the degeneracy of the energy levels.

\subsection{Energy-level degeneracy}
Up to combinatorial factors $\begin{pmatrix} N_{\rm v} \\ n \end{pmatrix}$ and $\begin{pmatrix} N_{\rm p} \\ m \end{pmatrix}$, that account for the choice for placing the vertex and plaquette excitations on the lattice, the full degeneracy of an energy level $E_{n,m}$ [see Eq.~\eqref{eq:kqdenergy}] is thus given by 
%
%
\begin{align}
\nonumber &D_G(\genus, n, m) \\
&= \nonumber \sum_{A_1, ..., A_n \in \text{Ch}}\sum_{B_1, ..., B_m \in \text{Fl}} \text{dim}(\genus; A_1, ..., A_n, B_1, ..., B_m)\\
&\nonumber \times \prod_{j=1}^{n}
n^\text{Ve}_{A_j}\prod_{k=1}^{m}
n^\text{Pl}_{B_k}, \\
&=\nonumber\sum_{A  \in  \mathcal{Z}(\vecg)}S_{\mathbf{1},A}^{2-2\genus-(n+m)}  \\ 
&\times \prod_{j=1}^{n} \bigg( \sum_{A_j \in \text{Ch}}S_{A,A_j} 
n^\text{Ve}_{A_j}\bigg)  \prod_{k=1}^{m} \bigg( \sum_{B_k \in \text{Fl}} S_{A,B_k} 
n^\text{Pl}_{B_k}\bigg).
\label{eq:kqddeg1}
\end{align}
%
%
The multiplicities $n^\text{Pl}_{B_k}$ are simply $1$ if $B_k \in$ Fl and $0$ otherwise. They are made explicit in view of simplifying the expression. Using Eq.~\eqref{eq:Smatrixrelations}, one further has 
%
%
\be  
\sum_{A_j \in {\rm Ch}}S_{A,A_j} n^\text{Ve}_{A_j}  = \sum_{A_j \in \underline{\rm Ch}}S_{A,A_j} n^\text{Ve}_{A_j} - S_{A, \mathbf{1}} = n^\text{Ve}_{A} - S_{A, \mathbf{1}}, 
\label{eq:raco}
\ee
%
%
for chargeons, and similarly for the fluxons. 
Therefore, we obtain:
%
%
\be 
D_G(\genus, n, m) =  \sum_{A  \in  \mathcal{Z}(\vecg)}S_{\mathbf{1},A}^{2-2\genus} 
\bigg(\frac{n^\text{Ve}_{A}}{S_{\mathbf{1}, A}}-1\bigg)^{n}
\,
\bigg( \frac{n^\text{Pl}_{A}}{S_{\mathbf{1}, A}}- 1\bigg)^{m}. 
\label{eq:kqddeg2}
\ee
%
%
%
To further simplify the expressions, we use Eqs.~\eqref{eq:nJVe}-\eqref{eq:nJPl},  and we introduce the integers $q_A^\text{Ve}$ and $q_A^\text{Pl}$ defined by
\beqn
q_A^\text{Ve} &=& \frac{n^\text{Ve}_{A}}{S_{\mathbf{1}, A}}=|G| \, \delta_{A \in \underline{\text{Ch}}} = S_{\mathbf{1},\mathbf{1}}^{-1} \, \delta_{A \in \underline{\text{Ch}}}, \label{eq:qA1}\\
q_A^\text{Pl}    &=& \frac{n^\text{Pl}_{A}}{S_{\mathbf{1}, A}}=\frac{|G|}{d_A} \, \delta_{A \in \underline{\text{Fl}}} = S_{\mathbf{1},A}^{-1} \, \delta_{A \in \underline{\text{Fl}}}.
\label{eq:qA2}
\eeqn
The choice of the letter $q$ for these quantities will be explained in Sec.~\ref{sec:pf}. Then, we obtain the final compact expression
%
%
\be 
D_G(\genus, n, m) =  \sum_{A  \in  \mathcal{Z}(\vecg)}S_{\mathbf{1},A}^{2-2\genus} 
(q_A^\text{Ve}-1)^{n}
(q_A^\text{Pl}- 1)^{m}. 
\label{eq:deg_final}
\ee
%
%
In particular, we recover the well-known formula of the ground-state degeneracy (GSD)
%
%
\be 
D_G(\genus, 0, 0) =  \sum_{A  \in  \mathcal{Z}(\vecg)}S_{\mathbf{1},A}^{2-2\genus},
\label{eq:gsd}
\ee
%
%
coming from conformal field theory~\cite{Verlinde88}, see Eq.~\eqref{eq:gsd0}.

\subsection{Hilbert space dimension}
We now want to show that our formula for the degeneracies Eq.~\eqref{eq:deg_final} correctly yields the Hilbert space dimension of the system. To demonstrate this, we sum over all possible numbers of plaquette or vertex excitations: 
%
%
\begin{align}
  \dim \mathcal{H} & = \sum_{n=0}^{N_{\rm v}}\sum_{m=0}^{N_{\rm p}} \begin{pmatrix} N_{\rm v} \\ n \end{pmatrix}
  \begin{pmatrix}N_{\rm p} \\ m\end{pmatrix} 
  D_G(\genus, n, m), \nonumber\\
   &= \sum_{A\in \mathcal{Z}(\vecg)}S_{\mathbf{1}, A}^{2-2\genus} \,\, \left( q_A^\text{Ve}\right)^{N_{\rm v}} 
  \left( q_A^\text{Pl}\right)^{N_{\rm p}}, \nonumber \\
  &= S_{\mathbf{1},\mathbf{1}}^{2-2\genus-N_{\rm p}-N_{\rm v}}, \nonumber \\
  &= |G|^{N_{\rm l}},
  \label{Hilbert}
\end{align}
%
%
where we used the fact that $1/S_{\mathbf{1},\mathbf{1}}=\mathcal{D}=|G|$ [see Eq.~\eqref{eq:totqudim}], as well as  Eq.~\eqref{eq:Euler}. To go from the second to the third line, we used the fact that $n^\text{Ve}_{A}$ and $n^\text{Pl}_{A}$ are only simultaneously nonvanishing for $A=\mathbf{1}$. This expression matches with Eq.~\eqref{eq:dimHKQD}. 

Although vertex and plaquette excitations can be treated mostly independently, Eq.~\eqref{Hilbert} shows that the Hilbert space does not factorize into plaquette and vertex parts (even if $\genus=0$ or $1$). This is due to the global constraint that all plaquette and vertex excitations together with the anyon stemming from the fusion of the $\genus$ handles must fuse to the vacuum (see Fig.~\ref{fig:fusiontreekqd}).

\section{Partition function}
\label{sec:pf}
%
\subsection{KQD for an arbitrary group}

The partition function at temperature $T$ of the KQD model can be derived directly from the formula for the spectral degeneracies. Indeed, one has
%
%
\beqn
Z &=& \sum_{n=0}^{N_{\rm v}}\sum_{m=0}^{N_{\rm p}} 
 \begin{pmatrix} N_{\rm p} \\ m\end{pmatrix}
 \begin{pmatrix} N_{\rm v} \\ n \end{pmatrix} 
 D_G(\genus, n, m) \: {\rm e}^{-\beta E_{n,m}},\nonumber \\ 
&=&\sum_{A\in \mathcal{Z}(\vecg)} S_{\mathbf{1},A}^{2-2\genus} \left(z_A^\text{Ve} \right)^{N_{\rm v}}\!\! \left(  z_A^\text{Pl} \right)^{N_{\rm p}},
\label{eq:kqdPF}
\eeqn
%
%
where $\beta=1/T$, and where we introduced
%
%
\beqn
z_A^\text{Ve} 
&=& \sum_{B\in \underline{\text{Ve}}} \frac{S_{B,A}}{S_{\mathbf{1},A}} \: d_B \: {\rm e}^{-\beta E_B^\text{Ve}}
= q_A^\text{Ve}-1+{\rm e}^{\beta J_{\rm v}}, \nonumber \\
z_A^\text{Pl} 
&=& \sum_{B\in \underline{\text{Pl}}} \frac{S_{B,A}}{S_{\mathbf{1},A}}  \:\:\:\:\:{\rm e}^{-\beta E_B^\text{Pl}}
= q_A^\text{Pl}-1+{\rm e}^{\beta J_{\rm p}},
\label{eq:zA}
\eeqn
%
%
with $E_B^\text{Ve}=-J_\text{v} \: \delta_{B,\mathbf{1}}$ and $E_B^\text{Pl}=-J_\text{p} \: \delta_{B,\mathbf{1}}$. Here, we used Eq.~\eqref{eq:Smatrixrelations} and the $q$'s defined in Eqs.~\eqref{eq:qA1} and \eqref{eq:qA2}. The two $z$'s play the role of partition functions for a single vertex ($z_A^\text{Ve}$) or a single plaquette ($z_A^\text{Pl}$) with the constraint that the total topological charge is $A$ (the true partition function for one vertex or plaquette would correspond to $A=\mathbf{1}$). They are similar to the partition function, $z=q-1+{\rm e}^{\beta J}$, of a single $q$-state classical spin $\sigma=1,...,q$, with Hamiltonian $H=-J\delta_{\sigma,1}$. Among anyons of the Drinfeld center, we see that chargeons and fluxons (unlike dyons) play a special role in the partition function. 

One can easily check that in the infinite temperature limit ($\beta = 0$), the partition function~\eqref{eq:kqdPF} yields the Hilbert space dimension [see Eq.~\eqref{Hilbert}], whereas, in the zero-temperature limit ($\beta=\infty$), one has
%
%
\be
\lim_{\beta \to \infty} Z \: {\rm e}^{\beta E_0}=\sum_{A\in \mathcal{Z}(\vecg)} S_{\mathbf{1},A}^{2-2\genus},
\ee
%
%
which is the GSD [see Eq.~\eqref{eq:gsd}], with $E_0$ given in Eq.~\eqref{eq:gse2}.

In the thermodynamic limit , the vacuum term is dominant in the sum of Eq.~\eqref{eq:kqdPF} because it maximizes both $q_A^\text{Ve}$ and $ q_A^\text{Pl}$  ($q_\mathbf{1}^\text{Ve}=q_\mathbf{1}^\text{Pl}=|G|$), and one gets
%
%
\be
\label{eq:kqdlimitlarge}
\underset{N_{\rm v} \to \infty}{\lim_{N_{\rm p} \to \infty}} Z= |G|^{2\genus-2} 
\left(|G|-1+{\rm e}^{\beta J_{\rm v}}\right)^{N_{\rm v}} \!\left(|G|-1+{\rm e}^{\beta J_{\rm p}}\right)^{N_{\rm p}}\!\!.
\ee
%
%
Contrary to the finite-size partition function Eq.~\eqref{eq:kqdPF}, this infinite-size partition function fully factorizes into a term for vertices, a term for plaquettes, and a term for the genus of the surface. In this limit, apart from a global constant $|G|^{2\genus-2}$ of topological origin, it is similar to the partition function of two decoupled sets of $N_\text{v}$ and $N_\text{p}$ independent spins with $q=|G|$ colors.

Thermodynamic consequences are easy to derive from Eq.~\eqref{eq:kqdlimitlarge} (see Ref.~\cite{Ritz24_2} for more details). The average energy is
\beqn
\langle E\rangle =  E_0 &+& \frac{N_{\rm v} \: J_{\rm v}}{{\rm e}^{\beta J_{\rm v}}(|G|-1)^{-1} +1} \nn \\
&+& \frac{N_{\rm p}\:  J_{\rm p}}{{\rm e}^{\beta J_{\rm p}}(|G|-1)^{-1} +1},
\eeqn
where $E_0$ is the ground-state energy \eqref{eq:gse2} and in which $|G|-1$ plays the role of a fugacity $e^{\beta \mu}$ in an effective Fermi-Dirac distribution $(e^{\beta J}e^{-\beta \mu}+1)^{-1}$. The corresponding specific heat $C=-\beta^2 \partial \langle E\rangle/\partial \beta$ features two maximas, known as Schottky anomalies, as a function of temperature. The entropy is
\beqn
S &=& N_{\rm v} \left[\ln(|G|-1+{\rm e}^{\beta J_{\rm v}})-\frac{\beta \: J_{\rm v}\: {\rm e}^{\beta J_{\rm v}}}{|G|-1+{\rm e}^{\beta J_{\rm v}}}\right] \nn \\
&+& N_{\rm p} \left[\ln(|G|-1+{\rm e}^{\beta J_{\rm p}})-\frac{\beta \: J_{\rm p}\: {\rm e}^{\beta J_{\rm p}}}{|G|-1+{\rm e}^{\beta J_{\rm p}}}\right] \nn \\
&+& (2\genus-2)\ln |G|.
\eeqn
In the high-temperature limit it goes to $\ln(\text{dim } \mathcal{H})$ with $\text{dim } \mathcal{H} = 2^{N_{\rm l}}$, as expected. In the low-temperature limit, it tends to $\ln |G|^{2\genus-2}$, which is strictly smaller than the expected $\ln(\text{GSD})$ with GSD given in Eq.~\eqref{eq:gsd}. This is due to the fact that the $N_{\rm v}, N_{\rm p} \to \infty$ limit does not commute with the $\beta \to \infty$ limit. This noncommutativity is related to a phase transition at $T=0^+$. In the thermodynamic limit, the (topological) order is only present at $T=0$ and is destroyed by any finite temperature, similarly to the one-dimensional Potts model (see the corresponding discussion in the Appendix D of Ref.~\cite{Ritz24_2}).

\medskip

In Appendix~\ref{app:examples}, we give the partition function for two examples of groups: the Abelian $\mathbb{Z}_N$ and the non-Abelian $S_3$. 

\subsection{Local commuting projectors Hamiltonians}
We observe that the structure of the partition function found for the KQD model [see Eq.~\eqref{eq:kqdPF}] is the same as for  the string-net model or its extended version with tails (see Appendix~\ref{app:sn} and Refs.~\cite{Ritz24_2,Soares25}). For these local commuting projectors Hamiltonians, the partition function schematically reads
%
%
\beqn 
Z &=& \sum_{A\in \mathcal{Z(C)}} Z_A^{(\genus)} \,\,  Z_A^{(\text{Ve})} \,\, Z_A^{(\text{Pl})}, \nn \\
&=&  \sum_{A\in \mathcal{Z(C)}} S_{\mathbf{1},A}^{2\genus-2}\,\,  (z_A^\text{Ve})^{N_\text{v}}\,\,  (z_A^\text{Pl})^{N_\text{p}}.
\label{eq:generalZ}
\eeqn
%
%
In general, it does not factorize as $Z^{(\genus)} \times Z^{(\text{Ve})} \times Z^{(\text{Pl})}$, except in the thermodynamic limit, where $A=\mathbf{1}$ dominates the sum [see Eq.~\eqref{eq:kqdlimitlarge}] and in the Abelian case (see the $\mathbb{Z}_N$ toric code in Appendix~\ref{app:examples}). 

The vertices and the plaquettes act as two sets of independent spins (with $q_A$ colors), i.e., they have a structure $Z_A=(z_A)^N$, except for a a global topological constraint that involves the $\genus$ handles and reflects the structure of the fusion tree. The individual partition functions have the structure
%
%
\be
z_A=q_A-1+{\rm e}^{\beta J},
\ee
%
%
where $q_A = n_A/S_{\mathbf{1},A} = \mathcal{D} \: n_A /d_A$ is the effective number of states of a single ``emergent spin,"which may not be an integer. The $n_A$'s are nonnegative integer, which gives the number of subtypes (or internal multiplicity) of the anyon type $A$ and $J$ stands for an energy $J_\text{v}$ or $J_\text{p}$. In general, $n_A$ is different from $d_A$.

When vertex excitations are forbidden (large-$J_\text{v}$ limit), as is usually the case in string-net models, one has \mbox{$z_A^\text{Ve}\underset{J_\text{v} \to \infty}{\simeq} e^{\beta J_\text{v}}$}, so that this quantity becomes independent of $A$ and it only contributes to an irrelevant global factor $e^{\beta J_\text{v} N_\text{v}}$ in the partition function \eqref{eq:generalZ}.

\section{Conclusion}
In this article, we have obtained the degeneracy of all energy levels of the KQD model [either in its original version, see Eq.~\eqref{eq:KQDham}, or in a refined one, see Eq.~\eqref{eq:Rham}], from which we have derived a closed form for the partition function \eqref{eq:kqdPF}. The main tool is a fusion tree describing the fusion channel degeneracy of vertex and plaquette excitations (see Fig.~\ref{fig:fusiontreekqd}), together with the identification of a non-topological internal multiplicity relevant for vertex excitations. Two detailed proofs of the degeneracy formula are given in Appendixes. For a given initial group $G$, the resulting partition function is found to be similar to that of two decoupled one-dimensional Potts models with $q=|G|$ states. In the thermodynamic limit, the topological order is lost at any non-zero temperature. It is preserved in a finite-size system at temperatures below a crossover temperature that scales as $T^*\propto 1/\ln L$.

The present work opens several perspectives. One could consider the case of nonclosed surfaces with boundaries, of closed but non-orientable surfaces, or even of inhomogeneous systems with an interface between two different models. Another direction, that we are currently following, is to compute other observables at finite temperature such as the entanglement entropy, the mutual information and the Wegner-Wilson loops. They should provide an alternative viewpoint on the loss of topological order at finite temperature. It would also be interesting to use the same machinery to study related models such as the twisted KQD model~\cite{Hu13} or the fermionic toric code~\cite{Gu14}. In the latter case, as it is still an exactly solvable model, we expect that a closed formula for the partition function could be obtained, but that, in contrast to the (bosonic) toric code, it should not correspond to that of any classical statistical mechanics model.
The present approach might be extended to other local commuting projector Hamiltonians, which, by construction, can only give rise to achiral topological phases~\cite{Hastings11}. For two-dimensional chiral phases, the spectrum is usually more complex (see, e.g., Kitaev Honeycomb model~\cite{Kitaev06}), and we do not know any model for which the partition function can be computed exactly.

\acknowledgements

We acknowledge fruitful discussions with Sofyan Iblisdir, Andrej Mesaros and Zohar Nussinov. SHS acknowledges support from EPSRC Grant No. EP/X030881/1.

\appendix

\section{Refined Hamiltonian}
\label{app:refined}
The KQD Hamiltonian ~\eqref{eq:KQDham} considered in the main part attributes the same energy to all vertex excitations $\in$ Ve and to all plaquette excitations $\in$ Pl. However, it is possible to distinguish between the different elements of Ve and Pl by introducing a refined Hamiltonian~\cite{Komar17}, which attributes a different energy to every \textit{energy eigenspace}. Here, we generalize the results of the main part to this refined Hamiltonian. 

\subsection{Hamiltonian and energy levels}
The refined Hamiltonian~\cite{Komar17} reads
%
%
\be
H_R=\sum_v \sum_{j=0}^M \alpha_j A_v^{\Gamma_j}+\sum_p \sum_{k=0}^M \beta_{k}B_p^{C_k},
\label{eq:Rham}
\ee
%
%
where $M+1$ is the total number of irreps, which is also the total number of conjugacy classes, of the group $G$. 
The operator $A^{\Gamma_j}_v$ is a generalization of the vertex operator defined in Eq.~\eqref{eq:trivirrep}, which now projects onto any irrep $\Gamma_j$ of $G$ : \be
A_v^{\Gamma_j} = \frac{|\Gamma_j|}{|G|} \sum_{h \in G} \chi_{\Gamma_j} (h) A_v^{h},
\label{eq:anyirrep}
\ee 
where $\chi_{\Gamma_j}(h)$ is the character of the irrep (for the group element $h$) and $|\Gamma_j|$ is its dimension. Note that here, we use an index $j=0,...,M$ to label the irreps $\Gamma_j$ of the group $G$, the trivial irrep being $\Gamma_0$ (called $\Gamma_1^G$ in the main text). 
These operators are orthonormal projectors, as can be proven using the grand othogonality theorem [Eq.~\eqref{eq:grand}] between irreps~\cite{Komar17}:
%
%
\be
A_v^{\Gamma_j} A_v^{\Gamma_k} = A_v^{\Gamma_j} \delta_{j, k}.
\label{eq:Aproj}
\ee
%
%
In a similar way, the operators $B_p^{C_k}$ are a generalized version of the plaquette operator $B_p$ [Eq.~\eqref{eq:Bp}] that project on any conjugacy class $C_k$ of $G$ (here also, we label conjugacy classes with an index $k=0,...,M$ and the trivial conjugacy class is $C_0$, instead of $C_e$ as in the main text):
%
%
\be
    B_p^{C_k}=\sum_{h\in C_k} B_p^h \label{eq:cc} .
\ee
%
%
These operators are also orthonormal projectors:
%
%
\be
B_p^{C_k} B_p^{C_m} = B_p^{C_k} \delta_{k,m} .
\ee
%
%
The Hamiltonian $H_R$ involves several coupling constants $\alpha_{j}$ and $\beta_{k}$ that allow one to distinguish the different energy eigenspaces $(C_0, \Gamma_j)\equiv \Gamma_j\in \underline{\textnormal{Ve}}$ and \mbox{$(C_k, \Gamma_0)\equiv C_k \in \underline{\textnormal{Pl}}$}. 

For simplicity, we assume that the coupling constants are chosen such that $\delta \alpha_j \equiv \alpha_j - \alpha_0\geqslant 0$ and \mbox{$\delta \beta_k \equiv \beta_k - \beta_0 \geqslant 0$}, so that each ground state $|\psi\rangle$ is such that $A_v^{\Gamma_j} |\psi\rangle=\delta_{j,0}|\psi\rangle$, and $B_p^{C_k}|\psi\rangle=\delta_{k,0}|\psi\rangle$. Hence, the ground-state energy is given by
%
%
\be
E_0 = N_\text{v} \, \alpha_{0} + N_\text{p} \, \beta_{0}.
\label{eq:gse}
\ee
%
%
The energy of an excited level is specified by giving the number of vertices ${\{n_j\}}$ in each non-trivial irrep $\Gamma_{j \geqslant 1}$, and the number of plaquettes ${\{m_k\}}$ in each non-trivial conjugacy class $C_{k \geqslant 1}$ ($n_0$ and $m_0$ are, respectively, the number of unexcited vertices and the number of unexcited plaquettes), and reads
%
%
\beqn
E_{\{n_j,m_k\}} &=& \sum_{j=0}^M \alpha_j \, n_j +  \sum_{k=0}^M \beta_k \, m_k, \\
&=& E_0 + \sum_{j=1}^M \delta \alpha_j \, n_j +  \sum_{k=1}^M \delta \beta_k \, m_k.
\label{eq:renergy}
\eeqn
%
%
The total number of vertices is the sum of $n_0$ and the number $n$ of excited vertices
%
%
\be
N_\text{v} = n_0+ \sum_{j=1}^M n_j = n_0 + n.
\label{eq:n}
\ee
%
%
Similarly, the total number of plaquettes reads 
%
%
\be
N_\text{p} = m_0+ \sum_{k=1}^M m_k = m_0 + m.
\label{eq:m}
\ee
%
%
One recovers the Kitaev Hamiltonian $H$ of Eq.~\eqref{eq:KQDham} for the specific choice of coupling constants
%
%
\be
\alpha_j = -J_\text{v} \delta_{j,0} \text{ and }\beta_k = -J_\text{p} \delta_{k,0},
\ee
%
%
corresponding to
%
%
\be
\delta \alpha_j =J_\text{v} \text{ and } \delta \beta_k = J_\text{p},
\ee
%
%
when $j$ and $k=1,...,M$.

\subsection{Degeneracies}
The number of possibilities of placing $n_1$ vertex excitations of type $\Gamma_1$, $n_2$ of type $\Gamma_2$, ..., $m_1$ plaquette excitations of type $C_1$, $m_2$ of type $C_2$, ...,  on the lattice is
%
%
\be
\frac{N_\text{v} !}{n_0! n_1! .... n_M!} \frac{N_\text{p} !}{m_0! m_1! .... m_M!}.
\ee
%
%
Up to this combinatorial factor, and assuming that there is no accidental degeneracy (i.e., that the coupling constants $\delta \alpha_j$ and $\delta \beta_k$ are incommensurate), the degeneracy of an energy level $E_{\{n_j,m_k\}}$ [see Eq.~\eqref{eq:renergy}] is obtained by applying Eqs.~\eqref{eq:kqdMS} and~\eqref{eq:kqdMSinternaldeg} to a situation with $n_j$ vertex excitations of type $\Gamma_j \equiv (C_0, \Gamma_j)$ and $m_k$ plaquette excitations $C_k \equiv (C_k, \Gamma_0)$ ($1 \leqslant j,k \leqslant M$):
\begin{align}
& D_G(\genus, \{n_j, m_k\}) 
=  \prod_{j=1}^{M} d_j^{n_j} \\ \nonumber
& \times \text{dim}\Bigg(\!\!\genus;\ 
\underbrace{\Gamma_1, .\!.\!., \Gamma_1}_{n_1},.\!.\!., 
\underbrace{\Gamma_M,.\!.\!., \Gamma_M}_{n_M},\underbrace{C_1, .\!.\!.,C_1}_{m_1},.\!.\!., \underbrace{C_M, .\!.\!., C_M}_{m_M}\!\!\Bigg) \\
&=\!\!\! \nonumber \sum_{A\in \mathcal{Z}(\vecg)} \!\!\! S_{\mathbf{1},A}^{2-2\genus-(m+n)} \prod_{j=1}^M \bigg(S_{\Gamma_j, A} \, d_{j}\bigg)^{n_j} \prod_{k=1}^M 
\bigg(S_{C_k, A}\bigg)^{m_k}
,
\end{align}
where $d_j$ is the dimension of the irrep $\Gamma_j$, $m,n$ have been defined in Eqs. \eqref{eq:n} and \eqref{eq:m}, and $\textbf{1} \equiv (C_0, \Gamma_0)$.

Here as well, we recover the GSD given in Eq.~\eqref{eq:gsd} when $n_j=\delta_{0,j}N_\text{v}$ and $m_k=\delta_{k,0}N_\text{p}$.

\subsection{Partition function}
We find that the partition function for the refined Hamiltonian $H_R$ has the same form as Eq.~\eqref{eq:kqdPF} but with
\beqn
z_A^\text{Ve} 
&=& \sum_{j=0}^M \frac{S_{\Gamma_j,A}}{S_{\mathbf{1},A}} d_j {\rm e}^{-\beta \alpha_j},\\
z_A^\text{Pl} 
&=& \sum_{k=0}^M \frac{S_{C_k,A}}{S_{\mathbf{1},A}}  {\rm e}^{-\beta \beta_k}.
\eeqn
Here, $\beta=1/T$ is the inverse temperature, not to be confused with $\beta_k = \beta_{C_k}$, the energy of a plaquette excitation of type $C_k$. 

When $\beta = 0$, using Eq.~\eqref{eq:Smatrixrelations}, the above quantities $z$ reduce to
\beqn
z_A^\text{Ve} 
&=& \sum_{B\in \underline{\text{Ve}}} \frac{S_{B,A}}{S_{\mathbf{1},A}} d_B = |G| \, \delta_{A \in \underline{\text{Ch}}} = q_A^\text{Ve},\\
z_A^\text{Pl} 
&=&\sum_{B\in \underline{\text{Pl}}} \frac{S_{B,A}}{S_{\mathbf{1},A}}  = \frac{|G|}{d_A} \, \delta_{A \in \underline{\text{Fl}}} =
q_A^\text{Pl} 
\eeqn
[see Eqs.~\eqref{eq:qA1} and \eqref{eq:qA2}]. Further using that 
\be
(q_A^\text{Ve})^{N_\text{v}}(q_A^\text{Pl})^{N_\text{p}}=|G|^{N_\text{v}+N_\text{p}}\delta_{A,\mathbf{1}},
\ee
one recovers, in this limit, the Hilbert space dimension given in Eq.~\eqref{Hilbert}.

Similarly, when $\beta \to \infty$, one has
\beqn
z_A^\text{Ve} 
&\simeq& {\rm e}^{-\beta \alpha_0},\\
z_A^\text{Pl} 
&\simeq&  {\rm e}^{-\beta \beta_0},
\eeqn
so that the GSD reads
\be
{\rm e}^{\beta E_0} Z = \sum_{A  \in  \mathcal{Z}(\vecg)}S_{\mathbf{1},A}^{2-2\genus},
  \ee
in agreement with Eq.~\eqref{eq:gsd} [$E_0$ is given in Eq.~\eqref{eq:gse}].

\section{Examples}\label{app:examples}
In this Appendix, we provide examples of quantum double models for two groups:  $\mathbb{Z}_N$ (Abelian) and  $S_3$ (non-Abelian). This will illustrate the relation between energy eigenspaces and anyons and in particular the difference between plaquette excitations and fluxons that exist in the case of a non-Abelian group. We also discuss the corresponding partition functions.

\subsection{$\mathbb{Z}_N$ group}
\subsubsection{$N=2$} 
For the special and simplest case $G=\mathbb{Z}_2$, the KQD model is known as the toric code model~\cite{Kitaev03}. It has two elements $0$ and $1$, with $0$ the trivial object, and $1\times 1 = 0$. As an Abelian group, every element is its own conjugacy class, and all irreducible representations are one dimensional. Thus, there are two conjugacy classes $C_0 =\lbrace 0 \rbrace$ and $C_1 = \lbrace 1 \rbrace$. The corresponding centralizers are $\mathcal{N}_0= \mathcal{N}_1 =\mathbb{Z}_2$. The two irreducible representations of $\mathbb{Z}_2$ are the trivial representation $\Gamma_0$ and the sign representation $\Gamma_{1}$. By pairing up conjugacy classes and irreducible representations, it is straightforward to see that the Drinfeld center $\mathcal{Z}(\text{Vec}_{\mathbb{Z}_2})$, which is also known as the toric code category, has four topological sectors, which are listed in Table~\ref{tab:tc}: the vacuum $\mathbf{1}$, a chargeon $e$ (where the $e$ stands for ``electric charge", in analogy with lattice gauge theory), a fluxon $m$ (where the $m$ stands for ``magnetic flux"), and a dyon $f$ (where $f$ stands for fermion), which is the combination of an $e$ and an $m$. All quantum dimensions are equal to $1$, and the total quantum dimension is $\mathcal{D}=|\mathbb{Z}_2|= 2$. 
\begin{table}[h!]
  \renewcommand{\arraystretch}{1.3}
\begin{center}
\begin{tabular}{|c |c| c| c| c| c| c| c| c| c|} 
\hline
$J=(C_k,\Gamma_j)$ & Set & $|C_k|$ & $|\Gamma_1|$ &  $d_J $ &  $n_{J}^{\textnormal{Ve}}$ & $n_{J}^{\textnormal{Pl}}$ \\ [0.5ex] 
\hline\hline
$\mathbf{1}=(C_0,\Gamma_0)$& Vac & 1 & 1 &  1 & 1 & 1  \\ 
\hline
$e=(C_0,\Gamma_{1})$& Ch & 1 & 1 & 1 &  1 & 0  \\ 
\hline
$m=(C_1,\Gamma_0)$& Fl & 1 & 1 & 1 &  0 & 1 \\ 
\hline
$f=(C_1,\Gamma_{1})$& Dy & 1 & 1  & 1 & 0 & 0 \\ 
\hline
\end{tabular}
\caption{Four anyons $J = (C_k,\Gamma_j)$ of $\mathcal{Z}(\text{Vec}_{\mathbb{Z}_2})$: $C_k$ are the conjugacy classes, and $\Gamma_j$ the irreps of $\mathbb{Z}_2$ with $k,j=0,1$, and $d_J$ the quantum dimensions. We also indicate the internal multiplicities $n_{J}^{\textnormal{Ve}}$ and $n_{J}^{\textnormal{Pl}}$.}
\label{tab:tc}
\end{center}
\end{table}

Using the general formula~\eqref{eq:kqdPF}, we obtain its partition function
%
%
\beqn
Z&=&{\rm e}^{\beta \frac{J_\text{v} N_\text{v}+J_\text{p} N_\text{p}}{2}} \times 2^{2\genus-2}\nonumber \\
&\times& \left[(2\cosh \frac{\beta J_\text{v}}{2})^{N_\text{v}} + (2\sinh \frac{\beta J_\text{v}}{2})^{N_\text{v}}\right]\nonumber \\
&\times& \left[(2\cosh \frac{\beta J_\text{p}}{2})^{N_\text{p}} + (2\sinh \frac{\beta J_\text{p}}{2})^{N_\text{p}}\right].\label{eq:tcp}
\eeqn
%
%

Compared to the usual toric code Hamiltonian~\cite{Kitaev03}
%
%
\be
\tilde{H}=-\tilde{J}_\text{v} \sum_v \tilde{A}_v -\tilde{J}_\text{p} \sum_p \tilde{B}_p,
\ee
%
%
the Hamiltonian $H$ considered in this article [see Eq.~\eqref{eq:KQDham}] is written with projectors ($A_v$ and $B_p$) rather than with involutions ($\tilde{A}_v=2A_v-1$ and $\tilde{B}_p=2 B_p-1$). This shifts the eigenenergies by $(J_\text{v} N_\text{v}+J_\text{p} N_\text{p})/2$ and rescales the couplings ($\tilde{J}_\text{v}=J_\text{v}/2$ and $\tilde{J}_\text{p}=J_\text{p}/2$). Apart from these trivial rescalings, Eq.~\eqref{eq:tcp} matches the one given in Refs.~\cite{Castelnovo07_2, Nussinov08, Iblisdir10}. Note that a similar calculation was performed in Ref.~\cite{Gregor11} for the equivalent lattice gauge theory not restricted to the gauge-invariant subspace. Compared to the toric code model, this difference introduces a global multiplication factor into the partition function, which affects the ground-state degeneracy.

\subsubsection{$N\geqslant 2$} 
\begin{table}[h!]
  \renewcommand{\arraystretch}{1.3}
\begin{center}
\begin{tabular}{|c|c|c|c| c| c| c| c| c| c| c|} 
\hline
$J=(C_k,\Gamma_j)$ & Set & $|\text{Set}|$ & $|C_k|$ & $|\Gamma_j|$ &  $d_J $ &  $n_{J}^{\textnormal{Ve}}$ & $n_{J}^{\textnormal{Pl}}$ \\ [0.5ex] 
\hline\hline
$(C_0,\Gamma_0)$& Vac & 1 & 1 & 1 &  1 & 1 & 1  \\ 
\hline
$(C_0,\Gamma_j)$& Ch & $N-1$ & 1 & 1 & 1 &  1 & 0  \\ 
\hline
$(C_k,\Gamma_0)$& Fl & $N-1$ & 1 & 1 & 1 &  0 & 1 \\ 
\hline
$(C_k,\Gamma_j)$& Dy & $(N-1)^2$ & 1 & 1  & 1 & 0 & 0 \\ 
\hline
\end{tabular}
\caption{$N^2$ anyons $(C_k,\Gamma_j)$ of $\mathcal{Z}(\text{Vec}_{\mathbb{Z}_N})$. Here \mbox{$k,j=1,...,N-1$} labels the non-trivial conjugacy classes $C_k$ and irreps $\Gamma_j$. The vacuum corresponds to the trivial conjugacy class $C_0$ and irrep $\Gamma_0$.}
\label{tab:tczn}
\end{center}
\end{table}
The above example is easily extended to the group $\mathbb{Z}_N$ with $N\geqslant 2$. All anyons $J$ have a quantum dimension $d_J=1$ and the total quantum dimension is \mbox{$\mathcal{D}=|\mathbb{Z}_N|= N$}. The energy eigenspaces of the KQD model correspond exactly to the $N^2$ simple objects of the Drinfeld center $\mathcal{Z}(\text{Vec}_{\mathbb{Z}_N})$. Anyons are labeled by $J=(C_k,\Gamma_j)$ with $k,j=0,...,N-1$, where $C_0$ is the trivial conjugacy class and $\Gamma_0$ the trivial irrep (see Table~\ref{tab:tczn}). There is a unique vacuum $(C_0,\Gamma_0)$, $N-1$ chargeons $(C_0,\Gamma_j)$ with $j\neq 0$, $N-1$ fluxons $(C_k,\Gamma_0)$ with $k\neq 0$ and $(N-1)^2$ dyons $(C_k,\Gamma_j)$ with $j,k\neq 0$. Therefore
\beqn
z_A^\text{Ve} &=& 
N\delta_{k,0}-1+{\rm e}^{\beta J_{\rm v}}, \\
z_A^\text{Pl}    &=& 
N\delta_{j,0}-1+{\rm e}^{\beta J_{\rm p}},
\eeqn
and the partition function \eqref{eq:kqdPF} becomes
%
%
\beqn
Z&=& \sum_{k=0}^{N-1}\sum_{j=0}^{N-1} N^{2\genus-2}
\\
&\times & \left(N\delta_{k,0}-1+{\rm e}^{\beta J_{\rm v}} \right)^{N_{\rm v}} \left(N\delta_{j,0}-1+{\rm e}^{\beta J_{\rm p}} \right)^{N_{\rm p}}\nn \\
&=& N^{2\genus-2} \\
&\times& \left[(N-1+ e^{\beta J_\text{v}})^{N_\text{v}} + (N-1)(-1+e^{\beta J_\text{v}})^{N_\text{v}}\right]\nonumber \\
&\times& \left[(N-1+ e^{\beta J_\text{p}})^{N_\text{p}} + (N-1)(-1+e^{\beta J_\text{p}})^{N_\text{p}}\right].\nonumber\label{eq:tcpN}
\eeqn
%
%
 This partition function factorizes as $Z^{(\genus)} \times Z^{(\text{Ve})} \times Z^{(\text{Pl})}$ and there is a complete symmetry between the vertex and plaquette contributions. Apart from an overall $N^{2\genus-2}$ factor (where $N^{2\genus}$ is the GSD), it is identical to the partition function of two independent one-dimensional $N$-states Potts models with $N_\text{v}$ and $N_\text{p}$ sites (see, e.g., Appendix D in Ref.~\cite{Ritz24_2}). This generalizes from Ising to Potts results found for the partition function of several stabilizer code Hamiltonians, see Table I in Ref.~\cite{Weinstein19}.

\medskip

In all Abelian group examples, plaquette excitations are fluxons. However, this identification does not exactly hold in the case of a non-Abelian group, as shown in the next example.

\subsection{$S_3$ group}
To obtain non-Abelian anyons (i.e., anyons with quantum dimension larger than $1$) through the KQD construction, one needs to start from a non-Abelian (i.e., non-commuting) group. Here, we consider the group $S_3$~\cite{Preskill_notes,Doucot04,Beigi11}, which is the simplest non-Abelian group. It is the symmetry group of the equilateral triangle, comprising the identity $e$, two rotations (or 3-cycles) $y$ and $y^2$ of angle $\pm 2\pi/3$ and, three reflections (mirrors or 2-cycles) $x, xy$ and $xy^2$. The multiplication rules of the group are non-commutative, (e.g., one has $xy^2=yx$). Other relevant relations are $y^3=e$ and $x^2=e$. 
The elements of $S_3$ form three conjugacy classes: 
\begin{align}
 C_e &= \lbrace e \rbrace ,\\
 C_x &= \lbrace x, xy, xy^2 \rbrace,\\
 C_y &= \lbrace y, y^2 \rbrace.
\end{align}
The centralizers of each element are
\begin{align}
 \mathcal{N}_e &= S_3,\\
 \mathcal{N}_y &= \mathcal{N}_{y^2} = \lbrace e, y, y^2 \rbrace \simeq \mathbb{Z}_3, \\
\mathcal{N}_x &= \lbrace e, x \rbrace \simeq \mathcal{N}_{xy} \simeq \mathcal{N}_{xy^2} \simeq \mathbb{Z}_2.
\end{align}
Note that the centralizer of the elements of $C_x$ are not strictly the same, but isomorphic to each other. The group $S_3$ has three irreps, which we will denote $\Gamma^{S_3}_1$ (the trivial representation of dimension $1$), $\Gamma^{S_3}_{-1}$ (the alternating representation, of dimension $1$), and a two-dimensional representation $\Gamma^{S_3}_2$. We do not give the explicit expressions of these irreps here, but they can be found, e.g., in Ref.~\cite{Komar17}. In the following, we first discuss the anyons of the Drinfeld center $\mathcal{Z}(\text{Vec}_{S_3})$ and then the excitations of the lattice model.

\subsubsection{Topological sectors versus energy eigenspaces} 
$\mathcal{Z}(\text{Vec}_{S_3})$ has eight topological sectors, commonly denoted $A, B, C, D, E, F, G, H$ (see, e.g., \cite{Beigi11}). Among these, there is the vacuum $A$ with $d_A=1$, chargeons $B$ and $C$, with quantum dimensions $d_B=1$ and $d_C=2$, fluxons $D$ and $F$ with quantum dimensions $d_D= 3$ and $d_F = 2$, and dyons $E$, $G$, $H$ with quantum dimensions $d_E=d_G=d_H=2$. The total quantum dimension is $\mathcal{D}=|S_3|= 6$. The chargeon $C$ exists in two different subtypes, the fluxon $D$ has three subtypes, and the fluxon $F$ two. However, as $n^{\textnormal{Pl}}_{D} = 1 < d_{D}$ and  $n^{\textnormal{Pl}}_{F} = 1 < d_{F}$, some subtypes of the fluxons $D$ and $F$ are not elementary plaquette excitations, but arise as a combination of a vertex and a plaquette excitation (similarly to what one expects for dyons). These properties are summarized in Table~\ref{table:ds3}. 
\begin{table}[h!]
  \renewcommand{\arraystretch}{1.3}
\begin{center}
\begin{tabular}{|c |c| c| c| c| c| c| c| c| c|} 
\hline
$J=(C_g,\Gamma^{\mathcal{N}_g})$& set  & $|C_g|$ & $|\Gamma^{\mathcal{N}_g}|$ &  $d_J$ & $n_{J}^{\textnormal{Ve}}$ & $n_{J}^{\textnormal{Pl}}$ \\ [0.5ex] 
\hline\hline
$A=(C_e,\Gamma_1^{S_3})$& Vac & 1 & 1 &  1  & 1 & 1 \\ 
\hline
$B=(C_e,\Gamma_{-1}^{S_3})$& Ch  & 1 & 1 & 1 &  1 & 0  \\ 
\hline
$C=(C_e,\Gamma_2^{S_3})$& Ch  & 1 & 2 & 2 & 2 & 0 \\ 
\hline
$D=(C_x,\Gamma_1^{\mathbb{Z}_2})$& Fl  & 3 & 1  & 3 & 0 & 1  \\ 
\hline
$E=(C_x,\Gamma_{-1}^{\mathbb{Z}_2})$& Dy  & 3 & 1 & 3 & 0 & 0 \\ 
\hline
$F=(C_y,\Gamma_1^{\mathbb{Z}_3})$& Fl  & 2 & 1 & 2 & 0 & 1  \\ 
\hline
$G=(C_y,\Gamma_{\omega}^{\mathbb{Z}_3})$& Dy  & 2 & 1 & 2 & 0 & 0 \\ 
\hline
$H=(C_y,\Gamma_{\bar{\omega}}^{\mathbb{Z}_3})$& Dy & 2 & 1 & 2 & 0 & 0  \\ 
\hline
\end{tabular}
\caption{Eight anyons $J = (C_g,\Gamma^{\mathcal{N}_g})$ of $\mathcal{Z}(\text{Vec}_{S_3})$: $C_g$ are the conjugacy classes, $\Gamma^{\mathcal{N}_g}$ the irreps of the centralizer $\mathcal{N}_g$, and $d_J$ the quantum dimensions (which is also the number of subtypes $n_J$).  The total number of subtypes is $\sum_J d_J = 16$.}
\label{table:ds3}
\end{center}
\end{table}

Since $S_3$ has three conjugacy classes and three irreps, there are nine energy eigenspaces. In particular, the energy eigenspaces $(C_e, \Gamma_{-1}^{S_3})$ and $(C_e, \Gamma_{2}^{S_3})$ (which comes in two subtypes as $|\Gamma_{2}^{S_3}|= 2$) are in one-to-one correspondence with, respectively, the chargeons $B$ and $C$. However, the fluxons $F$ and $D$ can be related to different energy eigenspaces. 
In fact, let us call $F_1$ and $F_2$ the two subtypes of $F$. $F_1$ is an elementary plaquette excitation and corresponds to the energy eigenspace $(C_y, \Gamma_1^{S_3})$, while $F_2$ corresponds to the energy eigenspace $(C_y, \Gamma_{-1}^{S_3})$ and can be obtained as the fusion product $F_1 \times B$, in agreement with the fusion rules of $\mathcal{Z}(\text{Vec}_{S_3})$ (see Ref.~\cite{Beigi11}, and Fig. 6 of~\cite{Komar17}). Similarly, among the three subtypes of $D$, $D_1 \equiv (C_x, \Gamma_{1}^{S_3})$ is an elementary plaquette excitation, while the two other subtypes correspond to the energy eigenspace $D_2 \equiv (C_x, \Gamma_{2}^{S_3})$, which has dimension $2$, and arises as a fusion products of $C$ with $D_1$ [this is also in agreement with Eq.~\eqref{eq:dimJ}]. 

Finally, dyons $E$, $G$ and $H$ are generated by the fusion of a vertex and a plaquette excitation. For example, one has $C \times F_1 = G + H$, so that the two dyons, $G$ and $H$, are generated by the fusion of the two subtypes of $C$ (vertex excitations) with $F_1$ (plaquette excitation). 

To summarize, the set of vertex excitations is ${\rm Ve}={\rm Ch}=\{B,C\}$ whereas the set of plaquette excitations is  ${\rm Pl}=\{D_1,F_1\}$, and only forms a subset of the fluxons ${\rm Fl}= \{D,F\}$. All other excitations (not in Ve or Pl) excite both a vertex and a plaquette and are called site excitations Si $=\{D_2,E_1,E_2,F_2,G,H\}$. They correspond either to dyons Dy $=\{E,G,H\}$ or to other subtypes of fluxons $\{D_2,F_2\}$. An overview of the energy eigenspaces is given in Table~\ref{table:ds3_2}.
\begin{table}[h!]
  \renewcommand{\arraystretch}{1.3}
\begin{center}
\begin{tabular}{|c |c| c| c| c|} 
\hline
$(C_g,\Gamma^G)$ & Set  & $n_{V\times P}$ & $V\times P$  \\ [0.5ex] 
\hline\hline
$(C_e,\Gamma_1^{S_3})$ & Vac  & 1 & $\mathbf{1}\times \mathbf{1}=A$ \\ 
\hline
$(C_e,\Gamma_{-1}^{S_3})$ & Ve &  1 & $(C_e,\Gamma_{-1}^{S_3})\times \mathbf{1}=B$ \\ 
\hline
$(C_e,\Gamma_2^{S_3})$ & Ve  & 2 & $(C_e,\Gamma_2^{S_3})\times \mathbf{1}=C$ \\
\hline
$(C_x,\Gamma_1^{S_3})$ & Pl  & 1 & $\mathbf{1}\times (C_x,\Gamma_{1}^{S_3})=D_1$ \\
\hline
$(C_x,\Gamma_{-1}^{S_3})$ & Si & 1 & $(C_e,\Gamma_{-1}^{S_3})\times (C_x,\Gamma_{1}^{S_3})=E_1$ \\
\hline
$(C_x,\Gamma_{2}^{S_3})$ & Si  & 4 & $(C_e,\Gamma_{2}^{S_3})\times (C_x,\Gamma_{1}^{S_3})=D_2+E_2$ \\
\hline
$(C_y,\Gamma_1^{S_3})$ & Pl &  1 & $\mathbf{1}\times (C_y,\Gamma_{1}^{S_3})=F_1$ \\
\hline
$(C_y,\Gamma_{-1}^{S_3})$ & Si  & 1 & $(C_e,\Gamma_{-1}^{S_3})\times (C_y,\Gamma_{1}^{S_3})=F_2$ \\
\hline
$(C_y,\Gamma_{2}^{S_3})$ & Si  & 4 & $(C_e,\Gamma_{2}^{S_3})\times (C_y,\Gamma_{1}^{S_3})=G+H$ \\
\hline
\end{tabular}
\caption{Nine energy eigenspaces $(C_g,\Gamma^{S_3})$ of the KQD model for the group $S_3$. Any energy eigenspace can be decomposed as a (cartesian) product of a $V \in \underline{\text{Ve}}$ and a $P\in \underline{\text{Pl}}$. The corresponding number of subtypes is notated $n_{V\times P}$. The objects $A$, $B$, $D_1$, $E_1$, $F_1$ and $F_2$ have only one subtype, while $C$, $D_2$, $E_2$, $G$ and $H$ have two subtypes. The total number of subtypes is $\sum_{V, P}n_{V\times P} = 16=\sum_{J} d_J$.}
\label{table:ds3_2}
\end{center}
\end{table}

\subsubsection{Partition function}
For the group $S_3$, the partition function \eqref{eq:kqdPF} becomes
%
%
\beqn
Z&=& 6^{2\genus-2} \left\{ (e^{\beta J_\text{v}}+5)^{N_\text{v}} (e^{\beta J_\text{p}}+5)^{N_\text{p}} \right. \\
&+& (e^{\beta J_\text{v}}+5)^{N_\text{v}}(e^{\beta J_\text{p}}-1)^{N_\text{p}} (1+2^{2-2\genus})\nn \\
&+& (e^{\beta J_\text{v}}-1)^{N_\text{v}}\left[(e^{\beta J_\text{p}}+1)^{N_\text{p}}3^{2-2\genus}+(e^{\beta J_\text{p}}+2)^{N_\text{p}}2^{2-2\genus} \right]\nn \\
&+& \left. (e^{\beta J_\text{v}}-1)^{N_\text{v}}(e^{\beta J_\text{p}}-1)^{N_\text{p}} (3^{2-2\genus}+2^{3-2\genus})\right\}, \nn
\eeqn
%
%
where the successive lines correspond to the vacuum, Ch, Fl, and Dy contributions, respectively. This expression is not particularly illuminating. The most notable feature is that the factorization $Z^{(\genus)} \times Z^{(\text{Ve})} \times Z^{(\text{Pl})}$ observed in the Abelian case no longer occurs. This is due to the loss of symmetry between vertex and plaquette excitations. In the Abelian case, this symmetry is present due to the fact that \mbox{Ve = Ch}, \mbox{Pl = Fl}, and there is a well-known duality between Ch and Fl. In the non-Abelian case, Pl $\neq$ Fl and this symmetry is lost.

\subsection{Generic group $G$}
Here, we collect the recipe to obtain the partition function for a generic finite group $G$ using Eq.~\eqref{eq:kqdPF}. The first step is to obtain its conjugacy classes $C_g$. Then, for a given element $g\in G$, one also needs the centralizer $\mathcal{N}_g$ and the corresponding irreps $\Gamma_j^{\mathcal{N}_g}$. Next, by pairing a conjugacy class and an irrep of the corresponding centralizer, one obtains the anyons $J=(C_g,\Gamma_j^{\mathcal{N}_g})$ with quantum dimension $d_J =|C_g| |\Gamma_j^{\mathcal{N}_g}|$, from which the $S$-matrix elements are $S_{\mathbf{1},J}=d_J/|G|$. Among the anyons, the set \underline{Ch} contains all $(C_e,\Gamma_j^{\mathcal{N}_g})$ and the set \underline{Fl} contains all $(C_g,\Gamma_1^{\mathcal{N}_g})$. Finally, the quantities $q$ in Eq.~\eqref{eq:zA} are given in Eqs.~\eqref{eq:qA1} and~\eqref{eq:qA2}.

\section{Proof of the degeneracy formula from a lattice-gauge theory perspective}\label{app:benoit}
In this Appendix, we provide a lattice-gauge theory perspective on the structure of the Hilbert space independently of the Hamiltonian and prove the degeneracy formulas \eqref{eq:kqdMS} and \eqref{eq:kqdMSinternaldeg}.
We begin by decomposing the Hilbert space of the lattice-gauge theory model
into a direct sum of subspaces $\mathcal{H}(p_1,\cl(x_1);\cdots;p_f,\cl(x_f))$, which
correspond to introducing a finite number $f$ of {\it plaquette excitations}, where the
plaquette $p_i$ carries a flux that belongs to the conjugacy class
$\cl(x_i)$ of $G$. We show that these subspaces are stable under the action of the gauge group $\mathcal{G}$, so we can decompose them as direct sums of irreducible representations
of $\mathcal{G}$. Each of these can be interpreted in terms of a collection of
{\it vertex excitations} carrying electric charges, whose total number is denoted by $c$. They are labeled
by a collection of irreducible representations $\Gamma_{i}^{G}\equiv (E_i,\rho_i)$
of $G$ ($1 \leqslant i \leqslant c$), where $E_i$ is the vector space and $\rho_i(g)$ is the unitary transformation
on $E_i$ ($g \in G$) attached to $\Gamma_{i}^{G}$.
Each copy of such representation $\{\rho\}$ of $\mathcal{G}$ acting on $\mathcal{H}(p_1,\cl(x_1);\cdots;p_f,\cl(x_f))$
corresponds to an energy eigenspace with
$f$ plaquette excitations and $c$ vertex excitations, as defined in Sec.~\ref{subsec_energy_eigenspaces}.
The number of such copies, denoted by $\mu_{\genus}(\cl(x_1),\cdots,\cl(x_f),(E_1,\rho_1),\cdots,(E_c,\rho_c))$,
is first expressed in the gauge theory setting, on the sphere, Eq.~(\ref{eq:lattice_mult}), and then on a
positive genus surface, Eq.~(\ref{eq_single_tilde_orbit_multiplicity}), which involve only informations on the
group $G$ and its irreducible representations. The next step is to interpret these expressions in terms
of the number of fusion channels as illustrated in Fig.~\ref{fig:fusiontreekqd}. This number is defined in the category
$\mathcal{Z}(\vecg)$ as the dimension of a space of arrows from the tensor product of a collection of objects, encoding the
vertex and plaquette excitations and the topological genus of the surface, into the unit object.
To achieve this goal, we give a self-contained description of $\mathcal{Z}(\vecg)$, directly inspired from the book~\cite{Etingof_book}.
It was striking to us to notice that the resulting description of spaces of arrows between pairs of objects 
is exactly what we need to reinterpret the lattice gauge theory multiplicities Eqs.~\eqref{eq:lattice_mult} and
\eqref{eq_single_tilde_orbit_multiplicity} in the $\mathcal{Z}(\vecg)$ language.

\subsection{Classical lattice gauge theory}

\subsubsection{Definition}

We consider a lattice gauge theory with the underlying group $G$ on an arbitrary graph. 
We shall focus on graphs that form the skeleton of a smooth compact surface $\Sigma$ (by which we mean that the graph goes around each noncontractable cycle of the manifold and plaquettes are defined accordingly to fill in the surface. i.e., we are representing it as a CW complex, whose 0-cells are the vertices, the 2-cells are the plaquettes, and the 1-cells are the links on the graph).  A classical configuration is given by a collection $\{g\}$ of group elements $g_{ij}$ for each oriented pair $(i,j)$ of nearest-neighbor sites. The only constraint imposed here is that $g_{ij}g_{ji}=e$, where $e$ denotes the unit element in the group $G$. Let $\mathcal{C}$ denote the set of all such configurations. On this set, we define an action of the gauge group $\mathcal{G}$.  Elements of $\mathcal{G}$ are collections $\{h\}$ of group elements $h_{i}$ defined at each lattice site $i$, and the group multiplication on $\mathcal{G}$ is defined by local multiplication in $G$: $(\{h\}\{h'\})_i = h_i h'_i$. $\mathcal{G}$ acts on $\mathcal{C}$ via gauge transformations: $(\{h\}\{g\})_{ij}=h_i g_{ij} h_{j}^{-1}$. The physical motivation for the introduction of the gauge group $\mathcal{G}$ is that if 
$\{g'\}=\{h\}\{g\}$, the two configurations $\{g\}$ and $\{g'\}$ are supposed to be
physically indistinguishable. Therefore, the set of distinct available physical states
is the set of gauge orbits on $\mathcal{C}$ under the action of the group $\mathcal{G}$.
Any physical classical Hamiltonian is given by a real-valued function over $\mathcal{C}$
that is constant on each gauge orbit. 

\subsubsection{Description of gauge orbits}
\label{subsec_desc_gauge_orbits}

A key physical insight is that gauge orbits are best identified via the collection
of fluxes along closed paths on the lattice. Let us consider an  oriented path $\gamma$ on the graph, visiting sites $0,1,...,L-1,L$, such that sites $j$ and $j+1$ are nearest neighbors. The flux $\Phi(\{g\},\gamma)$ is defined by
\be
\Phi(\{g\},\gamma)=g_{01}\,g_{12}\,...\,g_{L-1,L}.
\ee

Under a gauge transformation $\{h\}$, these fluxes transform according to
\be
\Phi(\{h\}\{g\},\gamma)=h_{0}\,\Phi(\{g\},\gamma)\,h_{L}^{-1}.
\label{gauge_transf_flux}
\ee 

In particular, if $\{g\}$ and $\{g'\}$ are related by a gauge transformation, 
there exists a group element $h_o \in G$ such that
for {\em any} closed path beginning and ending at site $o$, chosen as origin:
\be
\Phi(\{g'\},\gamma) = h_o\,\Phi(\{g\},\gamma)\,h_{o}^{-1}.
\label{criterion_gauge_orbit}
\ee
It turns out that the converse is true, i.e., this condition implies that
$\{g\}$ and $\{g'\}$ are related by a gauge transformation. To see this, 
let us pick a path $\gamma_i$ starting at $o$ and ending at $i$ for any site $i$ on the graph.
We define $h_i$ by $h_i = \Phi(\{g'\},\gamma_i)^{-1}h_o\Phi(\{g\},\gamma_i)$. The condition~(\ref{criterion_gauge_orbit}) ensures that this element $h_i$ does not depend on the path chosen to go from $o$ to $i$, so it is well defined. Consider a pair $(i,j)$ of nearest neighbor sites. Then we may choose $\gamma_j=\gamma_i (ij)$ so that $\Phi(\{g\},\gamma_j)=\Phi(\{g\},\gamma_i)\,g_{ij}$ and $\Phi(\{g'\},\gamma_j)=\Phi(\{g'\},\gamma_i)\,g'_{ij}$. From the definitions of $h_i$ and $h_j$, we get $h_j=(g'_{ij})^{-1}\,h_i\,g_{ij}$ so  $g'_{ij}=h_i\,g_{ij}\,h_j^{-1}$.

An important consequence of Eq.~(\ref{criterion_gauge_orbit}) is that 
for each closed loop $\gamma$, the conjugacy class of $\Phi(\{g\},\gamma)$ is gauge invariant.
In most physical situations, the classical energy function is minimized when all
these fluxes are trivial, i.e., when $\Phi(\{g\},\gamma)=e$ ($e$ being the neutral element in $G$)
for any closed loop $\gamma$. A natural set of low-energy classical configurations
is obtained by considering the subset $\mathcal{C}_{\{p_1,p_2,...,p_f\}}$ of 
configurations $\{g\}$ in which only a finite collection of elementary plaquettes $\{p_1,p_2,...,p_f\}$ is allowed to carry nontrivial fluxes. 

Consider a closed path $\gamma$ starting and ending at $o$ and let us deform it into $\gamma'$ while keeping the origin $o$ fixed. As long as the deformation does not cross any plaquette in $\{p_1,p_2,...,p_f\}$, the flux does not change, so $\Phi(\{g\},\gamma')=\Phi(\{g\},\gamma)$, therefore $\Phi(\{g\},\gamma)$ depends only on the homotopy class $[\gamma]$ of $\gamma$ on the complement $\Sigma\setminus\{p_1,p_2,...,p_f\}$. Furthermore, because upon composing two paths $\gamma_1$ and $\gamma_2$ with the same origin,  $\Phi(\{g\},\gamma_1\gamma_2)=\Phi(\{g\},\gamma_1)\Phi(\{g\},\gamma_2)$, we see that, for a given $\{g\}\in \mathcal{C}_{\{p_1,p_2,...,p_f\}}$, these  fluxes around closed paths with origin at $o$ define a group homomorphism from the fundamental group of the complement of $f$ holes in $\Sigma$,
$\pi_{1}(\Sigma\setminus\{p_1,p_2,...,p_f\})$, into $G$.  For a graph on a surface $\Sigma$ with finite genus, this fundamental group is finitely generated. This leads to a tremendous simplification of the above criterion. If $\{g\}$ and $\{g'\}$ both belong to $\mathcal{C}_{\{p_1,p_2,...,p_f\}}$, then they are related by a gauge transformation if and only if the  condition~(\ref{criterion_gauge_orbit}) holds for any path $\gamma$ whose homotopy class $[\gamma]$ is one of the generators of $\pi_{1}(\Sigma\setminus\{p_1,p_2,...,p_f\})$.

To proceed further, let us assume that $\Sigma$ is orientable, with genus $\genus$. 
Its fundamental group $\pi_{1}(\Sigma)$ is generated by $2\genus$ generators, denoted
by $a_1,...,a_\genus,b_1,...,b_\genus$. They are subjected to the unique constraint:
\be
a_1 b_1 a_{1}^{-1}b_{1}^{-1} \cdots a_\genus b_\genus a_{\genus}^{-1}b_{\genus}^{-1} = 1.
\ee
Each plaquette carrying a nontrivial flux has the effect of punching a hole in $\Sigma$,
and each hole adds its own generator $c_j$ ($1 \leqslant j \leqslant f$) to 
$\pi_{1}(\Sigma\setminus\{p_1,p_2,...,p_f\})$. This larger set of $2\genus+f$
generators is still subjected to a single constraint, that now reads
\be
a_1 b_1 a_{1}^{-1}b_{1}^{-1} \cdots a_\genus b_\genus a_{\genus}^{-1}b_{\genus}^{-1} = c_1 \cdots c_f.
\ee

For each generator, we associate a closed path beginning and ending at site $o$, so that for
each configuration $\{g\}\in \mathcal{C}_{\{p_1,p_2,...,p_f\}}$, we have a collection of
$2\genus+f$ fluxes $\Phi_{a_i}$, $\Phi_{b_i}$ for $1 \leqslant i \leqslant \genus$
and $\Phi_{c_j}\equiv \Phi_j$ for $1 \leqslant j \leqslant f$. 
These fluxes satisfy the constraint
\be
\Phi_{a_1} \Phi_{b_1} \Phi_{a_{1}}^{-1}\Phi_{b_{1}}^{-1} \cdots 
\Phi_{a_\genus} \Phi_{b_\genus} \Phi_{a_{\genus}}^{-1}\Phi_{b_{\genus}}^{-1} = \Phi_1 \cdots \Phi_f.
\label{genus_g_flux_constraint}
\ee
An illustration of this constraint is provided in Fig.~\ref{fig:octagon}, in the
case $\genus=2$ and $f=3$. It expresses that the fact that outer contour, drawn on the outer
boundary for a fundamental domain of the surface and based at the origin $o$, can be smoothly deformed into the union of the $f$ red internal contours (also based at $o$), without changing the total flux around it during this deformation:
\begin{figure}[t]
\begin{tikzpicture}
\draw [ultra thick] (0,0)--(2.8,0)--(4.8,2)--(4.8,4.8)--(2.8,6.8)--(0,6.8)--(-2,4.8)--(-2,2)--(0,0);
\node at (1.7,-0.3) {$\Phi_{a_1}$};
\draw[thick] (1.2,0.2)--(1.4,0)--(1.2,-0.2);
\node at (4.2,0.9) {$\Phi_{b_1}$};
\draw[thick] (3.52,1)--(3.8,1)--(3.8,0.72);
\node at  (5.2,3.5) {$\Phi_{a_1}^{-1}$};
\draw[thick] (4.6,3.2)--(4.8,3.4)--(5,3.2);
\node at (4,6.1) {$\Phi_{b_1}^{-1}$};
\draw[thick] (3.8,5.52)--(3.8,5.8)--(4.08,5.8);
\node at (1.1,7) {$\Phi_{a_2}$};
\draw[thick] (1.6,6.6)--(1.4,6.8)--(1.6,7);
\node at (-1.4,5.9) {$\Phi_{b_2}$};
\draw[thick] (-0.72,5.8)--(-1,5.8)--(-1,6.08);
\node at  (-2.4,3.2) {$\Phi_{a_2}^{-1}$};
\draw[thick] (-1.8,3.6)--(-2,3.4)--(-2.2,3.6);
\node at (-1.2,0.6) {$\Phi_{b_2}^{-1}$};
\draw[thick] (-1.28,1)--(-1,1)--(-1,1.28);
\draw[thick,red] (-0.6,3.4) circle [radius=0.7];
\node[red] at (-0.6,3.4) {$\Phi_1$};
\draw[thick,red] (0,0)--(-0.6,2.7);
\draw[thick,red] (-0.1,3.2)--(0.1,3.4)--(0.3,3.2);
\draw[thick,red] (1.4,3.4) circle [radius=0.7];
\node[red] at (1.4,3.4) {$\Phi_2$};
\draw[thick,red] (0,0)--(1.4,2.7);
\draw[thick,red] (1.9,3.2)--(2.1,3.4)--(2.3,3.2);
\draw[thick,red] (3.4,3.4) circle [radius=0.7];
\node[red] at (3.4,3.4) {$\Phi_3$};
\draw[thick,red] (0,0)--(3.4,2.7);
\draw[thick,red] (3.9,3.2)--(4.1,3.4)--(4.3,3.2);
\node[thick] at (0,-0.2) {$o$};
\draw[thick, blue] (0,0) to [out = 20, in=270] (4.45,3.4) to [out = 90, in=0] (1.4,6) to [out = 180, in=90] (-1.65,3.4) to [out = 270, in=120] (0,0);
\draw[thick, blue] (1.6,6.2) -- (1.4,6) -- (1.6,5.8);
\end{tikzpicture}
\caption{The octagon shown here is the fundamental domain for a genus 2 surface. Pairs of edges carrying inverse Aharonov-Bohm fluxes, such as $\Phi_{a_1}$ and $\Phi_{a_1}^{-1}$, are identified, with reversed orientation. The red circles inside correspond to $f=3$
plaquette excitations, carrying fluxes $\Phi_i$, for $i=1, 2, 3$. These fluxes have to be
evaluated along closed paths starting and ending at the same origin $o$. One way to do this is to connect the inner red circles with finite segments ending at $o$. The blue contour can be continuously shrunk from the outer boundary of the fundamental domain until it coincides with the union of the red paths, without changing its flux. At the beginning of this process, the flux along this contour is equal to the left-hand side of Eq.~(\ref{genus_g_flux_constraint}), and to the right-hand side at the end.}
\label{fig:octagon}
\end{figure}
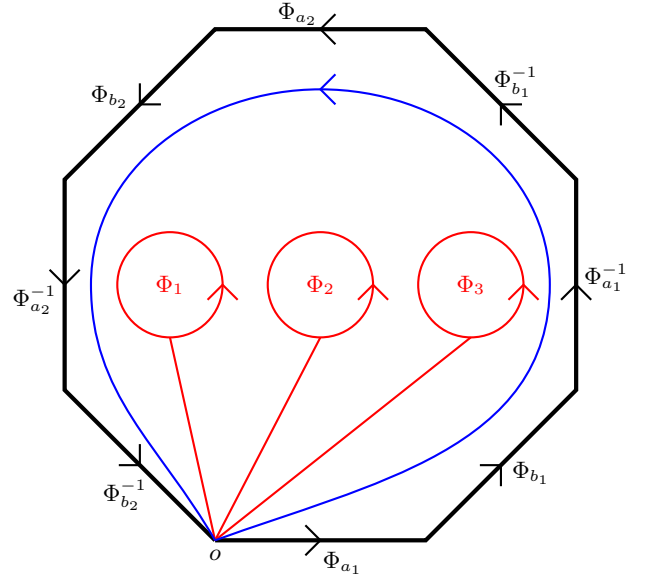

As we have seen, two configurations $\{g\}$ and $\{g'\}$ are gauge equivalent if and only if 
there exists $h \in G$ such that the associated $2\genus+f$ generating fluxes satisfy
\mbox{$\Phi'_{a_i}=h\,\Phi_{a_i}\,h^{-1}$}, \mbox{$\Phi'_{b_i}=h\,\Phi_{b_i}\,h^{-1}$}, for all $i$, and
\mbox{$\Phi'_j=h\,\Phi_j\,h^{-1}$}, for all $j$.

For later reference, it is convenient to denote by $\mathcal{F}_{\genus,f}$ the subset of 
$G^{2\genus+f}$ of ordered collections of $2\genus+f$ group elements satisfying the above constraint~(\ref{genus_g_flux_constraint}). $G$ acts on $\mathcal{F}_{\genus,f}$ via simultaneous
conjugacy of all fluxes, and we shall refer to this action as the $G_c$ action. Therefore, we see that the classical gauge orbits in $\mathcal{C}_{\{p_1,p_2,...,p_f\}}$ are in one-to-one correspondence with the $G_c$ orbits in $\mathcal{F}_{\genus,f}$. 

\subsection{Quantum lattice gauge models}

\subsubsection{Definition}

To get a quantum version of the model, we have to define a Hilbert space $\mathcal{H}$. A natural way to do this is to associate to any configuration $\{g\}$ a quantum state $|\{g\}\rangle$, and to assume that these form an orthonormal basis. The {\it classical} action of $\mathcal{G}$ on $\mathcal{C}$ gives rise naturally to a unitary representation of $\mathcal{G}$ on the quantum Hilbert space via $U(\{h\})|\{g\})\rangle=|\{h\}\{g\}\rangle$. The physical requirement of gauge invariance is
usually expressed by selecting only quantum states 
$|\Psi\rangle = \sum_{\{g\}\in\mathcal{C}} \psi(\{g\})|\{g\}\rangle$ such that
$U(\{h\})|\Psi\rangle=|\Psi\rangle$ for any $\{h\}\in\mathcal{G}$. 
Such states are characterized by wavefunctions $\psi(\{g\})$ that are invariant under arbitrary
gauge transformations (gauge singlets). Therefore, each classical gauge orbit gives rise
to a unique gauge-invariant quantum state. 

But as a very large group, $\mathcal{G}$ has many irreducible representations, besides the 
trivial one. Because $\mathcal{G}$ is a direct product of independent copies of $G$, one attached to each site, irreducible representations (irreps) of $\mathcal{G}$ are given by lists $\{\rho\}$ of irreducible representations $(E_i,\rho_i)$ of $G$ for each lattice site $i$. Here, $E_i$ denotes the finite-dimensional space in which the representation $\rho_i$ acts. The vector space attached to $\{\rho\}$ is the tensor product $E$ of the $E_i$'s, and an element $\{h\}$ acts on this space via the tensor product of the $\rho_i(h_i)$ operators. If $d_i$ is the dimension of $E_i$, the dimension $\dim(E)$ is equal to $\prod_{i}d_i$. In physical terms, each local representation $(E_i,\rho_i)$ of $G$ corresponds to a type of {\it electric charge} at site $i$. In particular the absence of charge corresponds to the trivial one-dimensional representation of $G$. 

In most physical models, these representation spaces encode some additional matter degrees of freedom.
But it is a very interesting feature of the KQD models that arbitrary representations
of the gauge group appear within the Hilbert space $\mathcal{H}$ of the pure gauge model. It is clear that
for each classical gauge orbit $\mathcal{O}=\mathcal{G}\{g\}$, the subspace $\mathcal{H}_{\mathcal{O}}$
spanned by all basis vectors $|\{g'\}\rangle$ where $\{g'\}\in\mathcal{O}$ is stable under the action
of $\mathcal{G}$. In the KQD model these spaces are also invariant under the action of the 
chosen Hamiltonian~\eqref{eq:KQDham} because the plaquette operators $B_p$ are diagonal in the
$|\{g\}\rangle$ basis, and the vertex operators $A_v$ are linear combinations
of gauge transformations $U(\{h\})$, such that $h_i=e$, when $i \neq v$. These linear
combinations are chosen so that the Hamiltonian~\eqref{eq:KQDham} commutes with all gauge transformations.
Typically, they involve local projectors from $\mathcal{H}$ onto the various irreps
$(E_v,\rho_v)$ of $G$ for each vertex $v$.
It is therefore very useful for physical applications to
decompose this action of $\mathcal{G}$ on $\mathcal{H}_{\mathcal{O}}$. 
As usual, a key question is to find the multiplicity $\mu_{\mathcal{O}}(\{\rho\})$ 
of the irreducible representation $\{\rho\}$
in the decomposition of the gauge action of $\mathcal{G}$ on $\mathcal{H}_{\mathcal{O}}$.  
For this, we will need to recall some elementary notions in group theory.

\subsubsection{Decomposition of a permutation group representation}
\label{subsec_dec}

Let us consider a group $H$ acting by permutations on a set $X$. To set the notation, we have $(hh')x=h(h'x)$ for all $h$ and $h'$ in $H$ and $x$ in $X$. We consider the Hilbert space $\mathcal{H}_X$ in which an orthonormal basis is composed of states $|x\rangle$ in one-to-one correspondence with elements of $X$. The {\it classical} action of $H$ on $X$ gives rise naturally to a unitary representation of $H$ on $\mathcal{H}_X$ via $U(h)|x\rangle=|hx\rangle$. Let us consider an irreducible  representation $(E,\rho)$ of $H$. We wish to evaluate the multiplicity $\mu(\rho)$ in the above permutation representation of $H$.

Having in mind applications to the case where $X=\mathcal{O}$ and $H=\mathcal{G}$, we assume
that $X$ contains a unique orbit $Hx$ under the action of $H$.
In general, there are some elements $s$ in $H$ leaving $x$ invariant, i.e., $sx=x$. These elements form a subgroup $S_x$ of $H$ called the stabilizer of $x$. $H$ is a disjoint union of left cosets $h_i S_x$, and $hx$ depends only on the left coset to which $h$ belongs. Denoting by $\{h_1,...,h_{q}\}$ a set of representatives for the left cosets, the orbit $Hx$ may be written as $\{h_{1}x,...,h_{q}x\}$, all elements being distinct.
We therefore get the direct sum decomposition  $\mathcal{H}_{X}=\bigoplus_{i}\mathbb{C}|h_i x\rangle$. The stabilizer $S_x$ acts on $\mathbb{C}|x\rangle$ via its trivial unitary representation. These two facts show that the representation of $H$ on $\mathcal{H}_{X}$ is the  {\em induced} representation from $S_x$ to $H$ of the trivial irreducible representation of $S_x$. 

Applying Frobenius reciprocity~\cite{Sternberg_book}, the multiplicity $\mu(\rho)$ is equal to the multiplicity of the trivial irrep of $S_x$, in the restriction of $(E,\rho)$ to $S_x$.  This latter multiplicity is simply expressed as the dimension of the space of vectors $v$ in $E$ that are invariant under the action of elements of $S_x$, i.e., $\rho(s)(v)=v$ for all $s \in S_x$. We shall denote this dimension by $\dim(\mathrm{Inv}(E,\rho)_{S_x})$. The main result that we shall use on many occasions reads
\be
\mu(\rho)=\dim(\mathrm{Inv}(E,\rho)_{S_x}).
\label{general_mult}
\ee

An interesting special case occurs when the stabilizer $S_x=\{e\}$. In this case, the orbit through $x$ is in one-to-one correspondence with elements of $H$, and the action of $H$ on this orbit is called the regular representation of $H$ (via left multiplication). Since all elements of $E$ are left invariant by $e$, we have $\dim(\mathrm{Inv}(E,\rho)_{\{e\}})=\dim(E)$ and we recover the well-known fact that the multiplicity of $(E,\rho)$ in the regular representation of $H$ is equal to the dimension of $E$. 

\subsubsection{Stabilizer subgroups for the gauge action on $\mathcal{C}$}
\label{subsec_gauge_stab}

Let us now turn our attention to stabilizer subgroups $S_{\{g\}}$. Equation~(\ref{gauge_transf_flux}) shows that the gauge transformation $\{s\}$ belongs to $S_{\{g\}}$ if and only if for any site $i$ and any oriented path $\gamma_i$ starting from an origin site $o$ and ending at $i$, we have:
\be
s_i = \Phi(\{g\},\gamma_i)^{-1}s_o \,\Phi(\{g\},\gamma_i).
\label{stab_gauge_conf}
\ee
This is a very strong set of constraints. It first shows that knowing $s_o$ determines values of $s_i$ at all sites. Therefore, the stabilizer $S_{\{g\}}$ is in one-to-one correspondence with a subgroup $S_{\{g\},o}$ of $G$, seen as the local gauge subgroup acting only at the origin site $o$. In general, this subgroup $S_{\{g\},o}$ is smaller than $G$ iself. To determine it, the discussion below
Eq.~(\ref{criterion_gauge_orbit}) shows that $s_o\in S_{\{g\},o}$ if and only if
$\Phi(\{g\},\gamma) = s_{o}\,\Phi(\{g\},\gamma)\,s_{o}^{-1}$
for all closed loops $\gamma$ starting and ending at $o$. If ${\{g\}}\in \mathcal{C}_{\{p_1,p_2,...,p_f\}}$,
we have seen that this is equivalent to imposing this condition on fluxes attached to a
set of generators of $\pi_{1}(\Sigma\setminus\{p_1,p_2,...,p_f\})$. So we get:
\beqn
S_{\{g\},o} & = & \{s\in G|
\Phi_{a_i}=s\,\Phi_{a_i}\,s^{-1}, \Phi_{b_i}=s\,\Phi_{b_i}\,s^{-1},
\nonumber\\
& & \Phi_j=s\,\Phi_j\,s^{-1}, 1 \leqslant i \leqslant g, 1 \leqslant j \leqslant f\}.
\label{eq_explicit_local_stabilizer}
\eeqn

Let us apply this description of $S_{\{g\}}$ to the evaluation of
the multiplicity $\mu_{\mathcal{O}}(\{\rho\})$. From Eq.~(\ref{general_mult}),
we have:
\be
\mu_{\mathcal{O}}(\{\rho\}) = \dim(\mathrm{Inv}(E,\rho)_{S_{\{g\}}}).
\ee
From the previous discussion, we see that
the restriction of the representation $\{\rho\}$ of $\mathcal{G}$ to $S_{\{g\}}$ is isomorphic to the restriction of the tensor product representation $\rho_1 \otimes\cdots\otimes \rho_c$ from $G$ to $S_{\{g\},o}$.
This results from the fact, manifested in Eq.~(\ref{stab_gauge_conf}), that the map from $s_o$ to $s_i$ is an inner automorphism of the group $G$. Specifically, the element $\{s\}$ in $S_{\{g\}}$ acts as $\rho_{1}(s_{i_1}(s_o))\otimes\cdots\otimes \rho_{c}(s_{i_c}(s_o))$ on $E_1 \otimes \cdots \otimes E_c$.
Using Eq.~\eqref{stab_gauge_conf}, this is equal to
$U^{-1}(\rho_{1}(s_o)\otimes\cdots\otimes \rho_{c}(s_o))U$, where $U=\rho_{1}(\Phi(\{g\},\gamma_1))\otimes\cdots\otimes \rho_{c}(\Phi(\{g\},\gamma_c))$ does not depend on $s_o$. 
Therefore, we get the rather explicit expression:
\be
\mu_{\mathcal{O}}(\{\rho\}) = \dim\left(\mathrm{Inv}(\rho_1 \otimes\cdots\otimes \rho_c)_{S_{\{g\},o}}\right),
\label{eq_single_orbit_multiplicity}
\ee
where $S_{\{g\},o}$ is the subgroup of $G$ defined by Eq.~(\ref{eq_explicit_local_stabilizer}).

\subsubsection{State counting in KQD models on a sphere}
\label{subsec_state_count_gauge_sphere}

In this appendix, a prominent role will be played by the configurations in which a subset of plaquettes $\{p_1,p_2,\cdots,p_f\}$ carries nontrivial fluxes $\{\Phi_1,\Phi_2,\cdots,\Phi_f\}$. Specifying the conjugacy classes of these fluxes, setting $\cl(\Phi_j)=\cl(x_j)$ with $x_j \in G$ for $1 \leqslant j \leqslant f$, defines a subset $\mathcal{C}(p_1,\cl(x_1);\cdots;p_f,\cl(x_f))$  of $\maC$, which is stable under classical gauge transformations and therefore a stable subspace $\mathcal{H}(p_1,\cl(x_1);\cdots;p_f,\cl(x_f))$, which will be the focus of our attention in this section. To simplify notations, we shall often omit the explicit reference to plaquette labels $p_j$ and write $\mathcal{C}(\cl(x_1),\cdots,\cl(x_f))$ or $\mathcal{H}(\cl(x_1),\cdots,\cl(x_f))$. Likewise, it is natural to consider representations $\{\rho\}$ of $\mathcal{G}$ such that there is a finite subset of sites $\{i_1,i_2,\cdots,i_c\}$ that host nontrivial charges,  corresponding to irreducible representations $(E_1,\rho_1),\cdots,(E_c,\rho_c)$ of $G$. Let us denote by
$\mu_0(\cl(x_1),\cdots,\cl(x_f),(E_1,\rho_1),\cdots,(E_c,\rho_c))$ the multiplicity of
the representation $\{\rho\}$ of $\mathcal{G}$ acting on $\mathcal{H}(\cl(x_1),\cdots,\cl(x_f))$.
The 0 subscript in $\mu$ refers to the genus (here $\genus=0$) of the surface $\Sigma$.
Clearly $\mathcal{H}(\cl(x_1),\cdots,\cl(x_f))$ is the direct sum of subspaces $\mathcal{H}_{\mathcal{O}}$
associated with gauge orbits $\mathcal{O}$ contained in $\mathcal{C}(\cl(x_1),\cdots,\cl(x_f))$. 
As an immediate consequence of Eq.~(\ref{eq_single_orbit_multiplicity}), we have the 
general expression for multiplicities as a sum over gauge orbits:
 \begin{widetext}
\be
\label{eq:lattice_mult}
\mu_0(\cl(x_1),\cdots,\cl(x_f),(E_1,\rho_1),\cdots,(E_c,\rho_c))=
\sum_{\mathcal{G}\{g\} \subset \mathcal{C}(\cl(x_1),\cdots,\cl(x_f))}
\dim\left(\mathrm{Inv}(\rho_1 \otimes\cdots\otimes \rho_c)_{S_{\{g\},o}}\right).
\ee
\end{widetext}
This formula could lead to explicit numerical calculations, but it is of limited practical use, when the numbers of fluxes and charges become large. Fortunately, building on the large body of works on topological theories in 2+1 dimensions~\cite{Simon_book, Bakalov_book}, it is possible to translate it in a numerically very efficient form, that calls upon a basic construction in category theory, namely the Drinfeld center $\mathcal{Z}(\vecg)$ of the category $\vecg$ of $G$-graded  vector spaces~\cite{Etingof_book}. One of our goals is to
prove that
\begin{widetext}
\be
\mu_0(\cl(x_1),\cdots,\cl(x_f),(E_1,\rho_1),\cdots,(E_c,\rho_c))=
\dim\left(\HomZvecg(\1,\cl(x_1)\otimes\cdots\otimes\cl(x_f)\otimes(E_1,\rho_1)\otimes\cdots\otimes(E_c,\rho_c))\right).
\label{eq:multiplicity_sphere}
\ee
\end{widetext}
The meaning of the right-hand side will be defined later in our discussion of the
$\mathcal{Z}(\vecg)$ category.
This has the expected form for the degeneracy in a topological theory on the sphere with $f+c$ punctures, carrying the objects $\cl(x_1),\cdots,\cl(x_f),(E_1,\rho_1),\cdots,(E_c,\rho_c)$ of the modular tensor category $\mathcal{Z}(\vecg)$~\cite{Bakalov_book}. However, in the lattice gauge theory model, this counts a number of gauge multiplets, each corresponding to a representation of dimension $\dim(E)=\prod_{i=1}^{c}\dim(E_i)$ of the gauge group $\maG$. This gives rise to extra degeneracies compared to the expected formula~(\ref{eq:multiplicity_sphere}) for the TQFT based on the $\mathcal{Z}(\vecg)$ category.

\subsubsection{State counting in KQD models at positive genus}
\label{subsec_state_count_gauge_positive_genus}

To count states on the sphere, we started from the 
formula~(\ref{eq_single_orbit_multiplicity}) for the
muliplicity $\mu_{\mathcal{O}}(\{\rho\})$ attached to a single gauge orbit
$\mathcal{O}$. In the positive genus case, we have to take into account
Aharonov-Bohm fluxes around closed cycles on $\Sigma$, which complicates
slightly the state counting. To relate multiplicities to
dimensions of Hom-spaces in the $\mathcal{Z}(\vecg)$ category, we need to replace the
muliplicity $\mu_{\mathcal{O}}(\{\rho\})$ by a new one, $\mu_{\tilde{\mathcal{O}}}(\{\rho\})$,
that involves some particular collections of gauge orbits $\mathcal{O}$.
As discussed in Sec.~\ref{subsec_desc_gauge_orbits}, gauge orbits on a genus $\genus$
surface in the presence of $f$ plaquettes carrying nontrivial fluxes are
in one-to-one correspondence with the $G_c$ orbits in $\mathcal{F}_{\genus,f}$.
It will be convenient to introduce new variables 
$z_i = \Phi_{a_i} \Phi_{b_{i}}^{-1}\Phi_{a_{i}}^{-1}$.
They allow us to express the topological constraint~(\ref{genus_g_flux_constraint}) as
\be
\Phi_{b_{\genus}} z_\genus \cdots \Phi_{b_{1}} z_1 \Phi_1 \cdots \Phi_f = e.
\label{genus_g_new_flux_constraint}
\ee
Let us denote by $\tilde{\mathcal{F}}_{\genus,f}$ the subset of 
$G^{2\genus+f}$ of ordered collections of $2\genus+f$ group elements satisfying the above constraint~(\ref{genus_g_new_flux_constraint}). Using the definition
$z_i = \Phi_{a_i} \Phi_{b_{i}}^{-1}\Phi_{a_{i}}^{-1}$, we define a map $\pi$
from $\mathcal{F}_{\genus,f}$ to $\tilde{\mathcal{F}}_{\genus,f}$. As for
$\mathcal{F}_{\genus,f}$, $G$ also acts on $\tilde{\mathcal{F}}_{\genus,f}$ via simultaneous
conjugacy of all fluxes, and the map $\pi$ is compatible with these two $G_c$
actions on $\mathcal{F}_{\genus,f}$ and $\tilde{\mathcal{F}}_{\genus,f}$.
Let us consider an orbit $\tilde{\mathcal{O}}$ through the element
$(\Phi_{b_{1}},z_1,\cdots,\Phi_{b_{\genus}},z_\genus,\Phi_1 \cdots \Phi_f)$ in $\tilde{\mathcal{F}}_{\genus,f}$.
Its inverse image under $\pi$ defines a collection of $G_c$ orbits in $\mathcal{F}_{\genus,f}$,
that we will identify with a collection of gauge orbits $\mathcal{O}$, for which we shall
use the notation $\mathcal{O} \in \pi^{-1}(\tilde{\mathcal{O}})$. This defines a subspace:
\be
\mathcal{H}_{\tilde{\mathcal{O}}} = \bigoplus_{\mathcal{O} \in \pi^{-1}(\tilde{\mathcal{O}})}
\mathcal{H}_{\mathcal{O}},
\ee
which is preserved by gauge transformations. A key step is to   
find the multiplicity $\mu_{\tilde{\mathcal{O}}}(\{\rho\})$ 
of the irreducible representation $\{\rho\}$
in the decomposition of the gauge action of $\mathcal{G}$ on $\mathcal{H}_{\tilde{\mathcal{O}}}$. 

Let us denote by $S$ the subgroup of $G$ composed of elements that commute
with $\Phi_{b_{i}}$, $z_i$, $\Phi_j$ for all $i,j$. Orbits in $\pi^{-1}(\tilde{\mathcal{O}})$
are in one-to-one correspondence with solutions $(\Phi_{a_{1}},\cdots,\Phi_{a_{\genus}})$
of the $\genus$ equations $z_i = \Phi_{a_i} \Phi_{b_{i}}^{-1}\Phi_{a_{i}}^{-1}$, modulo
residual gauge transformations. The latter take the form of simultaneous conjugacy
of $\Phi_{a_i}$ ($1 \leqslant i \leqslant \genus$) under an element of $S$, referred to as the $S_c$
action on $G^\genus$. So orbits $\mathcal{O} \in \pi^{-1}(\tilde{\mathcal{O}})$
can be identified with such $S_c$ orbits in $G^\genus$ through solutions
of $z_i = \Phi_{a_i} \Phi_{b_{i}}^{-1}\Phi_{a_{i}}^{-1}$.
For these equations to have solutions, we have to assume that
$z_i$ is in the same conjugacy class as $\Phi_{b_{i}}^{-1}$, so that there exist
group elements $h_i$ such that $z_i = h_i \Phi_{b_{i}}^{-1} h_i^{-1}$.
Then, we have $\Phi_{a_i}=h_i t_i$ with $t_i \in T_i$, where $T_i$ is the
stabilizer of $\Phi_{b_{i}}$ for the conjugacy action of $G$ on itself.
The stabilizer of $(\Phi_{a_{1}},\cdots,\Phi_{a_{\genus}})$
under the $S_c$ action, defined as the set
$\{s\in S|s \Phi_{a_i} = \Phi_{a_i}s, 1 \leqslant i \leqslant \genus\}$ will be denoted
by $\mathrm{Stab}_{S_c}\{\Phi_{a_{i}}\}$.
Then, we have
\be
\mu_{\tilde{\mathcal{O}}}(\{\rho\})=\sum_{S_c \{\Phi_{a_{i}}\} \in  \pi^{-1}(\tilde{\mathcal{O}})} \dim\left(\mathrm{Inv}(\hat{\rho})_
{\mathrm{Stab}_{S_c}\{\Phi_{a_{i}}\}}\right),
\ee
where $\hat{\rho}$ stands for the representation $\rho_1 \otimes\cdots\otimes \rho_c$ of $G$.
The right-hand side can be expressed in terms of the character
$\chi_{\hat{\rho}}$ of the $\hat{\rho}$ representation, since:
\be
\dim\left(\mathrm{Inv}(\hat{\rho})_
{H}\right)=
\frac{1}{|H|}
\sum_{s\in H}\chi_{\hat{\rho}}(s),
\label{eq:inv_from_character}
\ee
for any subgroup $H$ of $G$.
This allows us to replace the sum over $S_c$ orbits by a sum
over solutions $\Phi_{a_i}=h_i t_i$ with $t_i \in T_i$, 
because moving along an $S_c$ orbit in $G^\genus$
conjugates the stabilizer, an operation which leaves the
character $\chi_{\hat{\rho}}(s)$ unchanged, and 
the number of elements in the $S_c$ orbit through $(\Phi_{a_{1}},\cdots,\Phi_{a_{\genus}})$
is equal to $|S|/|\mathrm{Stab}_{S_c}\{\Phi_{a_{i}}\}|$.
Introducing the discrete $\delta$ function on $G$ by
$\delta(g,g')=1$ if $g=g'$ and $\delta(g,g')=0$ otherwise, the condition
$s\in \mathrm{Stab}_{S_c}\{\Phi_{a_{i}}\}$ is equivalent
to the statement that $\prod_i \delta(s \Phi_{a_i},\Phi_{a_i}s)=1$, so:
\be
\mu_{\tilde{\mathcal{O}}}(\{\rho\})=\frac{1}{|S|}\sum_{t_i \in T_i} \sum_{s\in S}
\prod_i \delta(s h_i t_i,h_i t_i s)\,\chi_{\hat{\rho}}(s).
\label{eq_single_tilde_orbit_multiplicity}
\ee

Stable subspaces $\mathcal{H}(p_1,\cl(x_1);\cdots;p_f,\cl(x_f))$ or 
$\mathcal{H}(\cl(x_1),\cdots,\cl(x_f))$ (omitting plaquette labels) are 
defined in the same way as on the sphere. In these spaces, the global 
Aharonov-Bohm fluxes $\Phi_{a_i}$ and $\Phi_{b_{i}}$ are allowed to take
any value. In physical terms, we are not trying to measure them. As we shall see,
starting from Eq.~(\ref{eq_single_tilde_orbit_multiplicity}) the generalization
of the counting formula Eq.~(\ref{eq:multiplicity_sphere}) to arbitrary positive genus $\genus$ reads
\begin{widetext}
\be
\mu_\genus(\cl(x_1),\cdots,\cl(x_f),(E_1,\rho_1),\cdots,(E_c,\rho_c))= \sum_{Z_1,\cdots,Z_\genus}
\dim\left(\HomZvecg(\1,\otimes_{i=1}^{\genus}(Z_i \otimes Z_i^{*})\otimes X_{f}\otimes Y_{c})\right),
\label{eq:multiplicity_positive_genus}
\ee
\end{widetext}
where in the sum each $Z_i$ for $1 \leqslant i \leqslant \genus$
runs over all simple objects in the $\mathcal{Z}(\vecg)$ category,
\mbox{$X_{f}\equiv \cl(x_1)\otimes\cdots\otimes\cl(x_f)$} and
\mbox{$Y_{c}\equiv (E_1,\rho_1)\otimes\cdots\otimes(E_c,\rho_c)$} in $\mathcal{Z}(\vecg)$.
This is the expected expression for the degeneracy in a topological theory on a genus $\genus$ surface with $f+c$ punctures, carrying the objects $\cl(x_1),\cdots,\cl(x_f),(E_1,\rho_1),\cdots,(E_c,\rho_c)$ of the modular tensor category $\mathcal{Z}(\vecg)$~\cite{Bakalov_book}.
Each added handle labeled by $i$ 
materializes as the tensor product of one simple object $Z_i$ and its dual object
$Z_i^{*}$, and these simple objects have to be summed over to account for the topological
degeneracies associated with the presence of non-contractible cycles.
But as was the case on the sphere, we have to keep in mind that
each among the $\mu_\genus$ copies of the $(E,\rho)$ representation corresponds to a degenerate stable
subspace (under gauge transformations) of dimension $\dim(E)=\prod_{i=1}^{c}\dim(E_i)$. 

\subsection{Basic facts about $\mathcal{Z}(\vecg)$}
\label{sec_Zvecg}

\subsubsection{Description of $\mathcal{Z}(\vecg)$}
\label{subsec_desc_Zvecg}

If $G$ is a group, the category $\vecg$ is defined as follows. Its objects are finite-dimensional $G$-graded vector spaces (over $\mathbb{C}$) that can be written as direct sums of the form $V=\bigoplus_{g \in G} V_g$ over $\mathbb{C}$. An arrow from $V$ to $W$ is a collection of linear maps $f_g : V_g \rightarrow W_g$. The direct sum of two objects if defined via $(V \bigoplus W)_g=V_g \bigoplus W_g$, and it is clear that any object $V$ can be written as a finite sum of simple objects denoted by $\delta_g$. These are defined by $(\delta_g)_h = 0$ if $g \neq h$ and $(\delta_g)_g = \mathbb{C}$. $\vecg$ is a tensor category, with the tensor product defined by \mbox{$(V \otimes W)_{g}=\bigoplus_{h} (V_h \otimes W_{h^{-1}g)}$.}  The tensor product of simple objects (fusion rules) reads $\delta_g \otimes \delta_h = \delta_{gh}$.

The Drinfeld center $\mathcal{Z}(\vecg)$ is a category whose objects have the form $(X,\sigma)$ where $X$ is an object of $\vecg$ and $\sigma$ is a half-braiding, i.e., a collection of arrows $\sigma_{V}: V \otimes X \rightarrow X \otimes V$ defined for any object $V$ in $\vecg$, subject to two conditions. The first one is naturality, which means that for any arrow $f: V \rightarrow W$, we have: $(\id_X \otimes f)\circ \sigma_V = \sigma_W \circ (f \otimes \id_X)$. The second condition expresses compatibility with tensor products in that $\sigma_{V\otimes W}=(\sigma_V \otimes \id_W)\circ(\id_V\otimes\sigma_W)$. From this definition, it is a simple exercise to identify objects in $\mathcal{Z}(\vecg)$.

Naturality implies that the $\sigma_V$ arrows are completely determined by those associated with simple objects \mbox{$V=\delta_g$}. From the fact that $(\delta_g) \cong \mathbb{C}$ as a vector space, we have $\delta_g \otimes X \cong X \cong X \otimes \delta_g$, so $\sigma_{\delta_g}$ can be seen as a linear map $\sigma_g$ from $X$ to itself. Compatibility  with tensor products implies that $\sigma_{gh}=\sigma_{g}\sigma_{h}$ so the collection $\{\sigma_{g}\}_{g \in G}$ forms a linear representation $\sigma$ of $G$. Further characterization of such representation is provided by keeping track of the $G$-grading. The finer structure of $\sigma_g$ is described via a collection of linear maps \mbox{$\sigma_{g,k}: X_k \cong (\delta_g \otimes X)_{gk} \rightarrow  (X \otimes \delta_g)_{gk} \cong X_{gkg^{-1}}$.}  This shows in particular that simple objects of $\mathcal{Z}(\vecg)$ have all their $G$-grades lying in the same conjugacy class $\cl(x)$ of $G$. Let us now assume that this is the case. 
Compatibility with tensor product implies that $\sigma_{gh,k}=\sigma_{g,hkh^{-1}} \circ \sigma_{h,k}$. Let us denote by $\Stab(x)$ the stabilizer of $x$ for the conjugacy action of $G$ on itself, i.e., it is the set of elements $s$ such that $sxs^{-1}=x$. Then if $s,s'\in \Stab(x)$, $\sigma_{ss',x}=\sigma_{s,x} \circ \sigma_{s',x}$,  so we get a representation $\rho$ of $\Stab(x)$, acting on a vector space $E$. Because $\sigma_{g,x}$ is an isomorphism between $X_x$ and $X_{gxg^{-1}}$, we see that $X_y$ is isomorphic to $E$ for any $y \in \cl(x)$. Using physics notations, it is therefore convenient to write:
\be
X= \bigoplus_{y \in \cl(x)} \mathbb{C} |y\rangle \otimes E,
\ee
where the $G$-grading is carried by $y$.  This presentation shows that the representation $\sigma$ of $G$ can be seen as the {\em induced representation} from $\Stab(x)$ to $G$ of the representation $(E,\rho)$ of $\Stab(x)$. Simple objects of $\mathcal{Z}(\vecg)$ are obtained when $(E,\rho)$ is irreducible. We denote them by $(X,\sigma) \equiv (\cl(x),(E,\rho))$.

In this paper, a prominent role is played by two special families of objects in $\mathcal{Z}(\vecg)$. In the first one, we consider a nontrivial conjugacy class $\cl(x)$ and the trivial representation of $\Stab(x)$. The induced representation is the permutation representation associated with the conjugacy action of $G$ on $\cl(x)$, given by $\sigma(g)(|y\rangle)=|gyg^{-1}\rangle$, and we have an example of the situation discussed in Sec.~\ref{subsec_dec}. Such objects, denoted by $(\cl(x))$ are called {\em fluxons} in the main text. As discussed there, [see in particular Eq.~(\ref{eq:map})], they are closely related to plaquettes carrying flux $\Phi \in \cl(x)$ in the lattice gauge model. 
The second family is obtained by taking the trivial conjugacy class, $\cl(e)=\{e\}$, for which $\Stab(e)=G$. Therefore $X$ is isomorphic to $E$, and the representation $\sigma$ is equal to $\rho$. These objects, denoted by $(E,\rho)$ are called {\em chargeons} in the main text.
They are closely related to vertices carrying a nontrivial charge in the lattice gauge model.

\subsubsection{Tensor products in $\mathcal{Z}(\vecg)$}
\label{subsubsec_tensor_products}

The Drinfeld center construction implies that $\mathcal{Z}(\vecg)$ is a tensor category. Let us write explictely the tensor product of two objects. From the general construction~\cite{Etingof_book}, we have $(X_1,\sigma_1)\otimes(X_2,\sigma_2)=(X_1 \otimes X_{2},\sigma_1 \otimes \sigma_2)$, where $\sigma_{1,2}$ are viewed as representations of $G$ as explained above. It is instructive to work out the tensor products for pairs of special objects just described. We have $(E_1,\rho_1)\otimes(E_2,\rho_2)=(E_1 \otimes E_2,\rho_1 \otimes \rho_2)$, which is not a simple object in $\mathcal{Z}(\vecg)$ in general. Decomposing the tensor product on the right as a sum of irreducible representations of $G$ shows that $(E_1,\rho_1)\otimes(E_2,\rho_2)$ is a sum of simple objects of the same type (chargeons). 

Let us consider $(\cl(x))\otimes (E,\rho)=(\cl(x),(E,\rho_{\Stab(x)}))$, where $\rho_{\Stab(x)}$ denotes the restriction of $\rho$ from $G$ to $\Stab(x)$. This restriction is in general not irreducible, showing that $(\cl(x))\otimes (E,\rho)$ decomposes as a direct sum of simple objects of the form $(\cl(x),(E',\rho'))$ where $(E',\rho')$ is an irreducible representation of $\Stab(x)$. Let us now make the connection with the fusion coefficients
$N_{A B}^{J}$ that appear in Eq.~(\ref{eq:dimJ}) in the main text. There, $A\equiv (E,\rho)$ is a chargeon,
$B\equiv (\cl(x))$ is a fluxon, and $J$ is any simple object in $\mathcal{Z}(\vecg)$ that may appear in the decomposition
of $A \otimes B$. As we have just seen, $J$ has to be of the form $(\cl(x),(E',\rho'))$. Then, the fusion coefficient
$N_{A B}^{J}$ is equal to the multiplicity $\mu(\rho',\rho_{\Stab(x)})$
of $(E',\rho')$ in the decomposition of $(E,\rho_{\Stab(x)})$ as a
sum of irreducible representations of $\Stab(x)$. The internal multiplicity $n_J$ of an anyon $J$, defined in the main text, takes the form $n_J=\sum_{(E,\rho)} \mu(\rho',\rho_{\Stab(x)})\,\mathrm{dim}(E)$, the sum being over
the irreducible representations of $G$. By Frobenius reciprocity, $\mu(\rho',\rho_{\Stab(x)})$ is equal
to the multiplicity of $(E,\rho)$ in the induced representation from $\Stab(x)$ to $G$ of $(E',\rho')$. Therefore,
$n_J$ is equal to the dimension of this induced representation (as a vector space over $\mathbb{C}$), that
coincides with the quantum dimension $d_J$ of the simple object $J$ (see remark at the end of Sec.~\ref{subsubsec_duality} below).

The case of two fluxons is more complex.  Consider two conjugacy classes $\cl(x_1)$ and $\cl(x_2)$ in $G$, we have to take the tensor product of the conjugacy representations associated with these two classes. We get a $G$-graded vector space $V=\bigoplus_{g_i \in \cl(x_i)} \mathbb{C}|g_1,g_2\rangle$, the grading being given by $g_1 g_2$. Forming products of elements in $\cl(x_1)$ and $\cl(x_2)$ produces new elements spanning in general several conjugacy classes, so the tensor product of two fluxons is not a simple object in general. But we also emphasize that its decomposition into simple objects does not only involve fluxons. To show how this may arise, let us denote by $S$ the stabilizer of $(x_1,x_2)$ for the conjugacy action of $G$ on $\cl(x_1) \times \cl(x_2)$ and consider the subspace 
$V_{\maO_{(x_1,x_2)}}=\bigoplus_{\alpha} \mathbb{C}|h_{\alpha}x_1h_{\alpha}^{-1},h_{\alpha}x_2h_{\alpha}^{-1}\rangle$,
where $\{h_{\alpha}\}$ is a system of representatives of left $S$-classes in $G$.
It is stable under the conjugacy representation on $V$, and its nontrivial gradings belong to $\cl(x_1x_2)$. Let us consider $T=\Stab(x_1x_2)$. It is clear that $S \subset T$. Let us denote by $\{t_1,\cdots,t_d\}$ a system of representatives of left classes $tS$ in $T$.  Then $(V_{\maO_{(x_1,x_2)}})_{x_1x_2}=\bigoplus_{i=1}^{d}\mathbb{C}|t_ix_1t_i^{-1},t_ix_2t_i^{-1}\rangle$. But this shows that the conjugacy action of $T$ on this subspace is nontrivial as soon as $d \geqslant 2$, so its associated representation on $V_{\maO_{(x_1,x_2)}}$  does not define a fluxon object in general.

\subsubsection{Arrows in $\mathcal{Z}(\vecg)$}

An arrow from $(X,\sigma)$ to $(Y,\tau)$ in $\mathcal{Z}(\vecg)$ is a $G$-grading preserving linear map $f: X \rightarrow Y$ compatible with half-braidings~\cite{Etingof_book}. The set of such arrows forms a vector space, that is usually denoted by $\HomZvecg((X,\sigma),(Y,\tau))$. By naturality of half-braidings, it is sufficient to impose  $f \circ \sigma_g = \tau_g \circ f$, for all $g$ in $G$. Keeping track of the $G$-grading, we get the finer conditions $f_{gkg^{-1}} \circ \sigma_{g,k} = \tau_{g,k} \circ f_k$, for all $g,k$ in $G$.

A particular case of interest occurs when $(X,\sigma)$ is equal to the unit element $\1=(\cl(e),\id)$ of $\mathcal{Z}(\vecg)$. Then only $f_e$ can be different from zero, and it satisfies $f_{e}  = \tau_{g,e} \circ f_e$. Since $f_e$ is a linear map from $\mathbb{C}$ to $Y_e$, it is determined by $f_e(1)=v \in Y_e$, and the previous condition simply states that $\tau_{g,e}(v)=v$, for any $g$ in $G$. Denoting by $\tau_{Y_e}$ the representation of $G$ obtained by restricting $\tau_g$ to $Y_e$, we have the useful result:
\be
\HomZvecg(\1,(Y,\tau)) \cong \Inv_{G}(Y_e,\tau_{Y_e}).
\label{eq_Zvecg_fusing_to_vacuum}
\ee
Let us apply this to the specific case where $(Y,\tau)=\cl(x_1)\otimes\cdots\otimes\cl(x_f)\otimes(E_1,\rho_1)\otimes\cdots\otimes(E_c,\rho_c)$. As a vector space, $Y=\bigoplus_{y_j \in \cl(x_j)} \mathbb{C} |y_1,\cdots,y_f\rangle \otimes E$, where $E=E_1 \otimes \cdots \otimes E_c$ and the grading is given by the product $y_1 \cdots y_f$ in $G$. Denoting by $\hat{\rho}$ the tensor product $\rho_1 \otimes \cdots \otimes \rho_c$, the representation $\tau$ is defined by $\tau_g(|y_1,\cdots,y_f\rangle \otimes v)=|gy_1g^{-1},\cdots,gy_fg^{-1}\rangle \otimes \hat{\rho}_g(v)$. To describe $\HomZvecg(\1,(Y,\tau))$, we need to determine $\Inv_{G}(Y_e,\tau_{Y_e})$. A vector $|y_1,\cdots,y_f\rangle \otimes v$ belongs to $Y_e$ whenever $y_1 \cdots y_f=e$. To proceed further, we note that $Y$ decomposes as a direct sum of invariant subspaces $Y_{\maO}$, where $\maO$ labels an orbit for the conjugacy action of $G$ on $\cl(x_1)\times\cdots\times\cl(x_f)$. Under this action, the element $g$ sends $(y_1,\cdots,y_f)$ to $(gy_1g^{-1},\cdots,gy_fg^{-1})$. Let $\maO=G_{c}y$ be the orbit through $y\equiv(y_1,\cdots,y_f)$ and let $S_y$ denotes the stabilizer of $y$ under the conjugacy action of $G$. The same reasoning as in Sec.~\ref{subsec_dec} shows that the restriction of $\tau$ to $Y_{\maO}$ is the induced representation from $S_y$ to $G$ of the restriction $(E,\hat{\rho})_{S_y}$ of $(E,\hat{\rho})$
to $S_y$. Therefore, still using Frobenius reciprocity, one has
\be
\Inv_{G}(Y_{\maO},\tau_{Y_{\maO}}) \cong \Inv (E,\hat{\rho})_{S_y}.
\label{eq:invariant_count}
\ee
Combining this result with Eq.~(\ref{eq_Zvecg_fusing_to_vacuum}), we get the useful formula:
\begin{widetext}
\be
\HomZvecg(\1,\cl(x_1)\otimes\cdots\otimes\cl(x_f)\otimes(E_1,\rho_1)\otimes\cdots\otimes(E_c,\rho_c)) \cong
\bigoplus_{G_{c}y, y_j \in \cl(x_j),y_1 \cdots y_f=e} \Inv (\rho_1\otimes\cdots\otimes\rho_c)_{S_y}.
\label{eq:useful_formula}
\ee
\end{widetext}

\subsubsection{Duality in $\mathcal{Z}(\vecg)$}
\label{subsubsec_duality}

Recall that if $V$ is a vector space over
$\mathbb{C}$ its dual space $V^*$ is the space
of $\mathbb{C}$-linear maps from $V$ to $\mathbb{C}$.
Duality in the $\vecg$ category is directly inspired
from this definition. We simply have to specify the
$G$-grading of $V^*$, for any $G$-graded vector space.
It is natural to take $(V^*)_g = (V_{g^{-1}})^*$.
The motivation for this is that we therefore get:
\be
(V^* \otimes V)_e = \bigoplus_{g \in G} (V^*)_{g^{-1}} \otimes V_g =
\bigoplus_{g \in G} (V_g)^* \otimes V_g.
\ee
This allows one to define the required evaluation 
$\mathrm{ev}_V : V^* \otimes V \rightarrow \delta_e$
and and coevaluation  $\mathrm{coev}_V :\delta_e \rightarrow V^* \otimes V$ maps.
For example $\mathrm{ev}_V$ is taken to vanish on $(V^* \otimes V)_h$ when
$h \neq e$ and $\mathrm{ev}_V(\sum_{g} \alpha_g \otimes v_g)= \sum_{g} \alpha_g (v_g)$,
where $v_g \in V_g$ and $\alpha_g \in (V_g)^*$. Since we will not use these maps
in this appendix, we shall not review them further.

Let us now turn to duality in $\mathcal{Z}(\vecg)$. Let us consider an object $(X,\sigma)$ in this
category. Its dual object has the form $(X^*,\tau)$ where $X^*$ is the dual of $X$
in $\vecg$, as defined in the previous paragraph. To define the half-braiding $\tau$,
we are looking for $\tau_g$ that sends $(X^*)_h$ into $(X^*)_{ghg^{-1}}$, or
equivalently sends $(X_{h^{-1}})^*$ into $(X_{gh^{-1}g^{-1}})^*$. Because 
$\sigma_{g^{-1}}$ sends $X_{gh^{-1}g^{-1}}$ into $X_{h^{-1}}$, its transpose
$\sigma_{g^{-1}}^{T}$ sends $(X_{h^{-1}})^*$ into $(X_{gh^{-1}g^{-1}})^*$.
Therefore, we take $\tau_g = \sigma_{g^{-1}}^{T}$. It would remain
to check that evaluation and coevaluation maps $\mathrm{ev}_X : X^* \otimes X \rightarrow \delta_e$ and $\mathrm{coev}_X :\delta_e \rightarrow X^* \otimes X$ in $\vecg$
also give rise to evaluation and coevaluation maps for $(X,\sigma)$
in $\mathcal{Z}(\vecg)$, which we will omit here. In the case of a simple object
$(X,\sigma) \equiv (\cl(x),(E,\rho))$, it is easy to check that
$(X,\sigma)^* \equiv (\cl(x^{-1}),(E^*,\bar{\rho}))$, where 
$\bar{\rho}(g) = \rho^{T}(g^{-1})$ for $g \in \mathrm{Stab}(x)$.

Once evaluation and coevaluation maps have been defined, we may compute the quantum dimension of objects
in $\mathcal{Z}(\vecg)$, which turns out to coincide with the usual notion of dimension for vector spaces over 
$\mathbb{C}$. So we have: $\mathrm{dim}(\cl(x),(E,\rho))=|\cl(x)|\,\mathrm{dim}(E)$, in agreement with
formula~(\ref{eq:quantum_dimension_anyon}) in the main text.

\subsection{Proofs of counting formulae}
\label{sec_proofs}

In the case of the sphere, we start from
Eq.~(\ref{eq:lattice_mult}), whose right-hand side is the
same as the right-hand side of Eq.~(\ref{eq:useful_formula}), thanks
to the one-to-one correspondence between gauge orbits in
$\mathcal{C}$ and $G_c$ orbits in $\mathcal{F}_{0,f}$ stated in
Sec.~\ref{subsec_desc_gauge_orbits}. So we can identify the left-hand
sides of Eqs.~(\ref{eq:lattice_mult}) and (\ref{eq:useful_formula}) and this proves
the counting formula, Eq.~(\ref{eq:multiplicity_sphere}).

In the case of positive genus $\genus$, we also have to keep track of Aharonov-Bohm
fluxes around non-contractible cycles. As done already in Sec.~\ref{subsec_state_count_gauge_positive_genus}, we assume some knowledge on the
$\Phi_{b_i}$ fluxes, in particular that $\Phi_{b_i}\in \mathrm{cl}(u_i)$. Information
on $\Phi_{a_i}$ fluxes will be handled through $z_i = \Phi_{a_i} \Phi_{b_{i}}^{-1}\Phi_{a_{i}}^{-1}$.
Clearly $z_i \in \mathrm{cl}((u_i)^{-1})$. This suggests to consider the object
$(Y^{\{\nu_i\}},\tau^{\{\nu_i\}})$ defined as
\begin{widetext}
\be
(Y^{\{\nu_i\}},\tau^{\{\nu_i\}})=
\bigotimes_{i=1}^{\genus} \left((\mathrm{cl}(u_i),(F_i,\nu_i))\otimes (\mathrm{cl}(u_i),(F_i,\nu_i))^*\right) 
\otimes X_f \otimes Y_c.
\label{eq:def_Y_tau_nu}
\ee
\end{widetext}
Here, for each $i$ such that $1 \leqslant i \leqslant \genus$, we choose an irreducible representation
$(F_i,\nu_i)$ of $T_i$, where $T_i$ is the stabilizer of $u_i$ for the conjugacy action
of $G$ on itself. The objects $X_f$ and $Y_c$ are defined in the same way as after 
Eq.~(\ref{eq:multiplicity_positive_genus})
in Sec.~\ref{subsec_state_count_gauge_positive_genus}.
$Y^{\{\nu_i\}}$ is a $G$-graded vector space, whose component at the neutral
element reads
\be
Y^{\{\nu_i\}}_e = \bigoplus_{p \in \tilde{\mathcal{F}}_{\genus,f}(\{\mathrm{cl}(u_i)\},\{\mathrm{cl}(x_j)\})} 
\mathbb{C} |p\rangle \otimes \left(\otimes_{i=1}^{\genus} F_i \otimes F_{i}^{*}\right) \otimes E,
\ee
where $E=E_1 \otimes \cdots \otimes E_c$. We recall that elements 
in $\tilde{\mathcal{F}}_{\genus,f}$ have the form
$p=(\Phi_{b_{1}},z_1,\cdots,\Phi_{b_{\genus}},z_\genus,\Phi_1 \cdots \Phi_f)$ subjected
to the constraint~(\ref{genus_g_new_flux_constraint}). 
$\tilde{\mathcal{F}}_{\genus,f}(\{\mathrm{cl}(u_i)\},\{\mathrm{cl}(x_j)\})$ is the subset
of $\tilde{\mathcal{F}}_{\genus,f}$ defined by imposing further constraints:
$\Phi_{b_i}\in \mathrm{cl}(u_i)$, $z_i \in \mathrm{cl}(u_i^{-1})$, and
$\Phi_{j}\in \mathrm{cl}(x_j)$.
Equation~(\ref{eq_Zvecg_fusing_to_vacuum}) specializes into
\be
\HomZvecg(\1,(Y^{\{\nu_i\}},\tau^{\{\nu_i\}})) \cong 
\Inv_{G}(Y^{\{\nu_i\}}_e,\tau^{\{\nu_i\}}_{Y_e}).
\label{eq:special_fusing_to_vacuum}
\ee
Since conjugacy classes of $\Phi_{b_i}$, $z_i$ and $\Phi_j$ are invariant
along $G_c$ orbits in $\tilde{\mathcal{F}}_{\genus,f}$, $\tilde{\mathcal{F}}_{\genus,f}(\{\mathrm{cl}(u_i)\},\{\mathrm{cl}(x_j)\})$ can be
written as a disjoint union of $G_c$ orbits $\tilde{\mathcal{O}}$. 

Let us now focus on such an orbit. It gives rise to a vector space
$Y^{\{\nu_i\}}_{\tilde{\mathcal{O}}}$, that is stable under the representation
$\tau^{\{\nu_i\}}$. In light of Eq.~(\ref{eq:special_fusing_to_vacuum})
we wish to compute $\mathrm{dim}(\Inv_{G}(Y^{\{\nu_i\}}_{\tilde{\mathcal{O}}},\tau^{\{\nu_i\}}_{\tilde{\mathcal{O}}}))$. As in Eq.~(\ref{eq:inv_from_character}), this dimension is easily related
to the character of the $\tau^{\{\nu_i\}}_{\tilde{\mathcal{O}}}$ representation by
\be
\mathrm{dim}(\Inv_{G}(Y^{\{\nu_i\}}_{\tilde{\mathcal{O}}},\tau^{\{\nu_i\}}_{\tilde{\mathcal{O}}}))=
\frac{1}{|G|}\sum_{g \in G} \chi_{\tau^{\{\nu_i\}}_{\tilde{\mathcal{O}}}}(g).
\label{eq:inv_from_character_2}
\ee
We have therefore to specify the representation $\tau^{\{\nu_i\}}_{\tilde{\mathcal{O}}}$ acting
on $Y^{\{\nu_i\}}_{\tilde{\mathcal{O}}}$. As in Sec.~\ref{subsec_state_count_gauge_positive_genus},
we denote by $S$ the subgroup of $G$ composed of elements that commute
with $\Phi_{b_{i}}$, $z_i$, $\Phi_j$ for all $i,j$. The $G_c$ orbit $\tilde{\mathcal{O}}$
through $p=(\Phi_{b_{1}},z_1,\cdots,\Phi_{b_{\genus}},z_\genus,\Phi_1 \cdots \Phi_f)$ may be described
as $\tilde{\mathcal{O}}=\{(g_{\alpha})_{c}(p), \alpha \in A\}$, where $\{g_{\alpha}\}_{\alpha \in A}$
is a representative set of left $S$ cosets in $G$. The space $Y^{\{\nu_i\}}_{\tilde{\mathcal{O}}}$ 
is spanned by vectors of the form $|(g_{\alpha})_{c}(p)\rangle \otimes
v_1 \otimes \varphi_1 \cdots v_g \otimes \varphi_g \otimes w$, with $v_i \in F_i$
$\varphi_i \in F_{i}^{*}$, and $w \in E$. Since we are interested in the trace of
$\tau^{\{\nu_i\}}_{\tilde{\mathcal{O}}}$, we evaluate $\tau^{\{\nu_i\}}(g)$ only 
for group elements $g$ that satisfy $(gg_{\alpha})_c(p)=(g_{\alpha})_c(p)$, or 
equivalently $g \in g_{\alpha}Sg_{\alpha}^{-1}$.

To determine the action of $\tau^{\{\nu_i\}}(g)$ on 
$v_1 \otimes \varphi_1 \cdots v_g \otimes \varphi_g \otimes w$, we use the fact that
the half-braiding associated with the object $(\mathrm{cl}(u_i),(F_i,\nu_i))$ is the induced
representation, that will be denoted by $\mu_i$, of $(F_i,\nu_i)$ from $T_i$ to $G$.
To construct this representation, it is convenient to introduce a system $\{h_{j}^{(i)}\}_{j\in J^{(i)}}$ of left $T_i$ cosets. At this point, we note that the role played by $u_i$ in the
reasonings to follow is merely to define the object $(\mathrm{cl}(u_i),(F_i,\nu_i))$.
Because $\mathrm{cl}(u_i)=\mathrm{cl}(\Phi_{b_i})$, we will set $u_i = \Phi_{b_i}$
once the orbit $\tilde{\mathcal{O}}$ included in $\tilde{\mathcal{F}}_{\genus,f}(\{\mathrm{cl}(u_i)\},\{\mathrm{cl}(x_j)\})$ has been chosen.

Therefore, for each $\alpha \in A$, there exists a unique \mbox{$j(\alpha)\in J^{(i)}$} such that
\be
g_{\alpha}\,\Phi_{b_i}\,g_{\alpha}^{-1}=h_{j(\alpha)}^{(i)}\,\Phi_{b_i}\,(h_{j(\alpha)}^{(i)})^{-1}.
\label{eq:def_j(alpha)}
\ee
The trace condition $(gg_{\alpha})_c(p)=(g_{\alpha})_c(p)$ implies then that
\be
g h_{j(\alpha)}^{(i)} = h_{j(\alpha)}^{(i)}\,t^{(i)}(g,j(\alpha)),
\label{eq:g_acting_on_h_j(alpha)}
\ee
with $t^{(i)}(g,j(\alpha)) \in T_i$.
Let the two sides of this equation act on the element $|\Phi_{b_i} \rangle  \otimes v_i$
through the induced representation $\mu_i$. This equality gives rise to
\beqn
\nonumber
\mu_i(g)(|g_{\alpha}\,\Phi_{b_i}\,g_{\alpha}^{-1}\rangle \otimes v_i) & = &
|g_{\alpha}\,\Phi_{b_i}\,g_{\alpha}^{-1}\rangle \otimes w_i, \\  \label{eq:mu_i_action}
w_i & = & \nu_i(t^{(i)}(g,j(\alpha)))(v_i).
\eeqn

In a similar manner, we can specify the half-braiding associated with the dual object
$(\mathrm{cl}(u_i),(F_i,\nu_i))^*$. It is equivalent to the induced
representation, that will be denoted by $\bar{\mu}_i$, of $(F_{i}^{*},\bar{\nu}_i)$ from $T_i$ to $G$, 
where $\bar{\nu}_i(g)=\nu_i(g^{-1})^{T}$.
We replace Eq.~(\ref{eq:def_j(alpha)}) by
\be
g_{\alpha}\,z_i\,g_{\alpha}^{-1}=h_{\tilde{\jmath}(\alpha)}^{(i)}\,\Phi_{b_i}^{-1}\,
(h_{\tilde{\jmath}(\alpha)}^{(i)})^{-1}.
\label{eq:def_tilde_j(alpha)}
\ee
Likewise, Eq.~(\ref{eq:g_acting_on_h_j(alpha)}) becomes
\be
g h_{\tilde{\jmath}(\alpha)}^{(i)} = h_{\tilde{\jmath}(\alpha)}^{(i)}\, t^{(i)}(g,\tilde{\jmath}(\alpha)).
\label{eq:g_acting_on_h_tilde_j(alpha)}
\ee
The same reasoning used to obtain Eq.~(\ref{eq:mu_i_action}) now gives
\beqn
\nonumber
\bar{\mu}_i(g)(|g_{\alpha}\,z_i\,g_{\alpha}^{-1}\rangle \otimes \varphi_i) & = &
|g_{\alpha}\,z_i\,g_{\alpha}^{-1}\rangle \otimes \psi_i ,\\  \label{eq:dual_mu_i_action}
\psi_i & = & \bar{\nu}_i(t^{(i)}(g,\tilde{\jmath}(\alpha)))(\varphi_i).
\eeqn

Using Eqs.~(\ref{eq:mu_i_action}) and (\ref{eq:dual_mu_i_action}), we can express
the character $\chi_{\tau^{\{\nu_i\}}_{\tilde{\mathcal{O}}}}$ as
\begin{widetext}
\be
\chi_{\tau^{\{\nu_i\}}_{\tilde{\mathcal{O}}}}(g)=
\sum_{\alpha \in A}\delta_{S}(g_{\alpha}^{-1}gg_{\alpha})
\left(\prod_{i=1}^{\genus}\chi_{\nu_i}(t^{(i)}(g,j(\alpha)))\,
\chi_{\nu_i}(t^{(i)}(g,\tilde{\jmath}(\alpha))^{-1})\right) \chi_{\hat{\rho}}(g),
\label{eq:character_single_orbit_1}
\ee
\end{widetext}
where $\hat{\rho}=\rho_1 \otimes \cdots \otimes \rho_c$. We have introduced the function
$\delta_{S}$ over $G$ such that $\delta_{S}(g)=1$, if $g \in S$ and $\delta_{S}(g)=0$,
otherwise. The characters appearing in Eq.~(\ref{eq:character_single_orbit_1}) can be
simplified further. From Eq.~(\ref{eq:def_j(alpha)}), we deduce that
$g_{\alpha}=h_{j(\alpha)}^{(i)} t_{\alpha}^{(i)}$ with $t_{\alpha}^{(i)}\in T_i$.
Multiplying Eq.~(\ref{eq:g_acting_on_h_j(alpha)}) on the right by $t_{\alpha}^{(i)}$, we get the
useful relation:
\be
g_{\alpha}^{-1}\,g\,g_{\alpha} = (t_{\alpha}^{(i)})^{-1}\,t^{(i)}(g,j(\alpha))\,t_{\alpha}^{(i)}.
\label{eq:simplify_t(g,j)}
\ee
Because characters are invariant under conjugacy and $t_{\alpha}^{(i)}\in T_i$,
this implies that
\be
\chi_{\nu_i}(t^{(i)}(g,j(\alpha)))= \chi_{\nu_i}(g_{\alpha}^{-1}\,g\,g_{\alpha}).
\label{eq:simplify_character}
\ee

Similarly, writing $z_i = h_i \Phi_{b_{i}}^{-1} h_i^{-1}$, Eq.~(\ref{eq:def_tilde_j(alpha)})
implies that 
$g_{\alpha}h_i=h_{\tilde{\jmath}(\alpha)}^{(i)} \tilde{t}_{\alpha}^{(i)}$ with $\tilde{t}_{\alpha}^{(i)}\in T_i$.
Multiplying~(\ref{eq:g_acting_on_h_tilde_j(alpha)}) on the right by $\tilde{t}_{\alpha}^{(i)}$, we get:
\be
h_{i}^{-1}\,g_{\alpha}^{-1}\,g\,g_{\alpha}\,h_i = (\tilde{t}_{\alpha}^{(i)})^{-1}\,t^{(i)}(g,\tilde{\jmath}(\alpha))\,\tilde{t}_{\alpha}^{(i)}.
\label{eq:simplify_tilde_t(g,j)}
\ee
Invariance of characters under conjugacy implies then:
\be
\chi_{\nu_i}(t^{(i)}(g,\tilde{\jmath}(\alpha))^{-1})= \chi_{\nu_i}
(h_{i}^{-1}\,g_{\alpha}^{-1}\,g^{-1}\,g_{\alpha}\,h_i).
\label{eq:simplify_dual_character}
\ee

Equations~(\ref{eq:simplify_character}) and~(\ref{eq:simplify_dual_character})  
combined with Eq.~(\ref{eq:character_single_orbit_1}) show that the right-hand
side of Eq.~(\ref{eq:inv_from_character_2}) can be expressed as
\begin{widetext}
\be
\frac{1}{|G|}\sum_{g \in G} \chi_{\tau^{\{\nu_i\}}_{\tilde{\mathcal{O}}}}(g)=
\frac{1}{|S|} \sum_{s \in S}
\left(\prod_{i=1}^{\genus}\chi_{\nu_i}(s)\,
\chi_{\nu_i}(h_{i}^{-1}\,s^{-1}\,h_i)\right)\,\chi_{\hat{\rho}}(s).
\label{eq:simplify_sum_over_G_character}
\ee
\end{widetext}
To obtain this relation, we write $g_{\alpha}^{-1}\,g\,g_{\alpha}=s \in S$,
so $\chi_{\hat{\rho}}(g)=\chi_{\hat{\rho}}(g_{\alpha}sg_{\alpha}^{-1})=\chi_{\hat{\rho}}(s)$, and
elements in the sum do not depend on $\alpha \in A$. We then use the fact that
$|A|=|G|/|S|$. Recall that $S \subset T_i$ and
note that $h_{i}^{-1}\,s\,h_i \in T_i$l, for any $s \in S$, because
$z_i = (h_i)_c (\Phi_{b_{i}}^{-1})$ and $s_c (z_i)=z_i$ imply that
$(h_{i}^{-1}\,s\,h_i)_c (\Phi_{b_{i}}^{-1})= \Phi_{b_{i}}^{-1}$. However, we emphasize
that $s$ and $h_{i}^{-1}\,s\,h_i$ are not a priori related by an \emph{inner} automorphism
of $T_i$. Such relation between $s$ and $h_{i}^{-1}\,s\,h_i$ appears only after summing
over all irreducible representations of the $T_i$ subgroups. 

The completeness relation of characters (in the space of functions on the group $T$ that are
constant on its conjugacy classes) reads~\cite{Sternberg_book}
\be
\sum_{\nu}\chi_{\nu}(t)\chi_{\nu}(t'^{-1})=\sum_{t''\in T}\delta(t''t,t't''),
\label{eq:completeness_characters}
\ee
where the sum runs over the distinct irreducible representations of $T$.
Applying this relation to the $T_i$ groups appearing in Eq.~(\ref{eq:simplify_sum_over_G_character})
yields:
\begin{widetext}
\be
\sum_{\{\nu_i\}} \mathrm{dim}(\Inv_{G}(Y^{\{\nu_i\}}_{\tilde{\mathcal{O}}},\tau^{\{\nu_i\}}_
{\tilde{\mathcal{O}}}))= \frac{1}{|S|}\sum_{t_i \in T_i} \sum_{s\in S}
\prod_i \delta(s h_i t_i,h_i t_i s)\,\chi_{\hat{\rho}}(s).
\label{eq:sum_over_irreps}
\ee
\end{widetext}
The right-hand side of this equation is the same as in Eq.~(\ref{eq_single_tilde_orbit_multiplicity}),
therefore we get our key formula:
\be
\mu_{\tilde{\mathcal{O}}}(\{\rho\}) = \sum_{\{\nu_i\}} \mathrm{dim}(\Inv_{G}(Y^{\{\nu_i\}}_{\tilde{\mathcal{O}}},\tau^{\{\nu_i\}}_
{\tilde{\mathcal{O}}})).
\label{eq:key_formula_positive_genus}
\ee
This generalizes the case of a single gauge orbit $\mathcal{O}$ on the sphere.
In this case, combining Eq.~(\ref{eq_single_orbit_multiplicity}) with
Eq.~(\ref{eq:invariant_count}) and using the one-to-one correspondence between gauge orbits in
$\mathcal{C}$ and $G_c$ orbits in $\mathcal{F}_{0,f}$, we get:
\be
\mu_{\mathcal{O}}(\{\rho\}) = \mathrm{dim}(\Inv_{G}(Y_{\maO},\tau_{Y_{\maO}})).
\label{eq:key_formula_genus_zero}
\ee
The presence of noncontractible cycles on $\Sigma$ for $\genus \geqslant 1$ is accounted for
by the additional objects $U_i \equiv (\mathrm{cl}(u_i),(F_i,\nu_i))$ and their duals.
Let us now sum multiplicities as in Eq.~(\ref{eq:key_formula_positive_genus}) over
all orbits in $\tilde{\mathcal{F}}_{\genus,f}(\{\mathrm{cl}(u_i)\},\{\mathrm{cl}(x_j)\})$.
This gives:
\begin{widetext}
\be
\sum_{\tilde{\mathcal{O}} \subset\tilde{\mathcal{F}}_{\genus,f}(\{\mathrm{cl}(u_i)\},\{\mathrm{cl}(x_j)\})} \mu_{\tilde{\mathcal{O}}}(\{\rho\})=
\sum_{\{\nu_i\}} \mathrm{dim}\left(\HomZvecg(\1,\otimes_{i=1}^{\genus} (U_i \otimes U_{i}^{*})
\otimes X_f \otimes Y_c)\right).
\label{eq:counting_fixed_conjugacy_classes_b_cycles}
\ee
\end{widetext}
Finally, deleting all information on the conjugacy classes of $\Phi_{b_i}$ fluxes, which amounts
to summing all $\mathrm{cl}(u_i)$ conjugacy classes gives the desired counting formula,
Eq.~(\ref{eq:multiplicity_positive_genus}).

\section{Proof of the degeneracy formula through lattice manipulations}
\label{app:steve}
The purpose of this appendix~\footnote{As a warning to the reader, note that in this appendix only, the KQD model is presented on the dual lattice. See the detailed explanation at the very beginning of Appendix~\ref{sec:prelimgeom}.} is to give an alternative proof of the main result for the degeneracies~\eqref{eq:kqdMS} and \eqref{eq:kqdMSinternaldeg} presented in Sec.~\ref{sec:degen}.

The main strategy of the appendix is to make use of the topological invariance of the model.  Given a model to study, the Kitaev quantum double model for a group $G$ on a particular graph on a particular manifold,  the full spectrum of this model can be shown to be exactly the same as the spectrum for another Kitaev quantum double model for the same group on the same manifold
but with a different graph (possibly with a minor change of making a restriction of the fusion at certain vertices).   This equivalence is discussed in detail in Sec~\ref{sub:restructure}.    Given this ability to restructure, we then choose to work with a very simple graph for which the full solution to the model can be written down with very little work.  For the case of the manifold $S^2$  each vertex and each plaquette (save one) can be explicitly put in an eigenstate independently, with only a single plaquette remaining that needs to be handled with a bit of high powered group theory, invoking some of the properties of the representation theory of the quantum double (the necessary facts being introduced in Sec.~\ref{sub:repquant}).  The higher-genus case is only a bit harder.

\subsection{Some results from representation theory of quantum doubles}

\label{sub:repquant}

We present a number of results from the representation theory of quantum doubles. The key reference here is Gould~\cite{Gould_1993}.  Refs.~\cite{Gould2_1994, Beigi11} are also useful and not too hard to read.

Given a group $G$ we can construct the quantum double of the group $D(G)$. The irreducible representations of $D(G)$ are the simple objects in the Drinfeld center ${\cal Z}(\vecg )\simeq  {\cal Z}(\repg)$.

The elements of $D(G)$ have a basis $gh^*$ or $h^*g$ with $h,g \in G$.  Sometimes for clarity we will write $(g,h^*)$.

The irreps of $D(G)$ are described by a conjugacy class $C$ and an irreducible representation $\Pi$  of the centralizer $\mathcal{N}_c$ of a representative element $c$ of the class.   For clarity we will sometimes write $\Pi_{\mathcal{N}_c}$. These irreps of $D(G)$ are the anyon types in the Drinfeld center. 

The anyon types in ${\cal Z}(\vecg)$, or equivalently the irreps of $D(G)$, we will write as
\begin{equation} A = ({C}, \Pi_{\mathcal{N}_c}).  \label{eq:defACPi}
\end{equation}
When it does not cause confusion we will drop the subscript $\mathcal{N}_c$ and just write $({C}, \Pi)$.  Since we use the letter $C$ for a conjugacy class we will not use it to represent an anyon type. 

Note that the inverse element $\bar{A}$ has the inverse conjugacy class $\overline{C}$ and the inverse representation $\overline{\Pi}$ as well.  (The inverse conjugacy class of an element $c$ is the conjugacy class of the inverse element $\bar c  = c^{-1}$.)

We can write characters for these representations, analogous to how we write characters for the representations of a group, and many analogous identities will hold. Here we will write $\underline \chi$ to mean a character of $D(G)$. When we need a character of the group $G$ we write $\chi$ without the underline. 

Define~\cite{Gould_1993}:
\begin{eqnarray*} \underline \chi_{A}(gh^*) &=& {\underline \chi}_{
{({C},\Pi_{\mathcal{N}_c})}}(gh^*), \\
&=& \delta_{h \in {C}} \,\delta_{gh,hg} \,\text{tr}_{\Pi_{\mathcal{N}_c}}(k_h g k_h^{-1}),
\end{eqnarray*}
where $h = k_h c k_h^{-1}$ defines $k_h$ and $c$ is the fixed representative element of conjugacy class $C$.  The trace is taken in the representation $\Pi_{\mathcal{N}_c}$.
From usual group theory we also know that the  $\text{tr}_{\Pi}(g) = \text{tr}_{\Pi}^*(g^{-1}) =  \text{tr}_{\overline{\Pi}}^*(g)$,which will be useful.

From Gould~\cite{Gould_1993} we also have (and we can take this as a definition)
\[\underline \chi_{({C},\Pi)}(gh^*) = \underline \chi^*_{({C},\Pi)}(h^*g^{-1}).\]
Since $A \to \bar A$ in Eq.~\eqref{eq:defACPi} involves ${C} \to \overline {{C}}$ and $\Pi \to \overline{\Pi}$ we have

\beqn& ~~~\underline  \chi_{A}(gh^*) ~=~~ & \underline \chi^*_{\bar A}(g [h^{-1}]^*). \label{eq:chiinv} \\ 
& \uparrow & ~~~~~~~~\uparrow \nonumber \\
& h \in {C}   & ~~~~~~h^{-1} \in \overline {{C}} \nonumber
\eeqn

\vspace*{10pt}

Orthogonality of characters: For $A$ and $B$ being anyon types, i.e., irreps of $D(G)$, we have~\cite{Gould_1993} 
\beqn\frac{1}{|G|} \sum_{g,h \in G} \underline \chi_{A} \nonumber (g^{-1}h^*)\underline \chi_{B} (h^*g) &=&  \\ \frac{1}{|G|} \sum_{g,h \in G} \underline \chi^*_{A}(h^*g)\underline \chi_{B}(h^*g) &=& \delta_{A,B}.
\eeqn

Double conjugacy class:  We introduce the idea of a ``double conjugacy class"  for elements $gh^* \in D(G)$ (we've changed nomenclature from that of Gould~\cite{Gould_1993} who just calls this object a conjugacy class, but Gould's nomenclature is confusing since $G$ also has its own conjugacy classes).   Given an element $(g,h^*)$ in $D(G)$, the double conjugacy class is the set of elements $(pgp^{-1}, (php^{-1})^*)$ for all $p \in G$.  We call this double conjugacy class ${\cal C}_{gh^*}$ (note the calligraphic font to distinguish from the usual group conjugacy class $C$).   We define the inverse double conjugacy class $\overline{\cal C}$ such that $\overline {\cal C}_{gh^*} = {\cal C}_{g^{-1} h^*}$. And we write $|{\cal C}_{gh^*}|$ to be the total number of elements in the double conjugacy class.   

Crucially, given the above definitions of $\underline \chi$, the value of a character is the same for any $gh^*$ within a double conjugacy class.   That is, $\underline \chi$ is a class function. 

Note that by the orbit-stabilizer theorem we have
\begin{equation}
    \frac{|G|}{|{\cal C}_{gh^*}|} = 
 \left\{ \begin{array}{c} \mbox{Number of values of $p\in G$ such that } \\ \mbox{ $ph=hp$ and $pg = gp$} \end{array} \right. . \label{eq:orbstab1}
\end{equation}
 
Completeness relation: Given elements $g_1 h_1^* \in {\cal C}_1$ and $g_2 h_2^* \in {\cal C}_2$ we have the completeness relation~\cite{Gould_1993}
\begin{equation}
  \sum_{A \in {\cal Z}(\vecg)}   \underline \chi_{A}(g_1 h_1^*)   \underline \chi_{A} (g_2 h_2^*)  =  \delta_{\overline  {\cal C}_1, 
 {\cal C}_2}  \frac{|G|}{|{\cal C}_1|},  \label{eq:completeness}
\end{equation}
where the sum is over anyon types.    
Here, to have a  nonzero result, $g_1 h^*_1$ must be in the same double conjugacy class as $g^{-1}_2 h^*_2$. We also must have $g_1 h_1 = h_1 g_1$ and $g_2 h_2 = h_2 g_2$ or the characters vanish. 
Notice that the value on the right when it is nonzero is exactly that given  by Eq.~\eqref{eq:orbstab1}.  
Note that the equation would read the same if we write the arguments as $g_1 h^*$ rather than $h^* g_1$ in both characters. 

Fusion of representations: The anyon types obey fusion relations described by fusion multiplicity matrices $N_{AB}^{F}$ as in Eq.~\eqref{eq:NABC}.   These multiplicity matrices are given by~\cite{Gould_1993} 
\begin{equation}
\label{eq:Nabc}
 N_{AB}^{F} = \frac{1}{|G|} \sum_{g,h,k \in G}
\underline \chi^*_{F} (h^*g)
\, \underline \chi_{A} ((k^{-1} h)^*g)
\, \underline \chi_{B} (k^*g),
\end{equation}
which is very analogous to the fusion rules of usual group representations.    
Note that the arguments with the $*$ multiply 
in the arguments on the right hand side reading right to left:   $k^*
 (k^{-1} h)^* = h^*$.

A convenient rewriting is in the form
\begin{align} \label{eq:Nabc2}
& N_{AB}^{F} = \\ 
&\frac{1}{|G|} \sum_{\scriptsize \begin{array}{c} g \in G \\ \alpha,\beta,h\in G \end{array}}  \delta_{\alpha\beta,h}  \,\, 
 \underline \chi^*_{F}(h^*g) \underline \chi_{A}(\beta^*g) \underline \chi_{B}(\alpha^*g).\nonumber
\end{align}

From this expression it is easy to show the usual symmetry features of $N_{AB}^{F}$.  For  example, using Eq.~\eqref{eq:chiinv} we can immediately derive 
 \begin{equation}
      N_{AB}^{F} =  N_{{B \bar F}}^{\bar A},
 \end{equation}
and also using the fact that $N$ is real 
\begin{equation}
      N_{AB}^{F} =  N_{{\bar  B \bar A}}^{\bar F}. 
\end{equation}
To show symmetry under exchange of the two lower indices we use the fact that in the sum in Eq.~\eqref{eq:Nabc2} the argument is zero unless $\alpha g = g \alpha$ and $\beta g = g \beta$ (and also $h g = g h$ but we don't need this).    We then use the fact that 
\be
\underline \chi_{B}(\alpha^*g)  = \underline \chi_{B}((\beta^{-1} \alpha \beta)^* (\beta^{-1} g \beta)) = \underline \chi_{B}((\beta^{-1} \alpha \beta)^* g),
\ee
so we can replace $\alpha$ by $\alpha' = \beta^{-1} \alpha \beta$ in Eq.~\eqref{eq:Nabc2} only now the factor of $\alpha \beta$ in the $\delta$ function becomes \mbox{becomes $\beta \alpha' \beta^{-1} \beta = \beta \alpha'$} thus switching the order of multiplication, thus deriving
\begin{equation}
      N_{{AB}}^{F} =  N_{{BA}}^{F}. \label{eq:swsym}
 \end{equation}

Just as a sanity check, let's try $N_{{AB}}^{\mathbf{1}}$ with $\mathbf{1}$ meaning the vacuum particle type. 

Here $\underline \chi_{\mathbf{1}}(g^*h) = \delta_{he}$ so we have [using Eq.~\eqref{eq:Nabc}]
\begin{eqnarray}
N_{{AB}}^{\mathbf{1}} &=& \frac{1}{|G|} \sum_{g,h,k\in G} \delta_{he} \,\,\underline \chi_{A}((k^{-1} h)^* g) \underline \chi_{B}(k^*g),\quad 
\\ &=& \frac{1}{|G|} \sum_{g,k \in G}  \underline \chi_{A}({k^{-1}}^*g) \underline \chi_{B}(k^*g),  
\\ &=& \frac{1}{|G|} \sum_{g,k \in G} \underline \chi^*_{\bar {A}}(k^*g)  \underline \chi_{B}(k^*g) = \delta_{{\bar {A}} {B}}, 
\end{eqnarray}
by orthogonality.

Multiple fusions:  The fusion multiplicity relation Eq.~\eqref{eq:Nabc2}
generalizes to describe more than three particles in an obvious way:  
\begin{align} \label{eq:multifus}
& N_{{ABD}}^{F} = \\ 
& \frac{1}{|G|} \!\!\!\!\! \sum_{\scriptsize \begin{array}{c} g,h \in G \\ \alpha,\beta,\gamma\in G \end{array}}  \!\!\!\!\!\delta_{\alpha\beta \gamma,h}  \,\, 
 \underline \chi_{F}^*(h^*g) \underline \chi_{A}(\gamma^*g) \underline \chi_{B}(\beta^*g) \underline \chi_{D}(\alpha^*g),
\nonumber
\end{align}
with the obvious further generalization to more particle types.

\vspace*{10pt}

{\it Proof of Eq.~\eqref{eq:multifus}:}

We define fusion of multiple anyons in terms of fusing only two anyons at a time: 
\begin{equation}
   N_{ABD}^{F} = \sum_{{K} \in {\cal Z}(\vecg)} N_{{AB}}^{K} N_{{DK}}^{F} .
   \label{eq:fusionNNN}
\end{equation}
Substituting in Eq.~\eqref{eq:Nabc}
 twice we get
\begin{align}
   &   N_{{ABD}}^{F}  = \sum_{{K} \in {\cal Z}(\vecg)}  \frac{1}{|G|^2} \nonumber \\ &   \sum_{g,h,k\in G} \underline \chi_{K}^*(h^*g) \underline 
 \chi_{A}((k^{-1}h)^*g') \underline \chi_{B}(k^*g) \nonumber\\ 
 & \sum_{g',h',k'\in G} \underline \chi_{F}^*(h'^*g') \underline \chi_{D}((k'^{-1} h')^*g') \underline \chi_{K}(k'^*g)
\end{align}
We can now use the completeness relation Eq.~\eqref{eq:completeness} to perform the sum over $K$,
\begin{equation}
\sum_{ K \in {\cal Z}(\vecg)}  \underline \chi_{K}^*(h^*g) \underline \chi_{K}(k'^* g') 
= \delta_{{\cal C}_{h^* g}, {\cal C}_{k'^*g'}} \frac{|G|}{|{\cal C}_{h^* g}|}.
\label{eq:comp2}    \end{equation}
The $\delta$ function on the right is nonzero if 
\begin{equation}
k'=php^{-1}  ~~~~~\mbox{and}~~~~~~  g'=pg p^{-1}, \label{eq:kfix}     
\end{equation}
for some  $p$. 

Now we are going to sum over $k'$ and $g'$ and we get nonzero results when Eq.~\eqref{eq:kfix} is satisfied.   But instead of doing this sum, we sum over $p$, and this overcounts by $|G|/|{\cal C}_{h^*g}|$  (by exactly the cases where $p$ commutes with $h$ and $g$), thus accounting for the factor on the right of Eq.~\eqref{eq:comp2}.  We thus have 
\begin{align}
\label{eq:expabk}    &N_{{ABD}}^{F} = |G|^{-2} \times\\
 &   \!\!  \sum_{g,h,k,h',p \in G} \!\!\!\! \underline \chi_{A}((k^{-1} h)^*g) \underline\chi_{B}(k^*g)\underline \chi_{F}^*(h'^*g') \underline\chi_{D}((k'^{-1}h')^*g'),
 \nonumber \end{align}
where $g'=pgp^{-1}$ and $k'=php^{-1}$.    Thus the last two factors  are actually 
$$
\underline \chi_{F}^*(h'^* p g p^{-1}) \underline\chi_{D}((p h^{-1} p^{-1} h')^* p g p^{-1}).
$$
But $\underline \chi$ is a class function (only depends on the double conjugacy class of its argument), so we can conjugate the last two terms by $p$ to get instead the factors    
$$
\underline \chi_{F}^*((p^{-1} h' p)^*g ) \underline\chi_{D}((h^{-1} p h' p^{-1})^* g).
$$
In Eq.~\eqref{eq:expabk} there is a sum over $h'$ and $p$ and these quantities only occur in the combination $p h' p^{-1}$  in these two final factors.   Thus instead of summing over $h'$ and $p$ we instead obtain a factor of $|G|$ times a sum over  $x =p^{-1} h' p$.  Thus we obtain 
\begin{align}
   &N_{ABD}^{F} = |G|^{-1} \\
 &   \times  \!\! \sum_{g,h,k,x \in G} \chi_{F}^*(x^*g)  \underline \chi_{A}((k^{-1} h)^*g) \underline\chi_{B}(k^*g) \underline\chi_{D}((h^{-1} x)^*g),
 \nonumber 
\end{align}
which completes our proof of Eq.~\eqref{eq:multifus}.

Note that $N_{ABD}^{F}$ is fully symmetric in its lower arguments although it is not obvious in this form [see the discussion above Eq.~\eqref{eq:swsym}]. 

The derivation of similar formulas for fusing more anyons together follows in a similar way.

Flux and charge: In conventional gauge theory language a ``fluxon" is a conjugacy class $C$ with a trivial irrep $\Gamma_0$ (we emphasize that $\Gamma   _0$ is here a trivial irrep of the centralizer $\mathcal{N}_c$, not necessarily of the whole group $G$):
\be
\underline \chi_{(C,\Gamma_0)}(g h^*) = \delta_{h \in C} \, \delta_{gh,hg}.
\ee
A ``chargeon" is the identity conjugacy class $e$ with an irrep  $\R$ of $G$:
\be
\underline \chi_{(e,\R)}(g h^*) = \delta_{h,e} \,\, \chi_\R(g),
\ee
where the $\chi_\R$ on the right hand side is the regular group character in irrep $\R$.

\subsubsection{Many fluxons and many chargeons}

Here we consider fusion of $N$ chargeons and $M$ fluxons all fusing to the identity anyon:
\begin{align}  
& N^{\mathbf{1}}_{(e,\R_1)(e,\R_2)...(e,\R_N),(C_1,\Gamma_0)(C_2,\Gamma_0)...(C_M,\Gamma_0)} =  |G|^{-1} \label{eq:manyfus}\nonumber
\\  &\times  \!\! \!\! \sum_{\scriptsize \begin{array}{c} s_1 \ldots s_M \in G \\ g \in G \end{array}} 
 \!\! \left[\prod_{j=1}^N \chi_{\R_j}(g)\right] \left[\prod_{k=1}^M \delta_{s_k \in C_k} \delta_{s_k g,g s_k}\right] \delta_{s_1...s_M,e}.
\end{align}

From the main text we expect the degeneracy on a sphere with $N$ chargeons in irreps $\R_1...\R_N$    and with $M$ fluxons with conjugacy classes $C_1...C_M$  to be given by this expression times the nontopological factor $\prod_{j=1}^N d_{\R_j}$, where $d_\R$ is the dimension of the irrep $\R$. As described in Sec.~\ref{sec:prelimgeom} in the main text, chargeons are on vertices, but in this appendix they are on plaquettes and, conversely in the main text fluxons are on plaquettes, but in this appendix they are on vertices. 
As with the above argument near Eq.~\eqref{eq:swsym}, the order of fusion does not matter (the modular tensor category describing the anyons has commutative fusion) so the order of the $s_i$'s in the $\delta$ function at the end of this equation does not matter.

Below, in Sec.~\ref{sub:sphere}, we will indeed derive Eq.~\eqref{eq:manyfus} but some of the  dummy indices are named differently.

\subsection{Preliminary material about geometry}
\label{sec:prelimgeom}

  In this appendix (and in contrast to the main text), it is more convenient to write the Kitaev model on the dual lattice so it looks more similar to a string-net model~\cite{Levin05} of ${\cal Z}(\vecg)$ rather than  a gauge theory. The equivalence of the direct lattice and dual lattice representation is discussed in chapter 31 of Ref.~\cite{Simon_book}.

In general, we will consider the KQD model for an arbitrary graph forming the skeleton of an arbitrary genus $\genus$ manifold (we write the genus $\genus$ in this font so as not to confuse it with a group element $g$). However, the genus ${\genus} = 0$ (the sphere) is much easier, so we will do that case first in Sec.~\ref{sub:sphere} and then the more general case in Sec.~\ref{sub:higher}.

\begin{figure}[h]
    \includegraphics[width=2in]{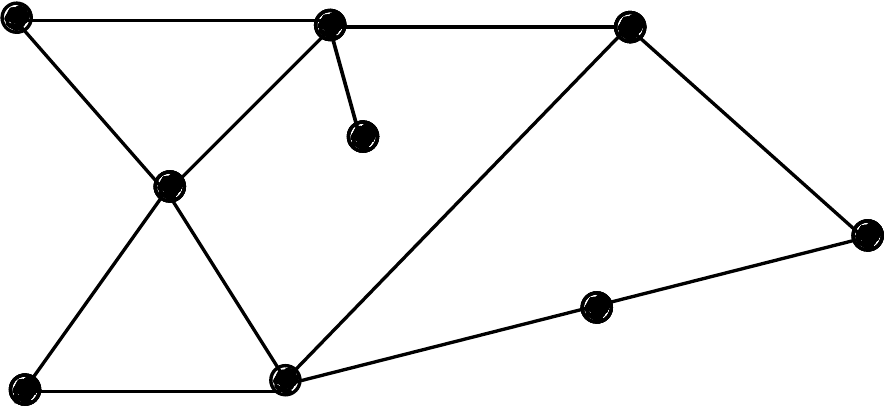}
    \caption{This example geometry has  
    9 vertices, 12 edges, 5 plaquettes
 and genus $\genus= 0$.  The 5th plaquette is the region outside of the figure and includes the point at infinity. 
 If there is no ambiguity we may leave off the dot at the vertex in some diagrams.}
 \label{fig:examplegeometry}
\end{figure}

We start by defining our geometry. Strictly speaking, our geometry is some two-dimensional CW complex.  Our geometry can have vertices of any valence (one or more). An example is given in Fig.~\ref{fig:examplegeometry}.

\begin{figure}[h]
    \centering
    \includegraphics[width=1.5in]
    {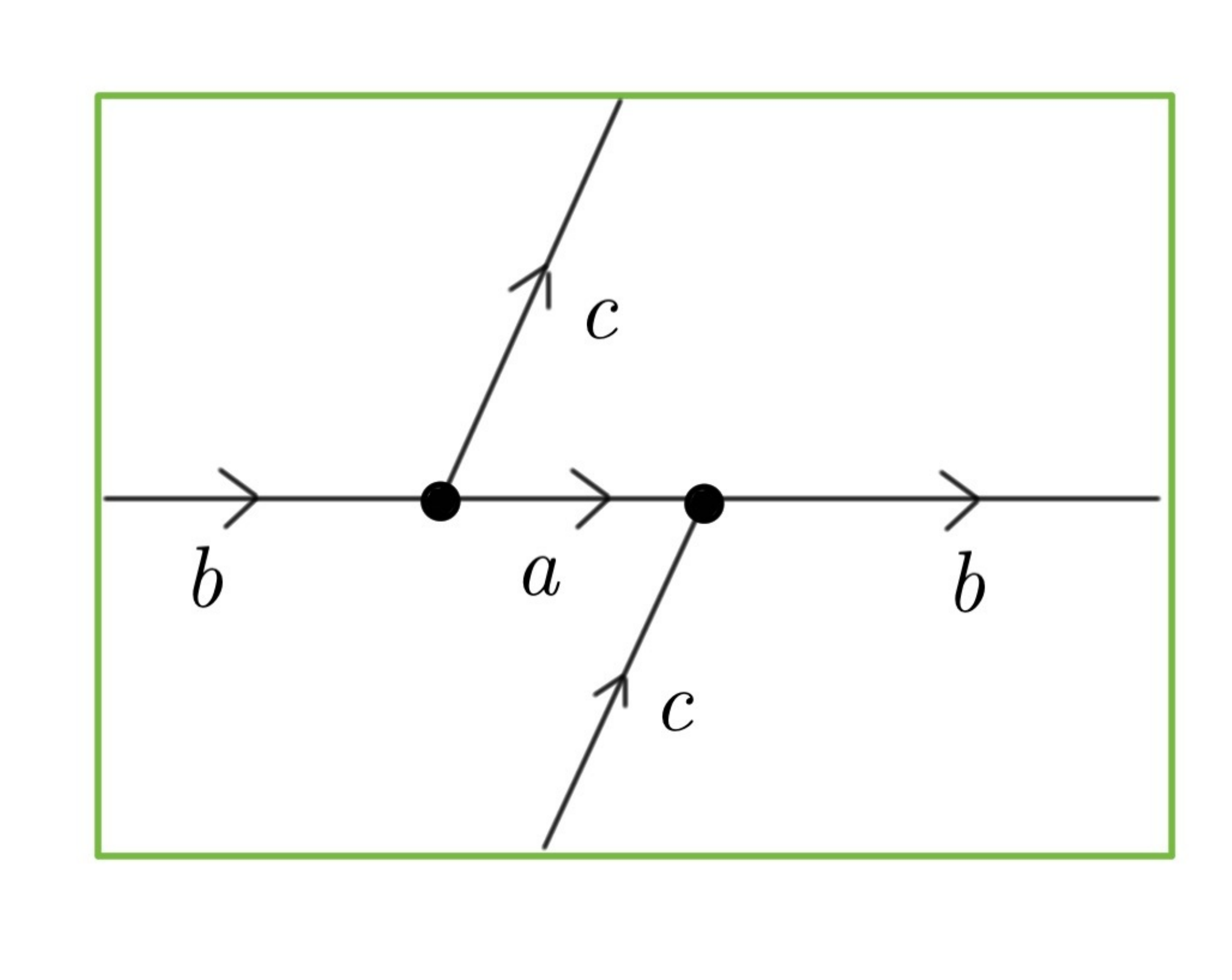}
    
    \vspace*{10pt}
    
    \includegraphics[width=1.5in]{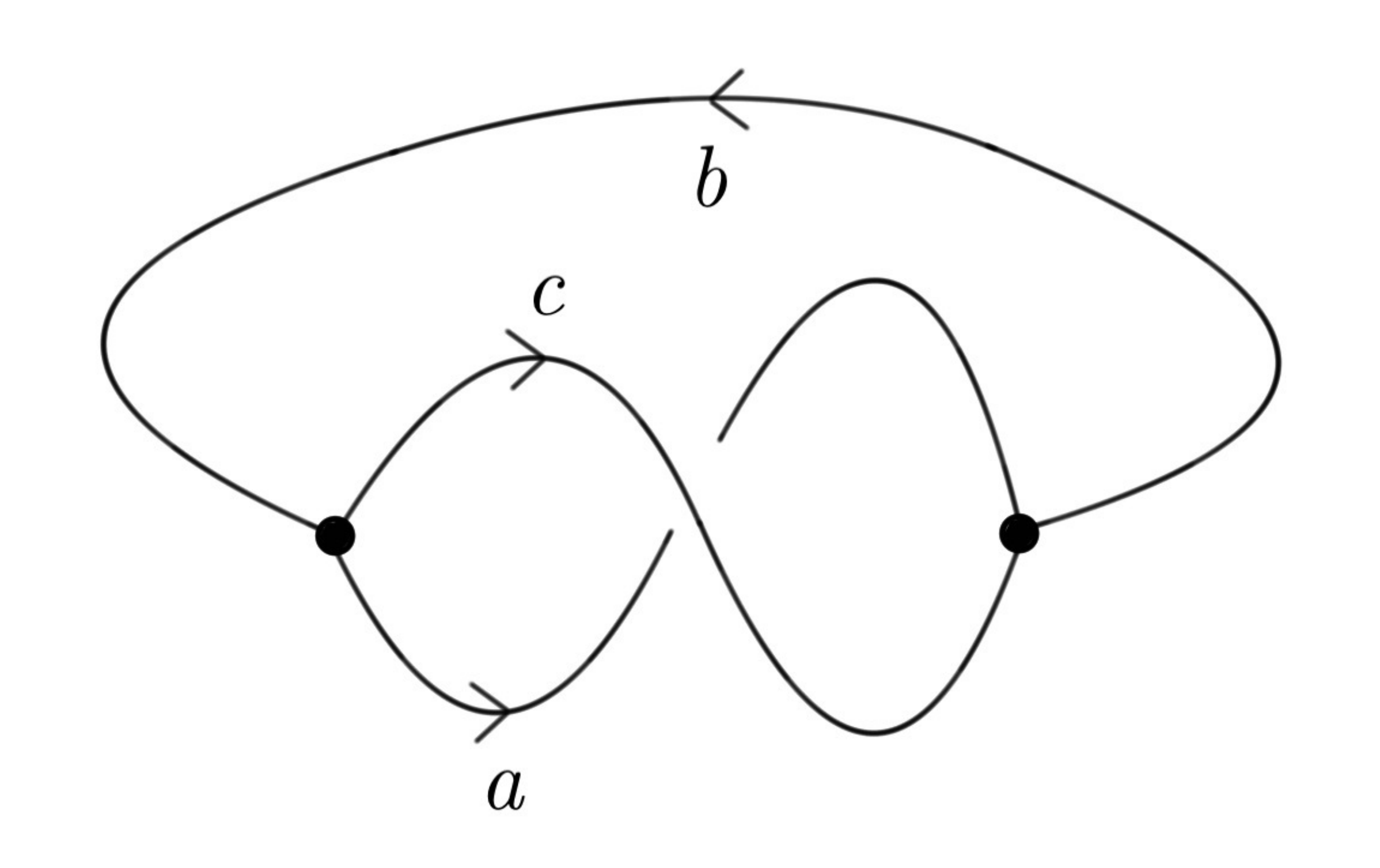}
    \caption{Geometry on a torus (top) specified using periodic boundary conditions. This picture has 2 vertices, 3 edges, and 1 plaquette. Bottom: same picture but without using periodic boundary conditions and instead allowing the lines to go out of the plane.  In the bottom figure it is harder to see the single plaquette.  Note that the bottom figure has the vertices correctly oriented when comparing with the top figure. }    
    \label{fig:periodic-boundary}
\end{figure}

\begin{figure}[h]
    \centering
    \includegraphics[width=1.5in]
    {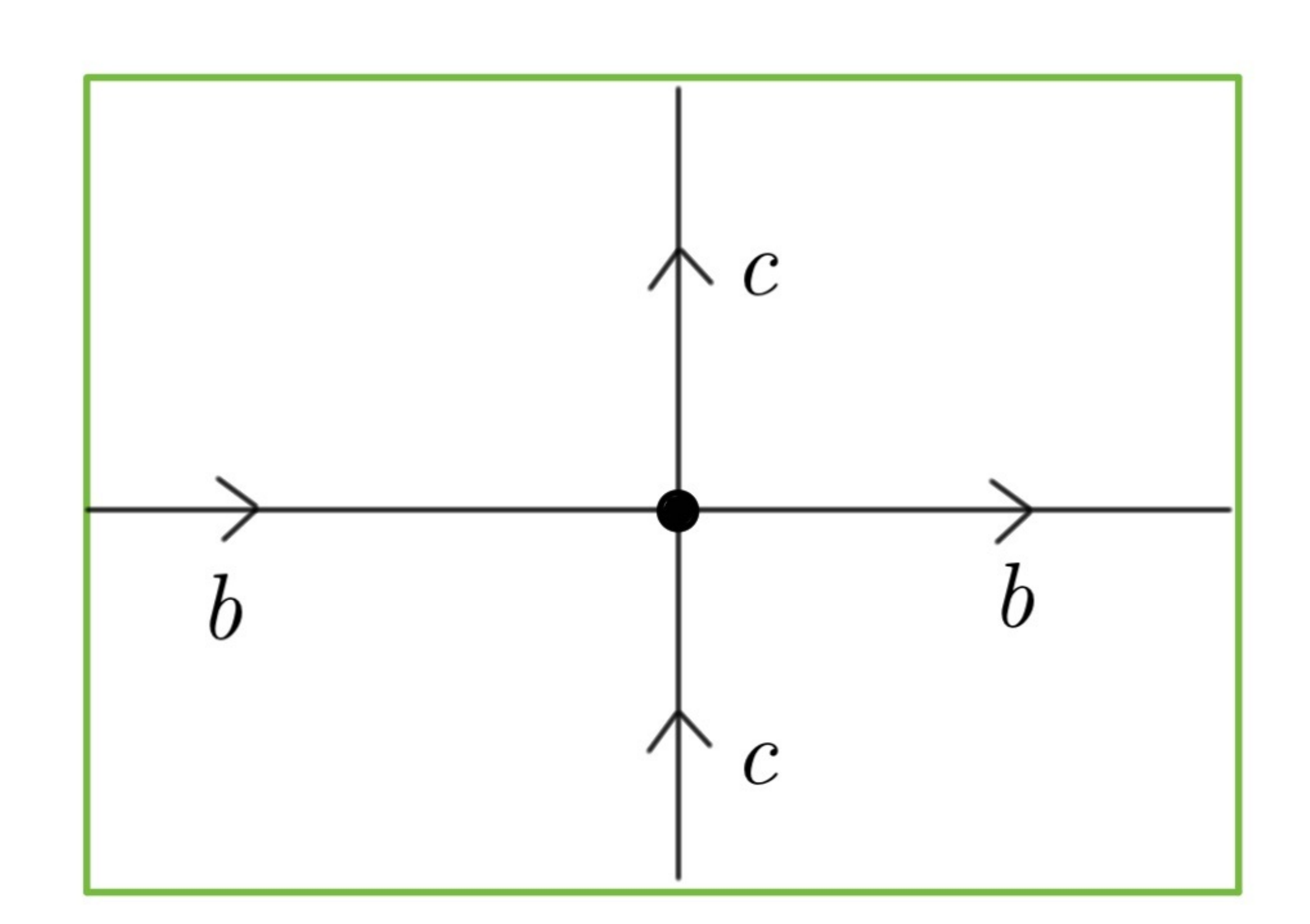}
    
    \vspace*{10pt}
    
    \includegraphics[width=1.5in]{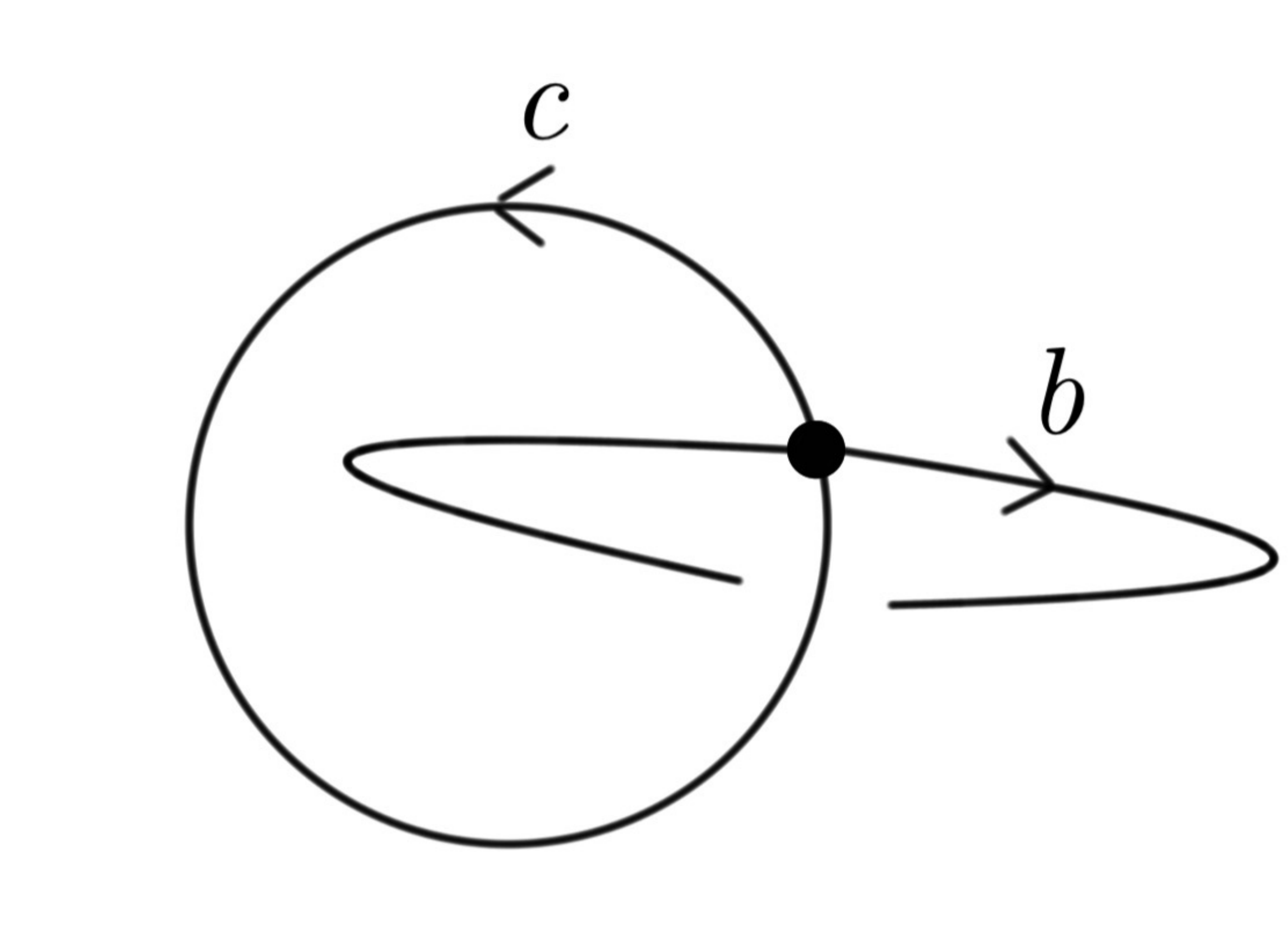}
    \caption{Same torus geometry as Figure \ref{fig:periodic-boundary} except here we use 1 vertex, 2 edges, and 1 plaquette.  }    
    \label{fig:periodic-boundary2}
\end{figure}

For higher genus, we need to:
 either (i)  specify periodic boundaries, or (ii)  allow edges to connect out of the plane of the board.   Examples of showing periodic boundary conditions on a torus are shown in Fig.~\ref{fig:periodic-boundary} and Fig.~\ref{fig:periodic-boundary2}.   In all cases we will assume that the manifold is oriented.

\subsection{Defining the Hilbert space and Hamiltonian}

We choose a group $G$ and assign a group element $g \in G$ to each oriented edge (we draw an arrow on the edge to indicate the orientation).  Reversing the arrow inverts the group element.  A basis for our Hilbert space is the set of all labels of all edges. 

\subsubsection{Vertex term}

The Hamiltonian has a vertex term and a plaquette term  (the notation mostly follows that of  Ref.~\cite{Simon_book}).   We start by defining the vertex term.  Going around the vertex in a clockwise fashion we multiply together the outgoing edge labels and determine the conjugacy class $C$ of the result.   Our vertex term assigns an energy to each conjugacy class. 
\begin{equation}
    \raisebox{-20pt}{\includegraphics[width=1in]{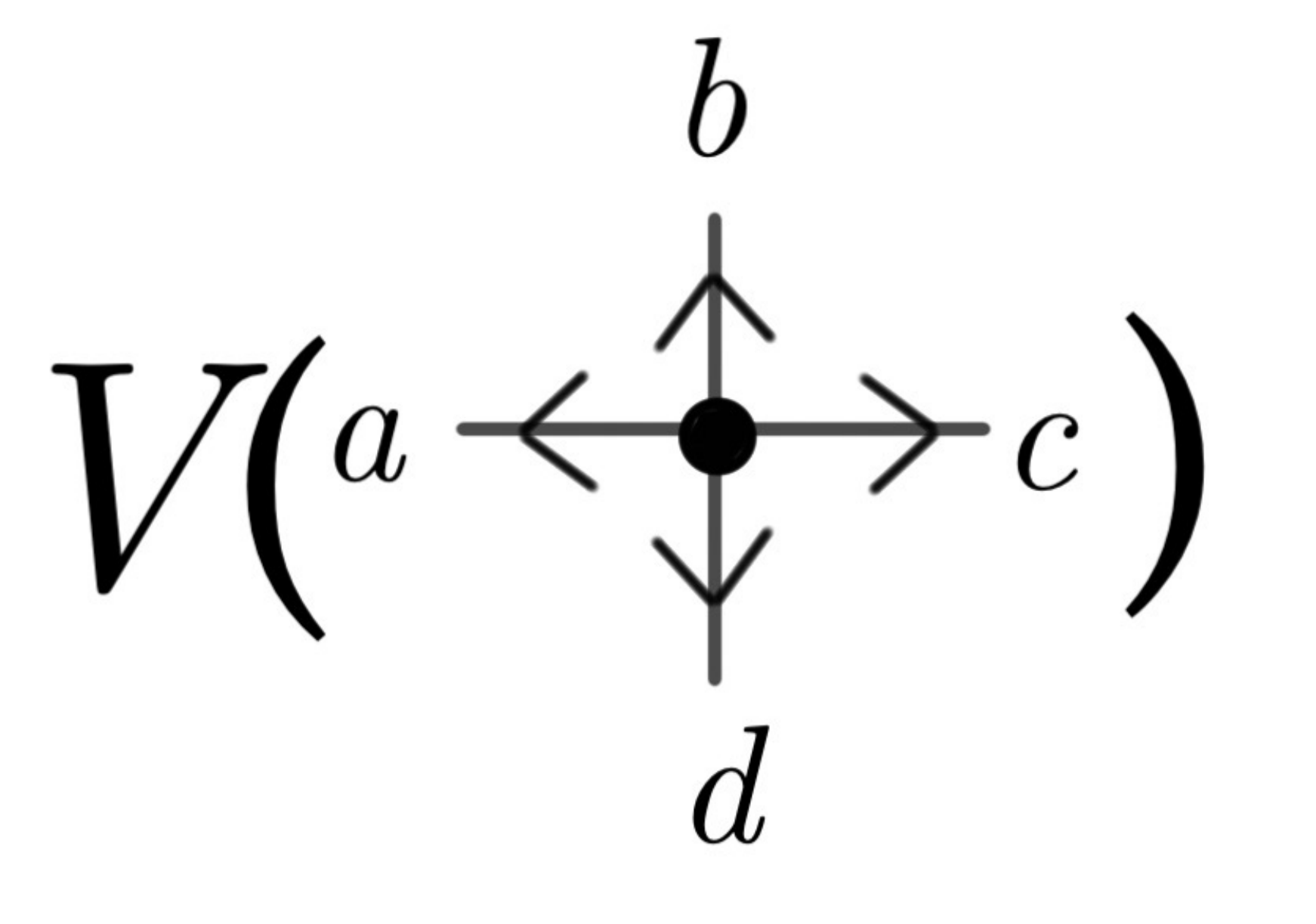}} = \sum_{C} \beta_{C} \,\, \delta_{{C},{ C}_{abcd}}   \label{eq:vertex}
\end{equation}
where the coefficients $\beta_{C}$ are arbitrary real numbers.  Here the sum is over $C$, the possible conjugacy classes of the group, and ${C}_{abcd}$ is the  conjugacy class of the product $abcd \in G$  (note that $bcda$ is in the same conjugacy class so we can start multiplying the product with any edge incident on the vertex as long as we read the outgoing edges clockwise around the vertex).  In other words we are assigning an energy $\beta_{C}$ to the vertex whenever the product of group elements $abcd$ is in conjugacy class $C$.    This rule can be applied to vertices with any number of edges incident.  For example, for three edges incident we would have $\delta_{{C},{C}_{abc}}$.
This vertex term is the $B$ term of the Hamiltonian Eq.~\eqref{eq:Rham}   (and this is a plaquette term in the main text since we are working on the dual lattice) which is a generalized version of the Hamiltonian Eq.~\ref{eq:KQDham} from the main text.    Note that we are free to choose a different set of $\beta_C$ for each vertex in the system. 

Often we take $\beta_{C}$ to  have lowest energy for the trivial conjugacy class ${C} = e$, but this is not required.     However, with this intuition in mind, we sometimes say that a vertex is ``violated" if its conjugacy class is not ${C}=e.$

\subsubsection{Non-violatable vertices}

\label{subsub:nonviolatable}

\begin{figure}
    \centering
\includegraphics[width=.8in]{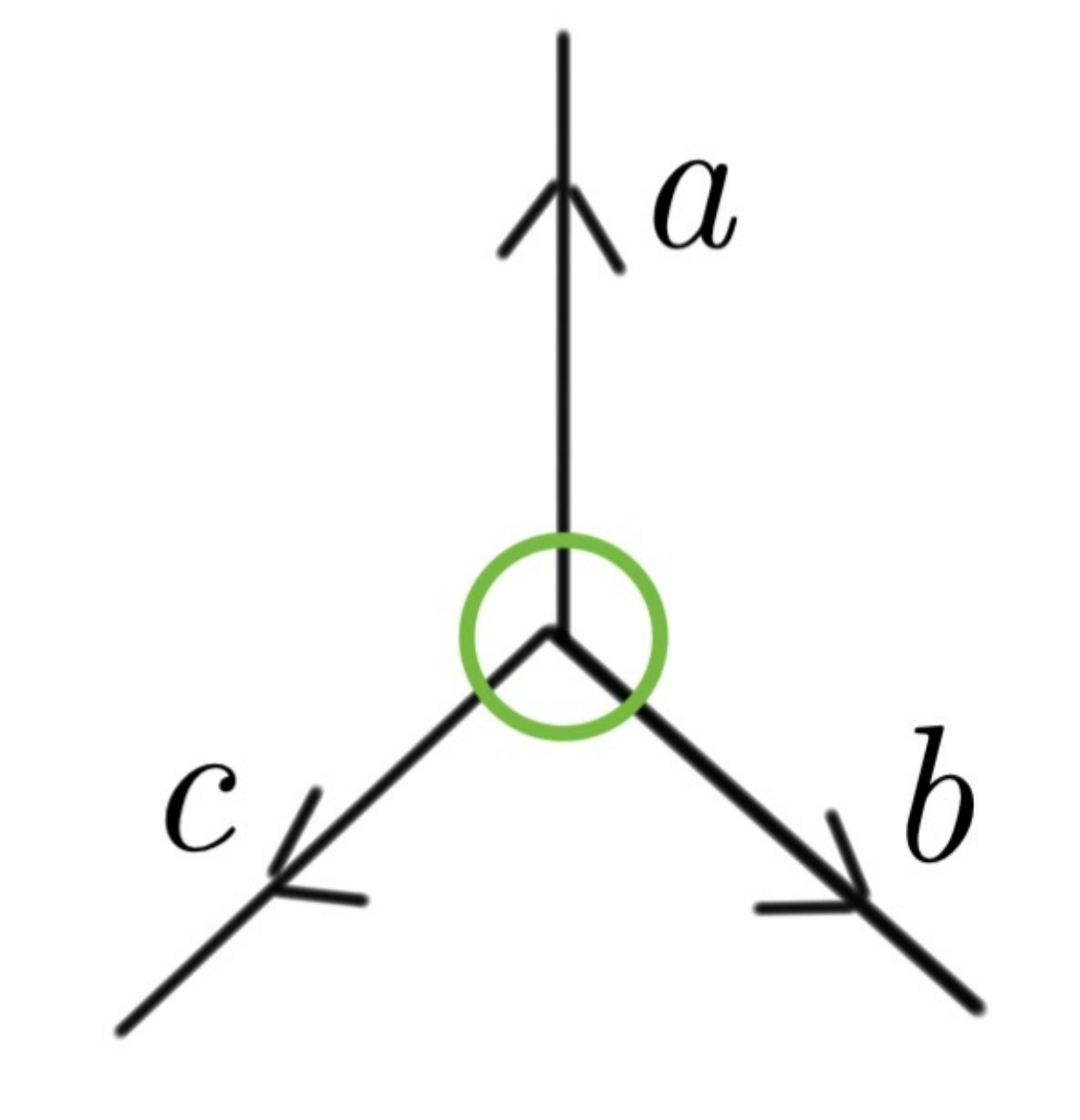} \raisebox{20pt}{implies $ a b c  = e$}~~~~~~~~~~~~~~~~~

\vspace*{10pt}

\includegraphics[width=.8in]{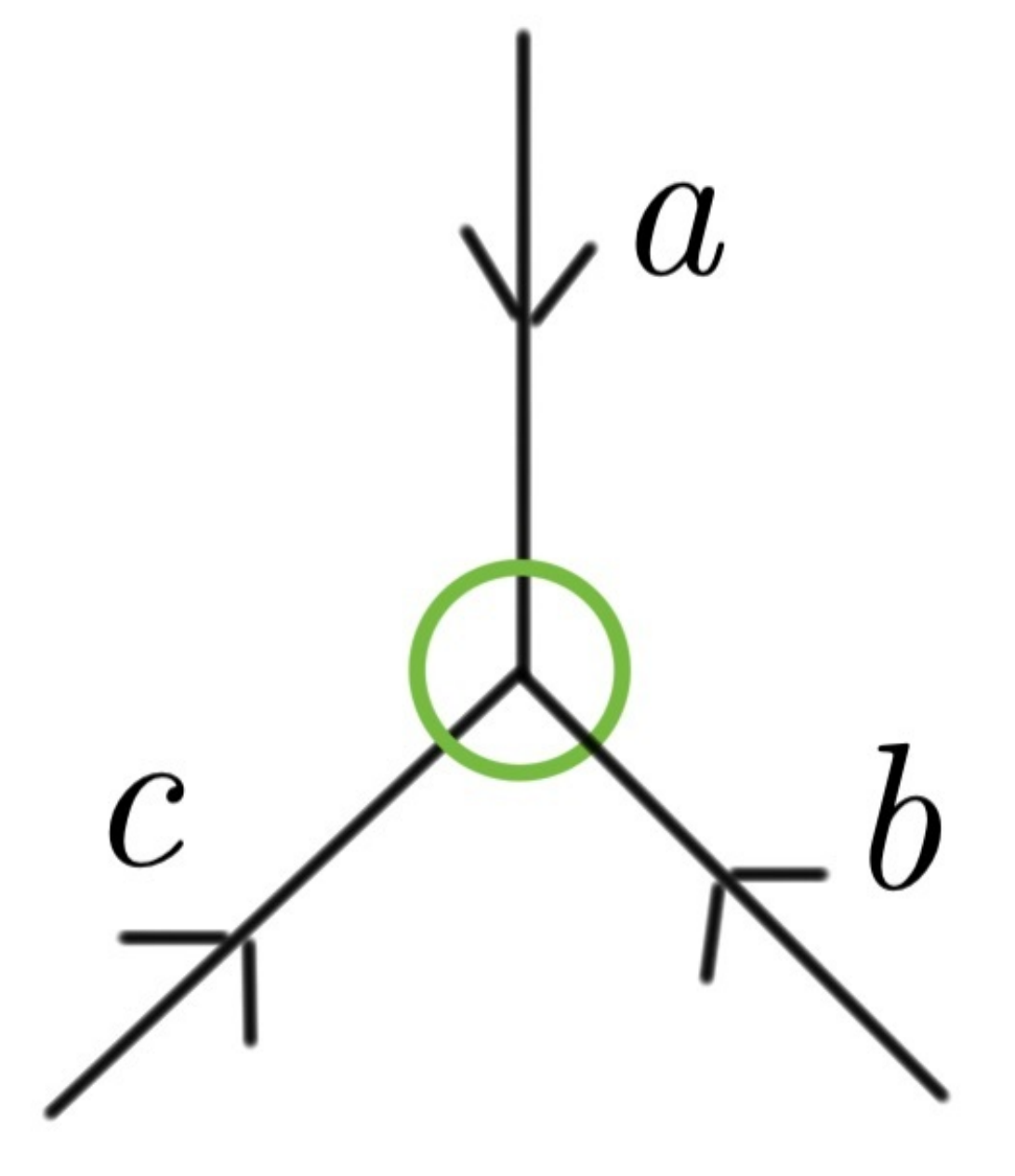} \raisebox{20pt}{implies $ a^{-1} b^{-1} c^{-1}  = e = c b a$}

\vspace*{10pt}

\includegraphics[width=.8in]{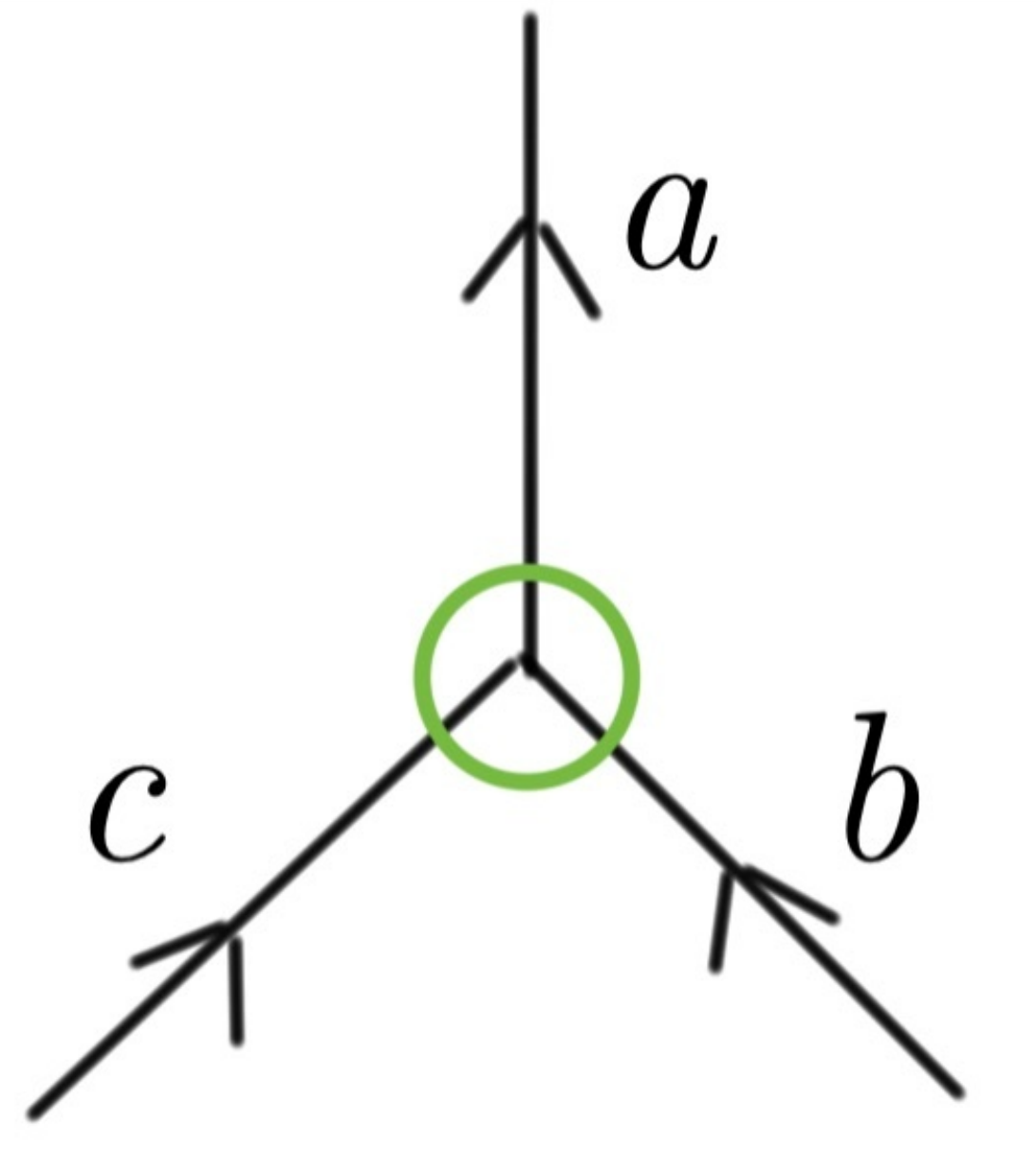} \raisebox{20pt}{implies $cb =a$}~~~~~~~~~~~~~~~~~

    \caption{A green circle around a vertex indicates the Hilbert space is restricted such that this vertex is always in the identity conjugacy class ( i.e., this vertex is never violated).}
    \label{fig:greencircle}
\end{figure}

It will sometimes be useful to restrict our Hilbert space so that certain vertices are always in the identity conjugacy class.   We notate this by drawing a green circle around the vertex as shown in Fig.~\ref{fig:greencircle}.  Such a vertex is never violated. 

\subsubsection{Plaquette term (chargeons)} 

There is also a plaquette term in the Hamiltonian.
Define $\hat{P}(h)$ as 
\begin{equation}
    \includegraphics[width=3in]{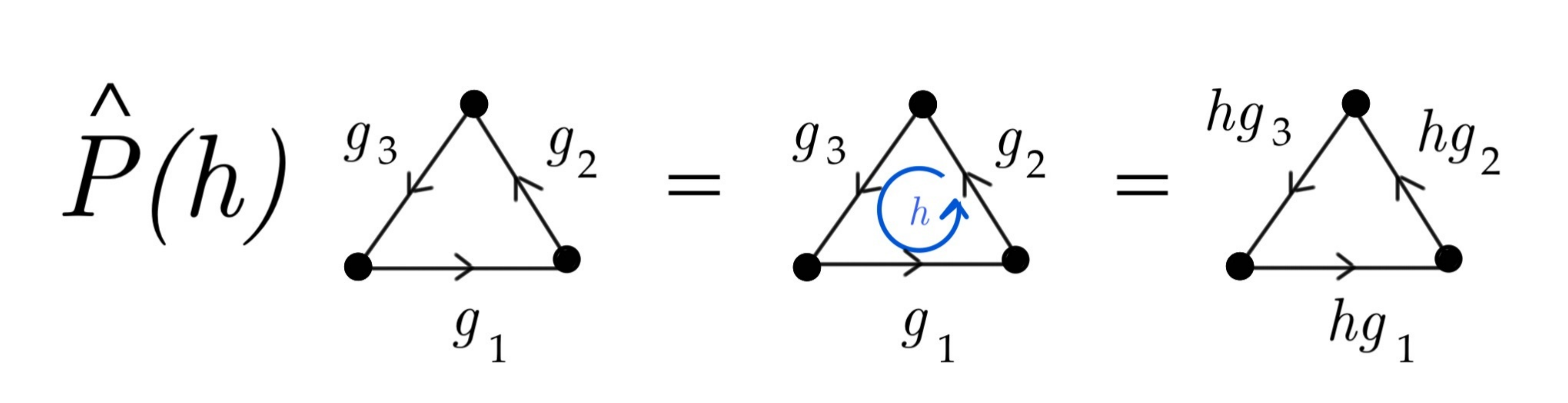}
\end{equation}
and analogously for plaquettes with larger number of sides. 

Note that these operators form a representation of the group 
\begin{equation} 
\hat{P}(h) \hat{P}(g) = \hat{P}(hg),
\end{equation}
We define a set of operators on the plaquette 
\begin{equation}
\hat{P}_{\R nn'} = \frac{d_n}{|G|} \sum_{h \in G} \rho_{nn'}^\R(h) \hat{P}(h), \label{eq:setPs}
\end{equation}
where $\rho_{nn'}^\R$ is an irreducible representation matrix of $G$ (labeled $\R$) of dimension $d_\R$  (so $n,n' \in 1 \ldots d_\R$).   Let us examine multiplication of these operators: 
\begin{align}
\hat{P}_{\R nn'} \hat{P}_{\R' m m'} &= \frac{d_{\R} d_{\R'}}{|G|} \sum_{g,h \in G} \rho_{nn'}^\R(g) \rho_{mm'}^{\R'}(h) \hat{P}(g) \hat{P}(h), \nonumber \\ 
 &= \frac{d_{\R} d_{\R'}}{|G|^2} \sum_{g, h \in G} \rho_{nn'}^\R(g) \rho_{mm'}^{\R'}(h) \hat{P}(gh). \nonumber
\end{align}

Letting $g h = k$, we have 
\begin{align}
& \hat{P}_{\R nn'} \hat{P}_{\R' m m'}
\\ &= \frac{d_{\R} d_{\R'}}{|G|^2} \sum_{g, k \in G} \rho_{nn'}^\R(g) \rho_{mm'}^{\R'}(g^{-1}k) \hat{P}(k), \nonumber 
\\ &= \frac{d_{\R} d_{\R'}}{|G|^2} \sum_{g, k \in G} \sum_{q=1}^{d_{\R'}}\rho_{nn'}^\R(g) \rho_{mq}^{\R'}(g^{-1}) \rho^{\R'}_{qm'}(k) \hat{P}(k) ,\nonumber \\
&= \frac{d_{\R} d_{\R'}}{|G| d_{\R'}} \sum_{k \in G} \sum_{q=1}^{d_{\R'}}\delta_{nq} \delta_{n'm} \delta_{{\R\R'}} \, \rho^{\R'}_{qm'}(k) \hat{P}(k), \nonumber
\end{align}
where going to the last line uses the grand orthogonality theorem Eq.~\eqref{eq:grand}. 

Thus we obtain 
\begin{equation}
    \hat{P}_{\R nn'} \hat{P}_{\R' mm'} = \delta_{n'm} \delta_{{\R\R'}} \hat{P}_{\R nm'}.
    \label{eq:markEq1}
\end{equation} 
If we trace on the lower indices $n,n'$ we get orthogonal projectors
\begin{equation}
    \hat{P}_\R = \sum_{n=1}^{d_\R} \hat{P}_{\R nn} = \frac{d_\R}{|G|} \sum_{g\in G} \chi_\R(g) \hat{P}(g), 
    \label{eq:PRdef}
\end{equation}
with $\chi_\R(g)  = {\rm Tr} \rho^\R(g)$ being the character of the representation. Using Eq.~\eqref{eq:markEq1} we establish that 
\begin{align} \nonumber
\hat{P}_\R \hat{P}_{\R'} &= \sum_{n=1}^{d_\R} \sum_{m=1}^{d_{\R'}} \hat{P}_{\R nn} \hat{P}_{\R' mm},  \\ &= \sum_{n=1}^{d_\R} \sum_{m=1}^{d_{\R'}} \delta_{nm} \delta_{\R\R'} \hat{P}_{\R nm} , \\ &= \delta_{\R\R'} \hat{P}_\R, \nonumber
\end{align}
confirming that these operators are orthogonal projectors [indeed, these are the projectors as defined by Ref.~\cite{Komar17}, see also Eq.~\eqref{eq:Aproj}].

The full plaquette term of the Hamiltonian is
\be
\hat{P} = \sum_\R \alpha_\R \hat{P}_\R,
\ee
for any real values of $\alpha_\R$ and the sum is over representations. This is the same as the $A$ term of the Hamiltonian Eq.~\eqref{eq:Rham}  (and this is the vertex term in the main text since we are working on the dual lattice) which is just a generalized version of the Hamiltonian used in the main text, Eq.~\ref{eq:KQDham}.  Note again that we are free to choose the $\alpha_\R$ to be different on each plaquette of the system. We often choose the trivial rep to be the lowest energy one but we do not have to do so.

\subsection{Related group theory: Rep basis for a loop}

\subsection*{Applying the plaquette $\hat{P}_{\R nn'}$ operator to a single-loop plaquette}

Here we consider an edge labeled $g$ in a loop, which we draw as $\circlearrowleft^g = |g\rangle$ and consider applying $\hat{P}_{\R nn'}$.

Viewing this loop as a plaquette we have 
\begin{equation}
\hat{P}(h) \resizebox{25pt}{!}{$\circlearrowleft^{g}$} = 
\resizebox{25pt}{!}{$\circlearrowleft^{g}$} 
= \resizebox{32pt}{!}{$\circlearrowleft^{h g}$} 
  \!\!\!\!\!\!\!\!\!\!\!\!
  \!\!\!\!\!\!
  \!\!\!\!\!\!
  \!\!\!\!\!\!
   \!\!\!\!\!\!
   \!\!\!\!\!
  {\color{blue} 
\raisebox{1pt}{\resizebox{10pt}{!}{$\circlearrowleft$} }
\!\!\!\!\!\! \raisebox{3pt}{\mbox{\tiny $h$} }}
\label{eq:Pheq}
    \end{equation}
where here we use the graphical notation where we  draw a (blue) $h$ loop and then fuse it into the $g$ edge.

So given the definition, 
\be
\hat{P}_{\R nn'} = \frac{d_\R}{|G|} \sum_{h\in G} \rho_{nn'}^{\R}(h) \hat{P}(h).
\ee
we can work with an irreducible representation basis and things simplify. 

Define a basis of states with $\R$ an irreducible representation and $n,n' \in 1 \ldots d_\R$:
\begin{equation}
|\R n n'\rangle = \sqrt{\frac{d_\R}{|G|}} \sum_{g\in G} \rho_{nn'}^\R(g) |g\rangle.\label{eq:Rnndef}
    \end{equation}  
This is a kind of ``Fourier transform'' between group elements and irreps.  Since $\sum_\R d_\R^2 = |G|$, there are $|G|$ different states here, so these states span the Hilbert space of the one edge. 

Let us check that they are orthonormal as wavefunctions
\begin{align}
\langle \R  nn'|\R'mm'\rangle &= \sqrt{\frac{d_\R d_{\R'}}{|G|}} \sum_{g,g'\in G} \rho_{nn'}^\R(g) \rho_{mm'}^{\R'}(g') \langle g|g'\rangle, \nonumber \\
& = \frac{\sqrt{d_\R d_{\R'}}}{|G|} \sum_{g\in G} \rho_{nn'}^\R(g) \rho_{mm'}^{\R'}(g) ,\nonumber \\ &= \frac{\sqrt{d_\R d_{\R'}}}{|G|} \frac{|G|}{d_\R} \delta_{\R\R'} \delta_{nm} \delta_{n'm'} ,\nonumber
\\ &= \delta_{nm} \delta_{n'm'} \delta_{\R\R'},
\end{align}
again having used the grand orthogonality theorem Eq.~\eqref{eq:grand}. Now let's see how $\hat{P}_{\R ab}$ acts on $|\R'nn'\rangle$
\be
\hat{P}_{\R ab}  = \frac{d_\R}{|G|} \sum_{h\in G} \rho^\R_{ab}(h)\hat{P}(h),
\ee
so that 
\begin{align}
\hat{P}_{\R ab}|\R' nn'\rangle &= \frac{d_\R}{|G|}\sqrt{\frac{d_{\R'}}{|G|}} \sum_{h,g\in G} \rho^\R_{ab}(h)\rho^{\R'}_{nn'}(g)\hat{P}(h) |g\rangle ,\\
 &= \frac{d_\R}{|G|}\sqrt{\frac{d_{\R'}}{|G|}} \sum_{h,g\in G} \rho^\R_{ab}(h)\rho^{\R'}_{nn'}(g)  |hg\rangle ~~~.
\end{align}
Defining $x = hg$ we have 
\begin{align}
& \hat{P}_{\R ab}|\R'nn'\rangle = \frac{d_\R}{|G|}\sqrt{\frac{d_{\R'}}{G}} \sum_{h,x\in G} \rho^\R_{ab}(h)\rho_{nn'}^{\R'}(h^{-1}x)|x\rangle, \\
&= \frac{d_\R}{|G|}\sqrt{\frac{d_{\R'}}{G}} \sum_{h,x\in G} \sum_{m=1}^{d_{\R'}} \rho^\R_{ab}(h)\rho^{\R'}_{nm}(h^{-1})\rho_{mn'}^{\R'}(x)|x\rangle, \\
&= \frac{d_\R}{|G|} \sum_{m=1}^{d_{\R'}} \sum_{h\in G} \rho^\R_{ab}(h)\rho^{\R'}_{nm}(h^{-1})|\R'mn' \rangle, \\
&= \frac{d_\R}{|G|} \sum_{m=1}^{d_{\R'}} \frac{\delta_{am}\delta_{bn} \delta_{\R\R'}|G|}{d_\R}|\R'mn' \rangle,
\end{align}
so
\begin{equation}
\label{eq:funnyop} \hat{P}_{\R ab}|\R'nn'\rangle = \delta_{\R\R'}\delta_{bn}|\R'an'\rangle.
\end{equation}
Note that the $n'$ index is untouched.

Since the plaquette Hamiltonian is written in terms of
\be
\hat{P}_\R = \sum_{a=1}^{d_\R}\hat{P}_{\R aa},
\ee
we have
\begin{equation}
    \hat{P}_\R|\R'nn'\rangle = \sum_{a=1}^{d_\R}\hat{P}_{\R aa}|\R'nn'\rangle = \delta_{\R\R'}|\R nn'\rangle, \label{eq:Rnneig}
\end{equation}
so the $|\R nn'\rangle$ states are eigenstates of the projectors $\hat{P}_\R$.

\subsection{Restructuring lemmas}

\label{sub:restructure}

Our calculation relies on the fact that we  can restructure the graph (the geometry) without changing the spectrum.   This may seem obvious, but it is worth being precise about it. 

Here, we are trying to show that two geometries can be mapped precisely to each other.  We start with some easy lemmas (or ``moves"), which we can use to prove the principle more generally. 

\vspace*{10pt}

{$\bullet$ Move 1:  \it Moves on vertices where no violations are allowed.} 

Here we consider the case mentioned above in Sec.~\ref{subsub:nonviolatable}
 where we restrict the Hilbert space such that ``violations" are not allowed at certain vertices (meaning that the incident edges must fuse to the identity). As above we notate this restriction by circling the vertex in green.   We claim the following equivalence 
$$
 \includegraphics[width=2in]{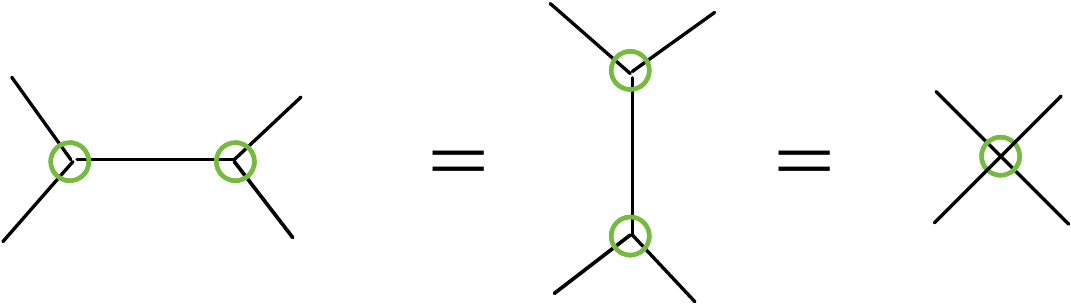}
$$
What we mean by ``$=$" here is that the spectrum of the system remains unchanged when we make these changes to the system geometry.  

 To show this equivalence, we again work in a basis where the edges are labeled. Let us consider the left-most diagram with edges labeled
 $$
 \includegraphics[width=1.25in]{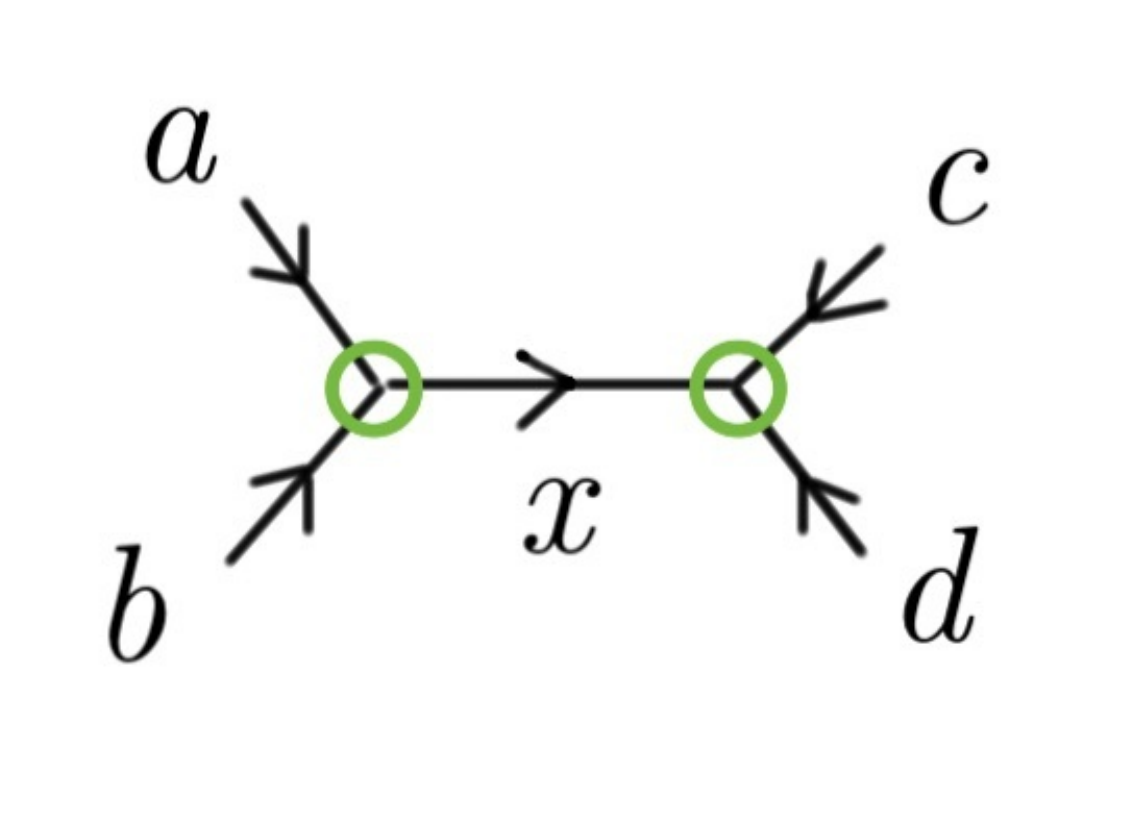}
 $$

 Here, the value of the edge $x = (ab) = (dc)^{-1}$ is completely fixed by the values of the other edges. So, specifying this edge in the Hilbert space is redundant and we can just as well write it as 
 $$
 \includegraphics[width=1in]{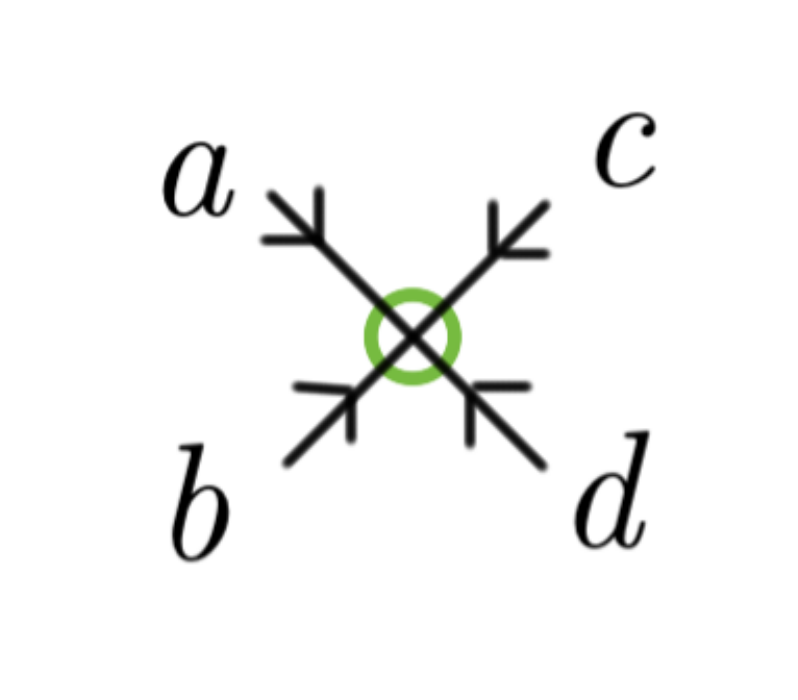}
 $$
where $abdc = e$. Similarly, we can consider the central diagram 
 $$
 \includegraphics[width=.85in]{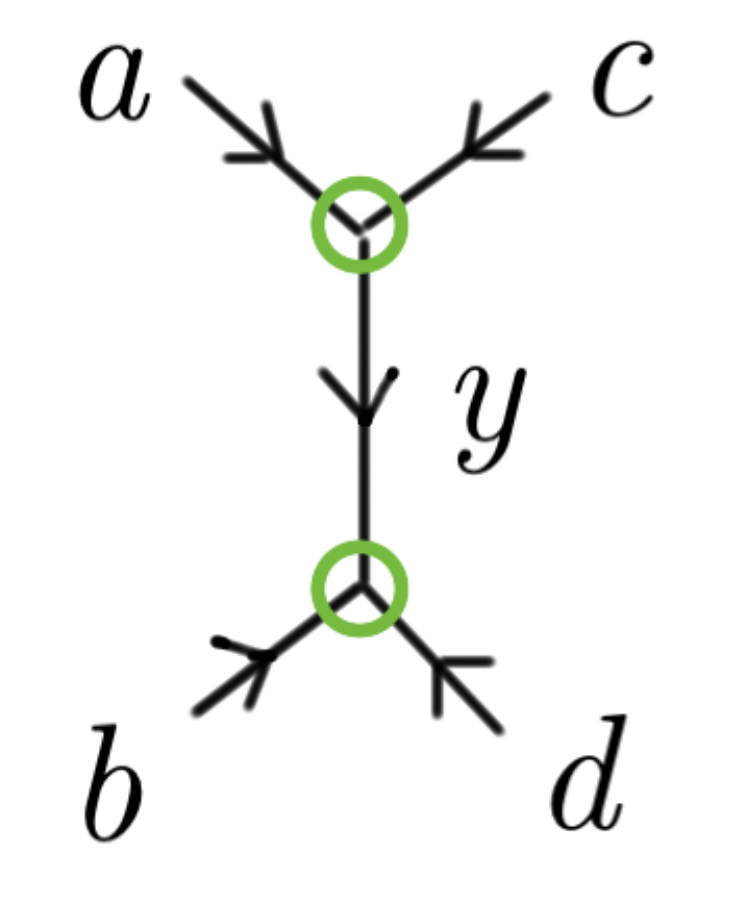}
 $$
and realize that the edge labeled $y$ is similarly redundant since $y = (ca) = (bd)^{-1}$.

\vspace*{10pt}

{$\bullet$ Move 2:  \it Moving a vertex violation away from a vertex.} 

Here, we claim that we can replace a vertex, such as in the left of the following figure, with an nonviolatable vertex, and a new violatable vertex on one of the attached edges as on the right of the figure. 
 $$
~~~~~\includegraphics[width=2.5in]{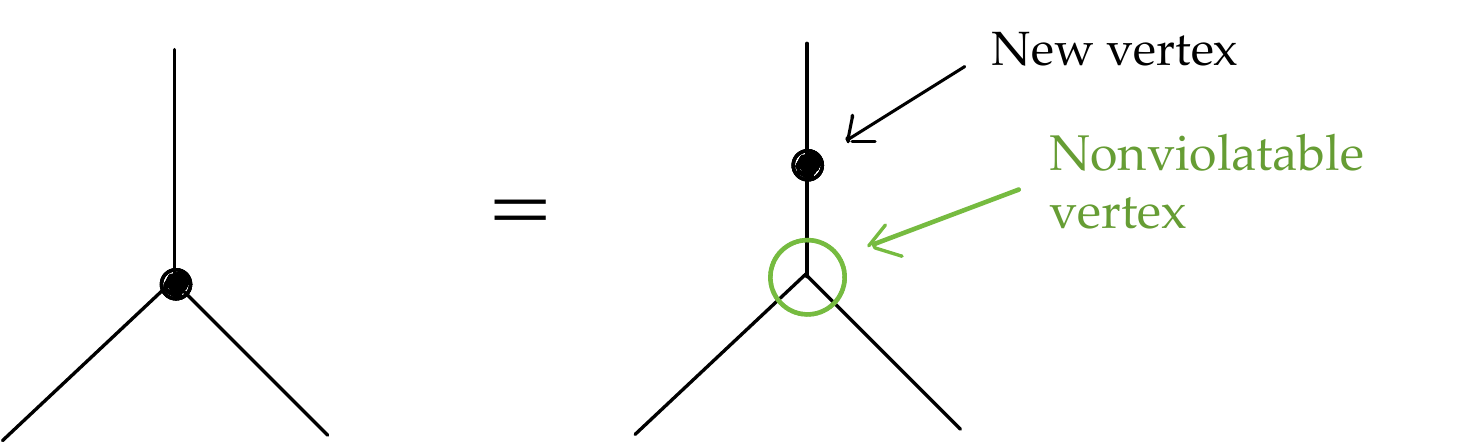}
 $$
This can be done with a vertex of any valence.  The argument here is similar to that of the previous move.   To see this we label the edges
$$
~~~~~\includegraphics[width=2.25in]{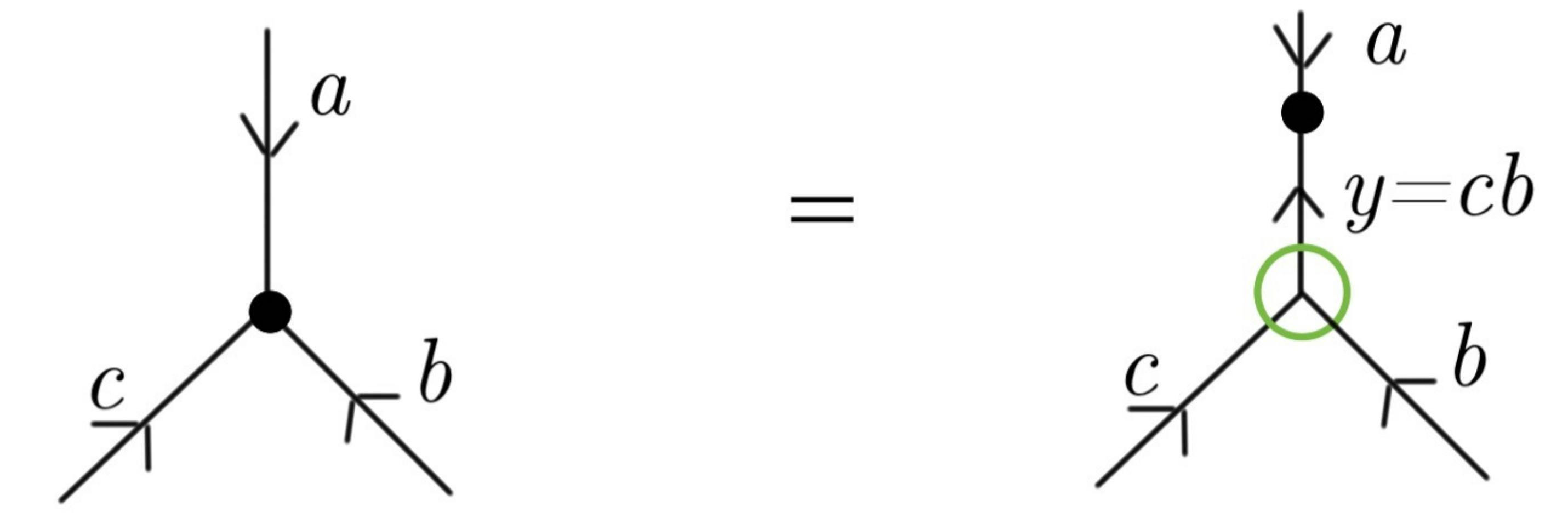}
 $$
and we see that the newly formed edge marked $y$ has a value that is completely defined by the other edges, and thus is redundant information.  Furthermore, we see that the conjugacy class of the violatable vertex is the same both before and after the move $C_{a^{-1} b^{-1} c^{-1}}$, as in the picture in both cases. 
Again, the meaning of equivalence of diagrams is that the spectrum remains the same before and after making the move. 

From these moves (and similar thinking) we can derive many other useful transformations such as 
$$ \includegraphics[width=1.75in]{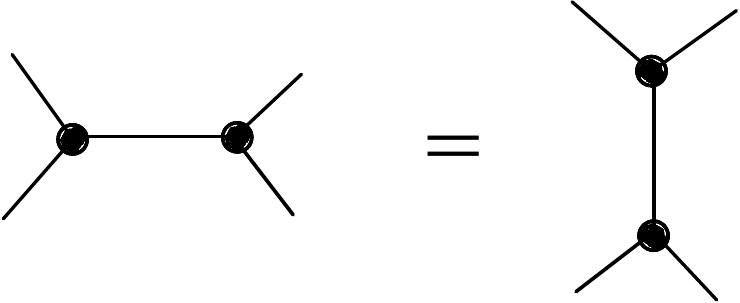}
$$
where now both vertices are allowed to be violated. We can derive this by using Moves 1 and 2 above in the following sequence
$$ \includegraphics[width=3in]{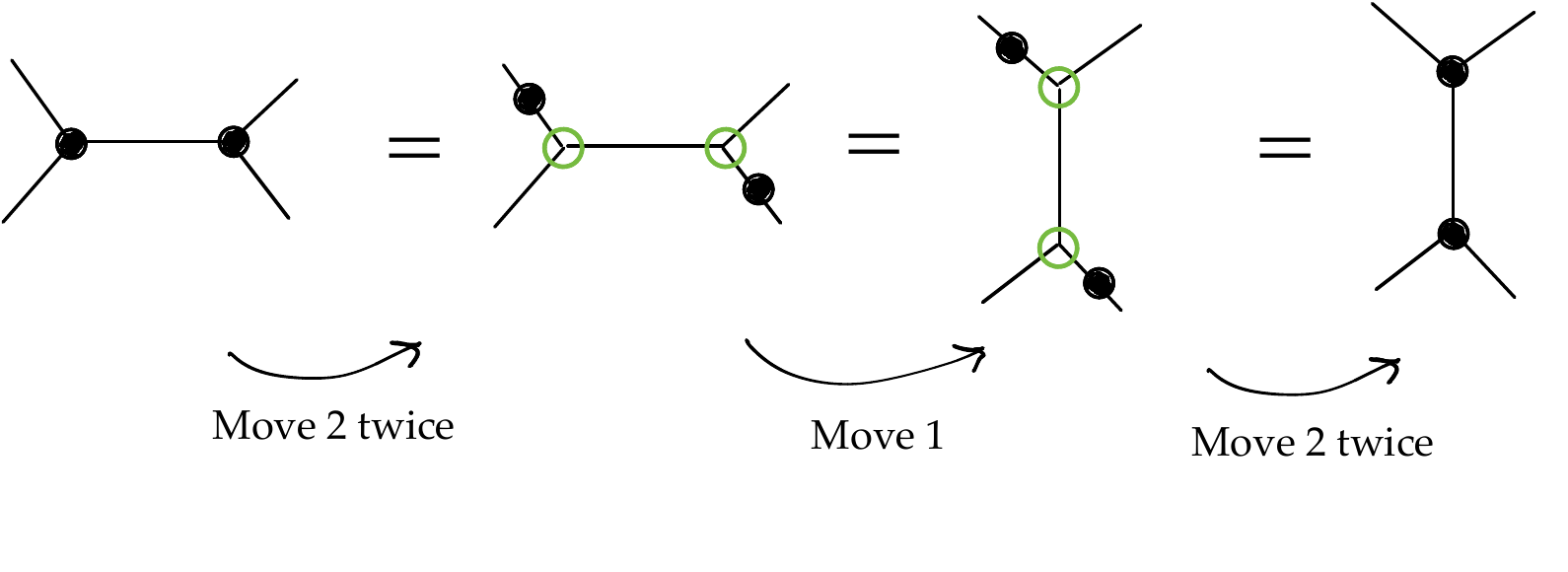}
$$
Similarly, we can derive transformations such as 
$$ \includegraphics[width=2in]{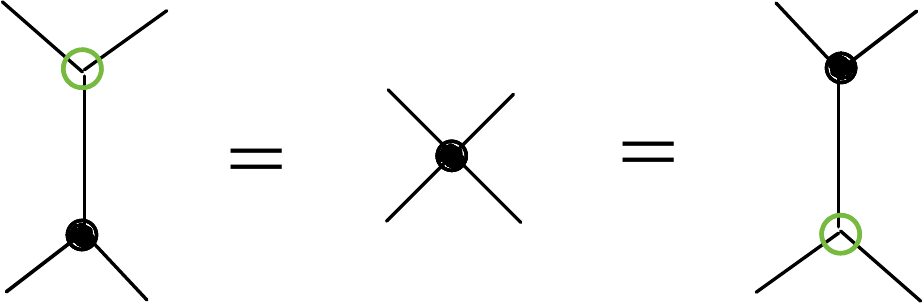}
$$
and
$$ \includegraphics[width=1.7in]{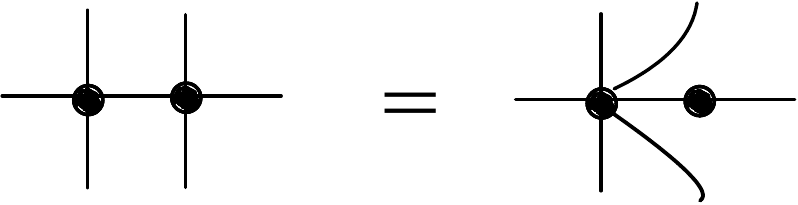}
$$
We can also freely add nonviolatable vertices
$$ \includegraphics[width=1.7in]{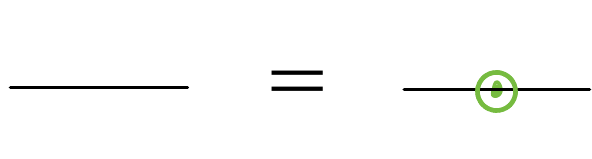}
$$
and even edges attached to nonviolatable tadpoles (single-valent vertices). 
$$ \includegraphics[width=1.7in]{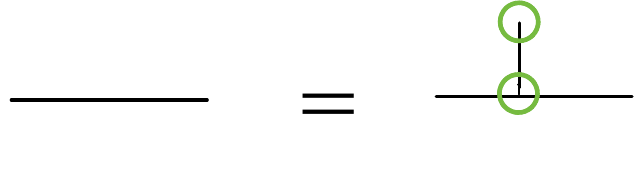}
$$
where we see that the added edge can only be in the identity channel here.

\vspace*{10pt}

For all types of transformations we have the following conserved quantities:

\begin{enumerate}
    \item The number $N_P$ of plaquettes is conserved. 
    \item The number of {\it violatable} vertices $N_V$ is conserved. 
    \item The genus $\genus$ of the manifold is unchanged.   
\end{enumerate}
The number of independent edges $N_E$ (i.e., the number of degrees of freedom and hence the dimension of the Hilbert space $|G|^{N_E}$) is also unchanged. This can be viewed as being a consequence of the Euler-Poincar\'e relation $2-2{\genus}= N_V-N_E+N_P$, and the above conserved quantities. Or, equivalently, we can check that each individual move conserves the number of independent edges. 

A few brief comments on these conserved quantities: for the number of plaquettes and number of violatable vertices, recall that the Hamiltonian most generally can have a different set of coefficients $\alpha_R$ for each plaquette and a different set of coefficients $\beta_{C}$ for each violatable vertex. Clearly, if the spectrum is to be unchanged, then the number of these plaquettes and
violatable vertices had better remain unchanged.   For the genus $\genus$ of the manifold we simply realize that all of these moves are local and cannot restructure the manifold's topology globally.      And crucially under all of these moves, the spectrum remains unchanged. 

\subsection{Simple canonical geometries}

\begin{figure}[h]
    \centering
    \includegraphics[width=\linewidth]{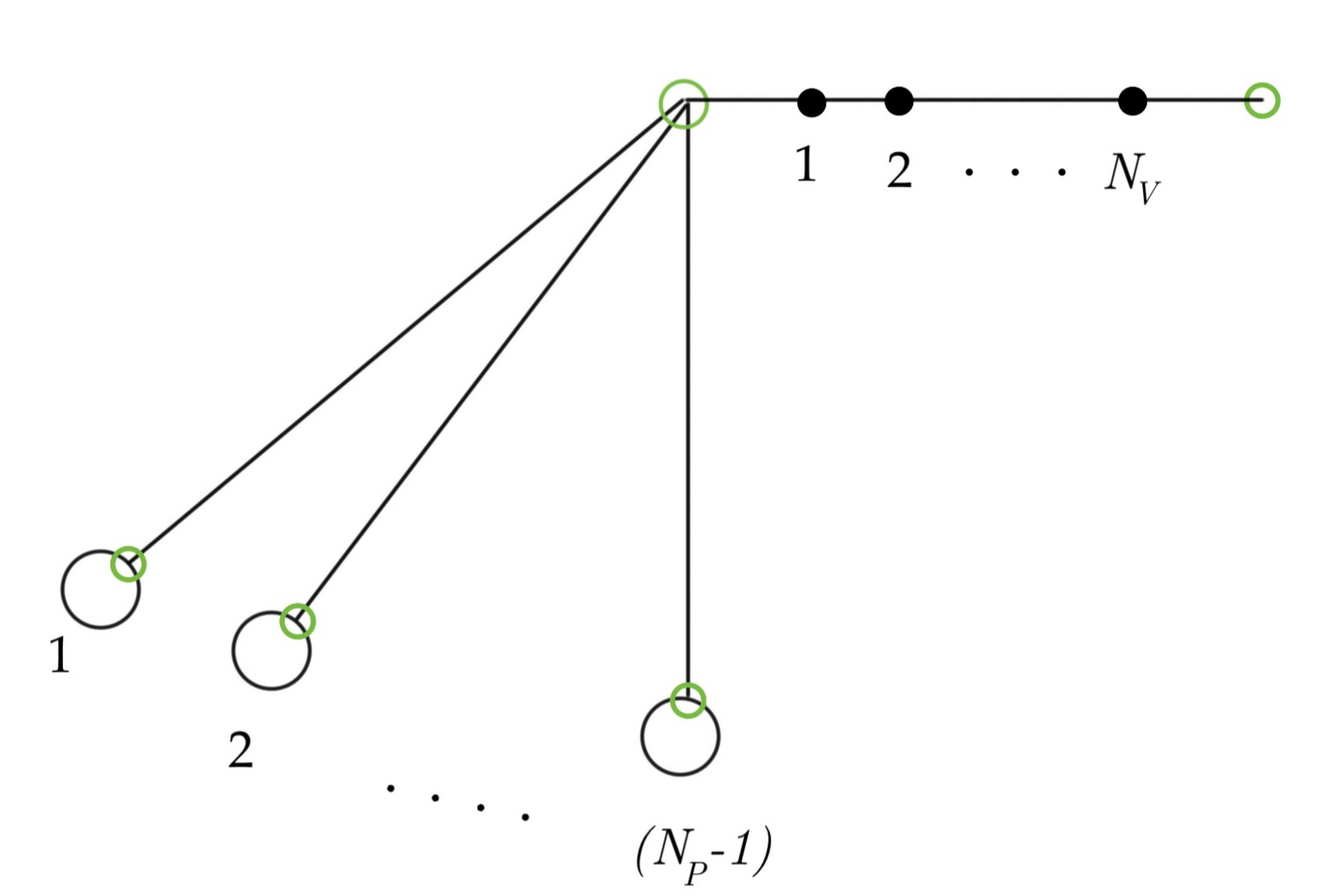}
    \caption{Canonical geometry for genus ${\genus}=0$.  We can reduce all ${\genus}=0$  geometries into a geometry of this form by using the above restructurings.   There are $N_V$ violatable vertices across the top.  $(N_P-1)$  of the plaquettes are loops in the lower left.  The $N_P^{\rm th}$ plaquette is the region outside of the figure, including the point at infinity. }
    \label{fig:canonical-geom-g0}
\end{figure}

Given our restructuring lemmas, it is useful to reduce all geometries to a fixed canonical very simple geometry.   There are many such simple geometries we might choose, but a good one is of the type shown  in Fig.~\ref{fig:canonical-geom-g0} for genus ${\genus}=0$.  In this figure 
there are $N_V$ violatable vertices across the top row (and no other vertex violations are allowed).  In the lower left there are $(N_P -1)$ loops, each corresponding to one plaquette.  The $N_P^{\rm th}$ plaquette is the region outside of the drawn figure, including the point at infinity.    

In this canonical geometry we can think of the $(N_P-1)$ plaquettes as being local objects and the $N_P^{th}$ plaquette is global.

\begin{figure}[h]
    \centering
    \includegraphics[width=\linewidth]{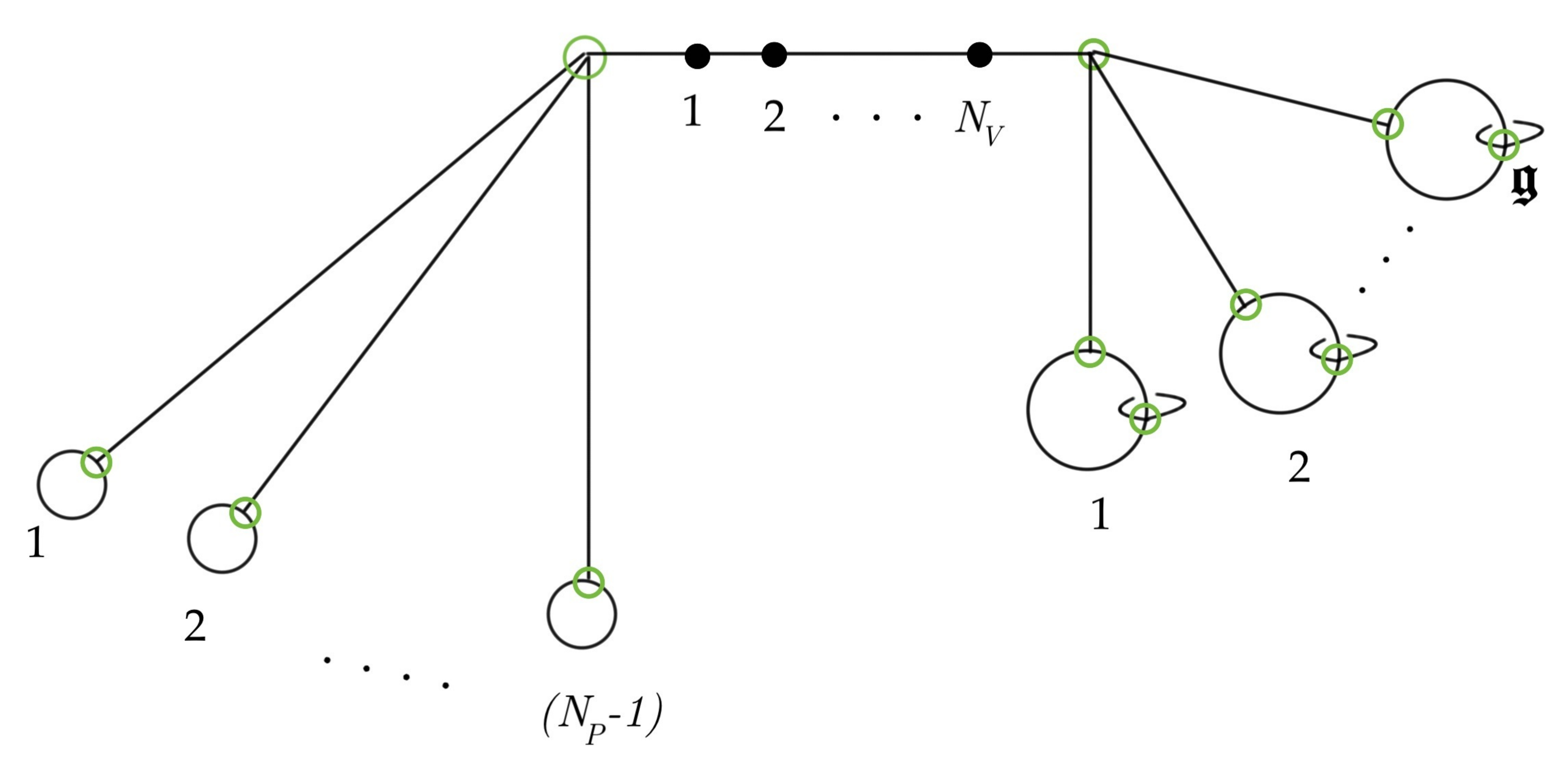}
    \caption{Canonical geometry for genus ${\genus} > 0$.   As above there are $N_V$ violatable vertices across the top, and $(N_P-1)$ plaquette loops in the lower left.  The $N_P^{\rm th}$ plaquette is still the region outside the diagram.  }
    \label{fig:canonical-geom}
\end{figure}

This canonical geometry can be extended to higher genus manifolds as in Fig.~\ref{fig:canonical-geom}.  Here each of the $\genus$ handles is presented as two connected loops over on the right hand side with an underpass between them as in the lower part of Fig.~\ref{fig:periodic-boundary2}.

\begin{figure}
    \centering
    \includegraphics[width=3in]{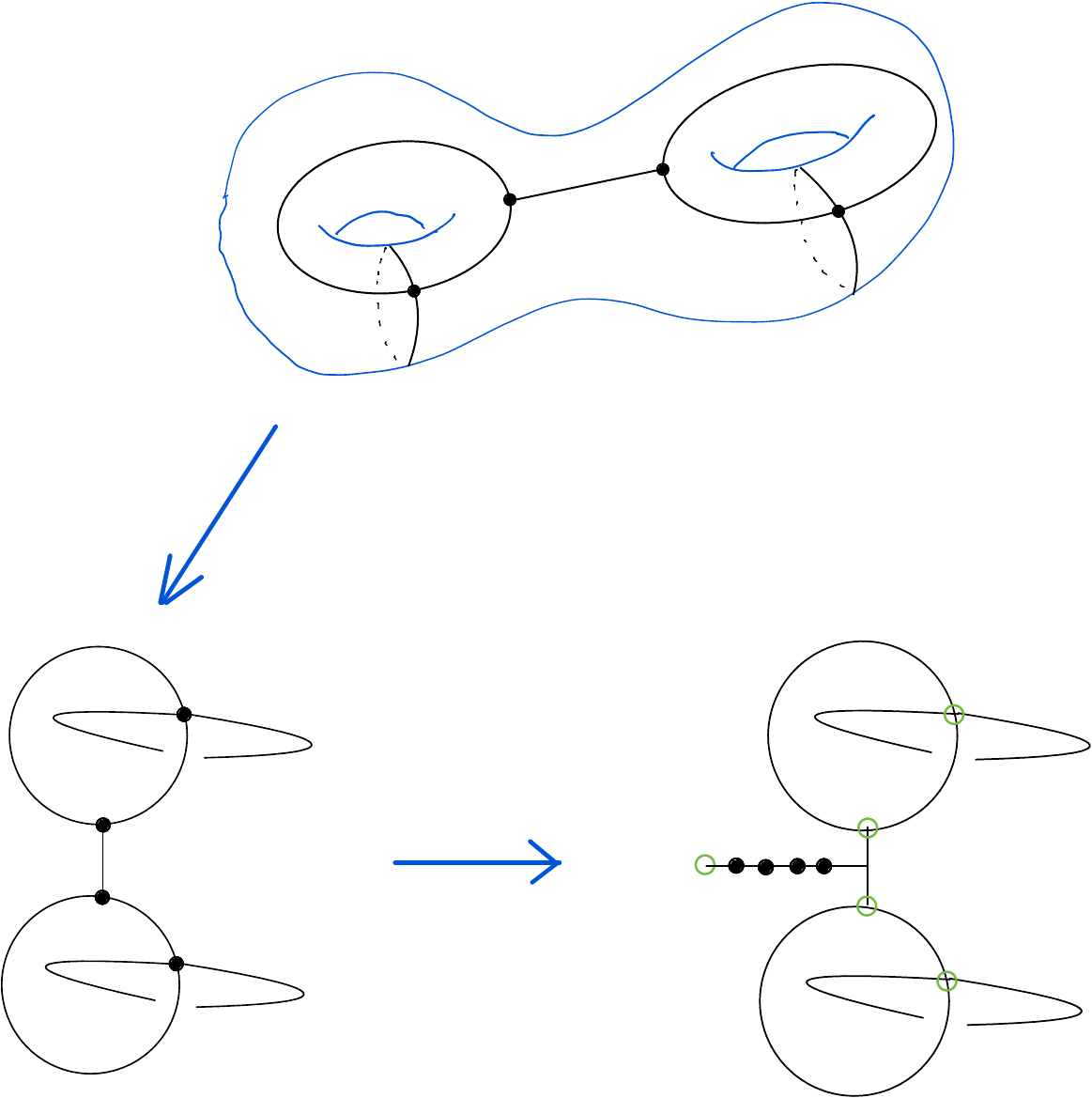}
    \caption{Two-handled torus with four violatable vertices and a single plaquette.  In going to the final figure on the bottom, we use restructuring moves to replace all of the original vertices with nonviolatable vertices and move the violatable vertices to the horizontal line in the middle.}
    \label{fig:g2example}
\end{figure}

Just to give an example of how such a structure arises, we consider the ${\genus}=2$ case with a single plaquette as shown in Fig.~\ref{fig:g2example}.

\subsection{Spectrum on a sphere}
\label{sub:sphere}

\begin{figure}[h]
\includegraphics[width=3in]{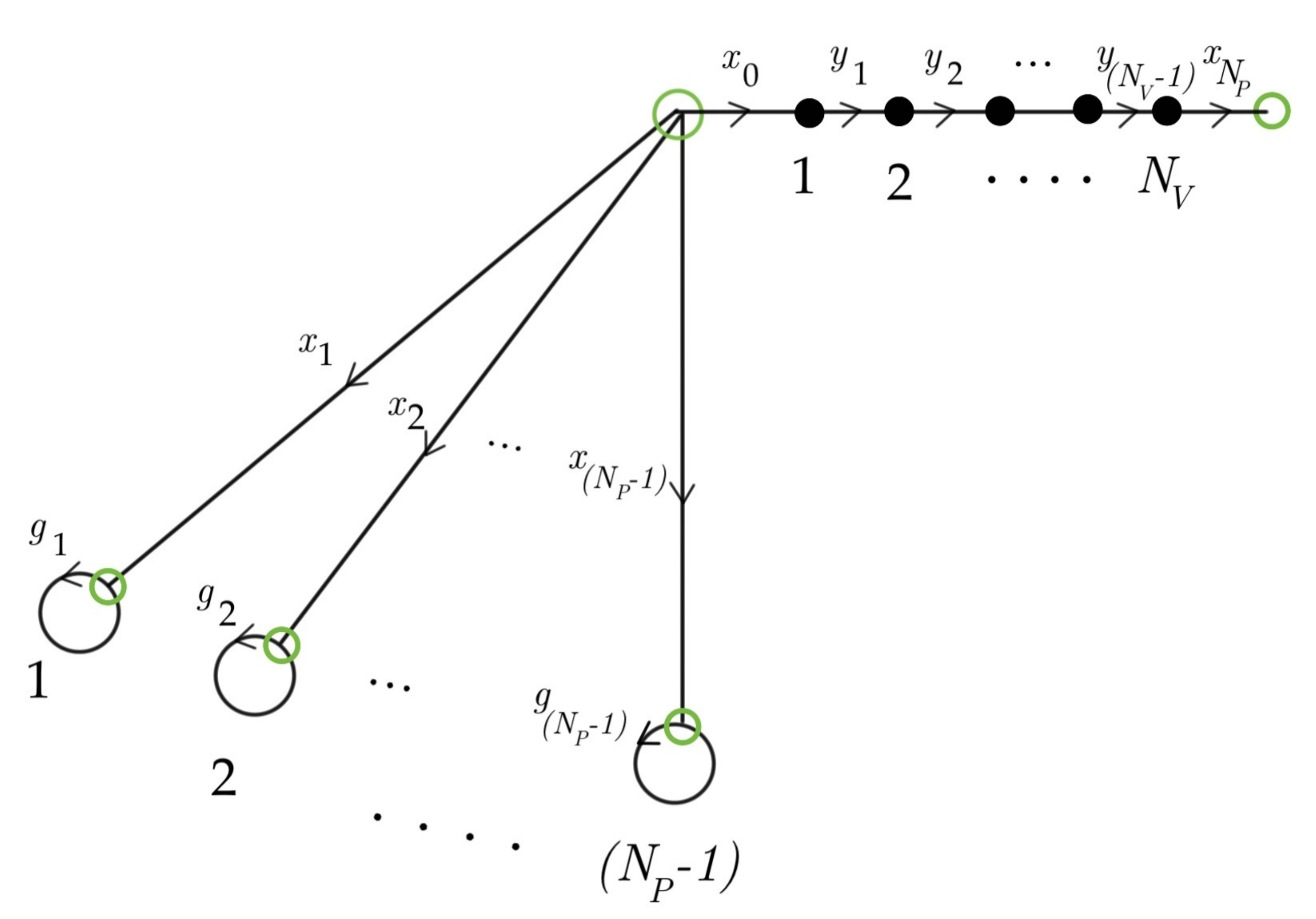}
    \caption{Geometry for a sphere as  in Fig.~\ref{fig:canonical-geom-g0} but here edges are labeled.}
    \label{fig:canonical-geom-g01}
\end{figure}

We will start by calculating the spectrum given that the genus of our surface is ${\genus}=0$, i.e., we are considering a sphere.   Our geometry will be that of Fig.~\ref{fig:canonical-geom-g0}, but here let us label the edges as in Fig.~\ref{fig:canonical-geom-g01}.    Due to the constraints from the green circles all of the edges labeled with $x_i$ must have only the identity label $x_i = e$, leaving only the $y_i$ degrees of freedom. 

\begin{figure}[h]
\vspace*{10pt}
~~~~\includegraphics[width=2.5in]{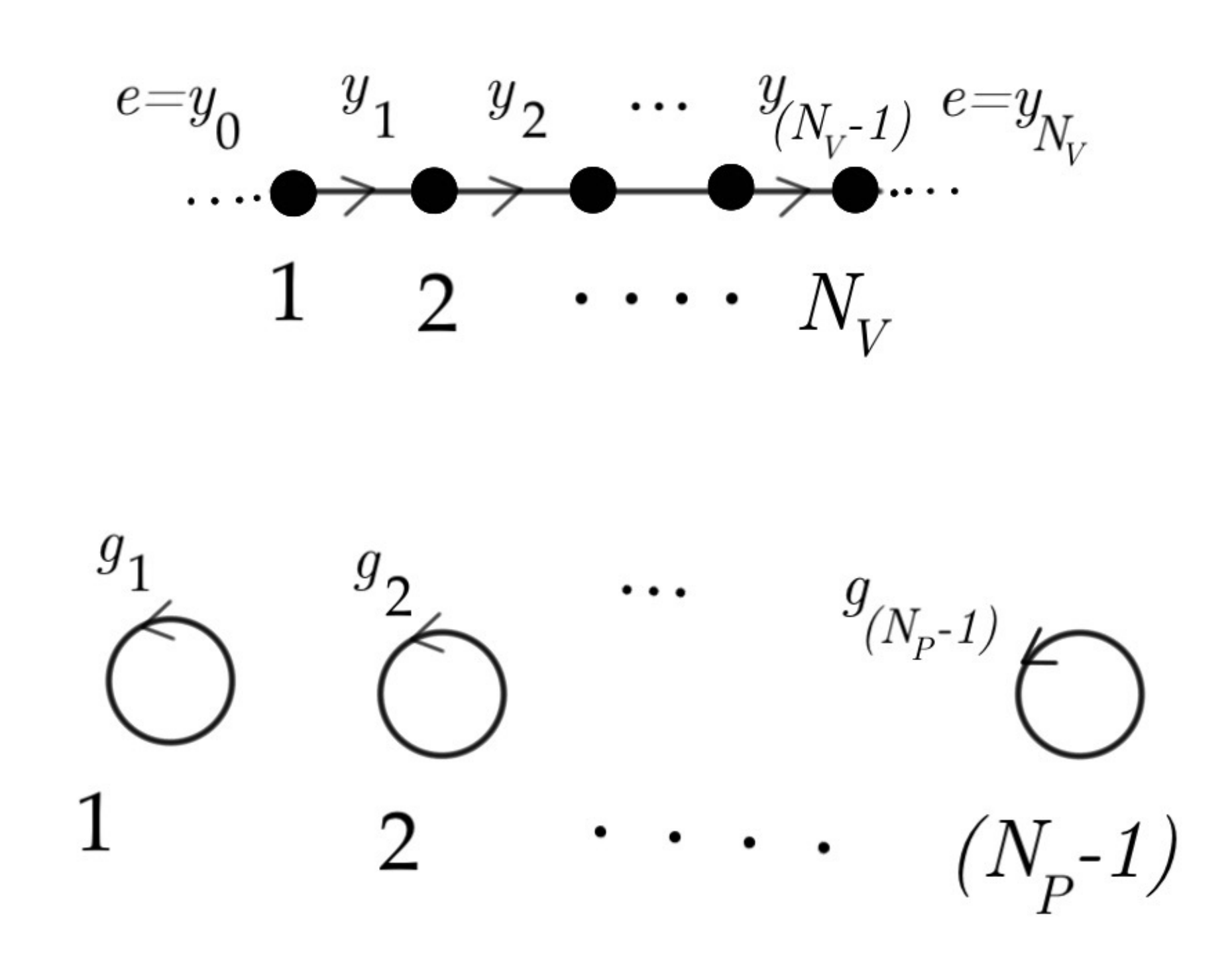}

\vspace*{-10pt}
    \caption{Geometry as  in Fig.~\ref{fig:canonical-geom-g01} but we have removed all edges which can only be labeled with the identity $e$.}
    \label{fig:canonical-geom-g02}
\end{figure}

Thus, our diagram simplifies into  that shown in Fig.~\ref{fig:canonical-geom-g02}. Here, there are $(N_P-1)$ isolated loops labeled $g_1, \ldots, g_{(N_P -1)}$ and there is a line of connected edges labeled $y_1, \ldots, y_{(N_V -1)}$ connecting the $N_V$ violatable vertices.  For simplicity of notation, we have also added edges $y_0$ and $y_{N_V}$, but set them equal to the identity $y_0 = y_{N_V} = e$.    The violatable vertices can each be assigned  a charge 
$$
 s_i  = y_{i-1}^{-1} y_i 
$$
such that the product
\begin{equation}
 s_1  s_2 \ldots s_{(N_V -1)}  s_{N_V} = e.
    \label{eq:sconstraint}
\end{equation}
The vertex $i$ therefore has an energy $\beta_{C_{s_i}}$.

We now aim to put the system in an eigenstate of the Hamiltonian. As we have already put the vertices in eigenstates, we now need to take care of the plaquettes.  For $i=1, \ldots, (N_P-1)$, we put the loop labeled $g_i$ in state $|\R_i n_i n_i'\rangle$ as defined in Eq.~\eqref{eq:Rnndef}.
As shown in Eq.~\eqref{eq:Rnneig}, these states are eigenstates of the plaquette projector, and thus the loop $i$ has energy $\alpha_{\R_i}$.     Their energies are independent of the $n, n'$ indices and so these wavefunctions can be superposed to form global eigenstates. It may appear that there is a degeneracy of $d_{\R_i}^2$, but notice that we have not yet imposed the constraint set by the last plaquette. This will actually reduce the degeneracy to $d_{\R_i}$ supplemented with a condition on the fusion of the fluxons and chargeons.

\begin{figure}[h]
~~~~~~~~~~~~~\includegraphics[width=3in]{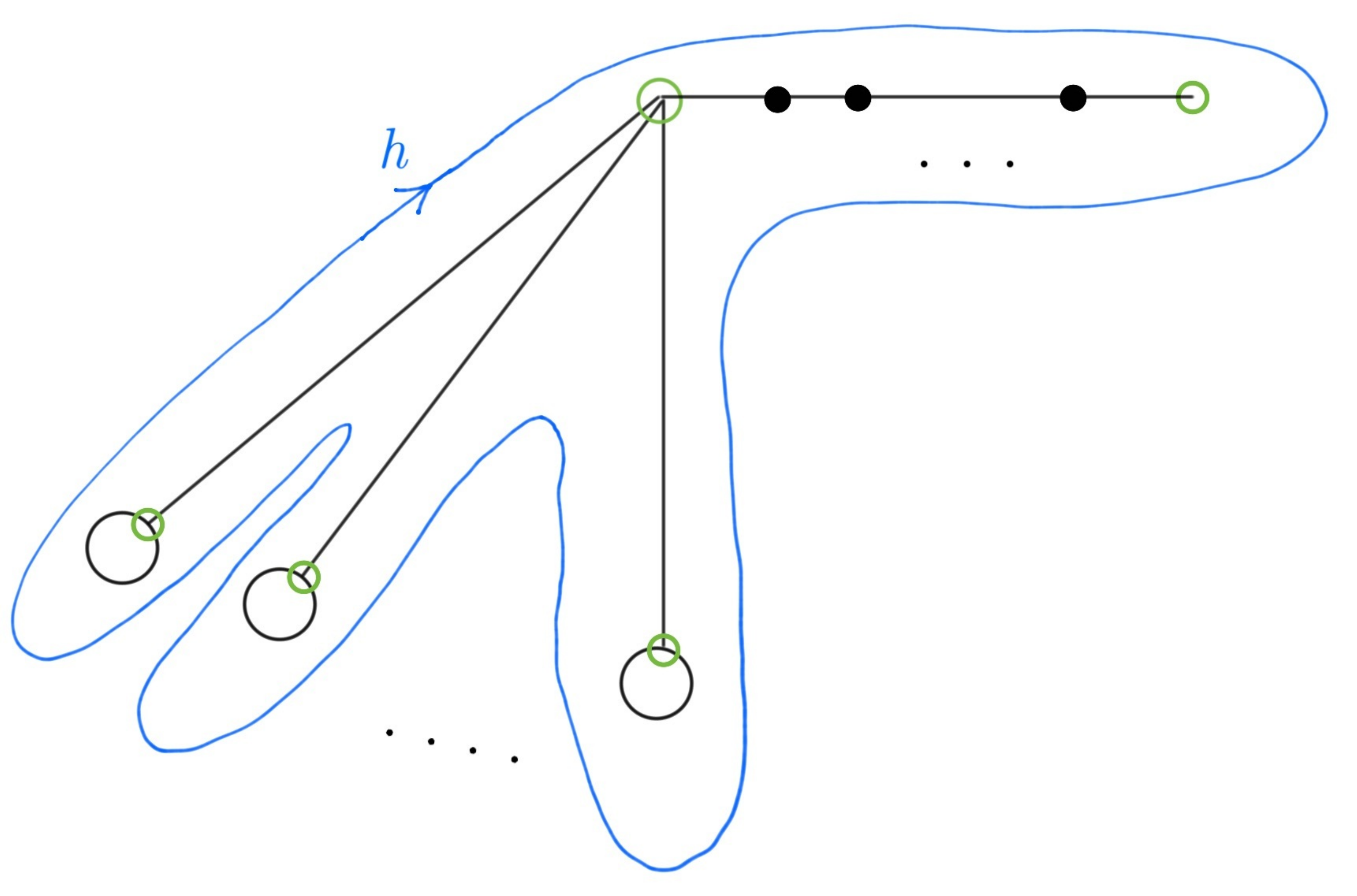}
    \caption{Geometry as  in Fig.~\ref{fig:canonical-geom-g01} showing the plaquette projector $\hat P(h)$ for the $N_P^{\rm th}$ plaquette which encircles the outside region including the point at infinity.  Note the orientation of the arrow is reversed since it ``encircles" the region outside.}
    \label{fig:canonical-geom-g0h}
\end{figure}

\begin{figure}[h]
~~~~~~~~~\includegraphics[width=2.5in]{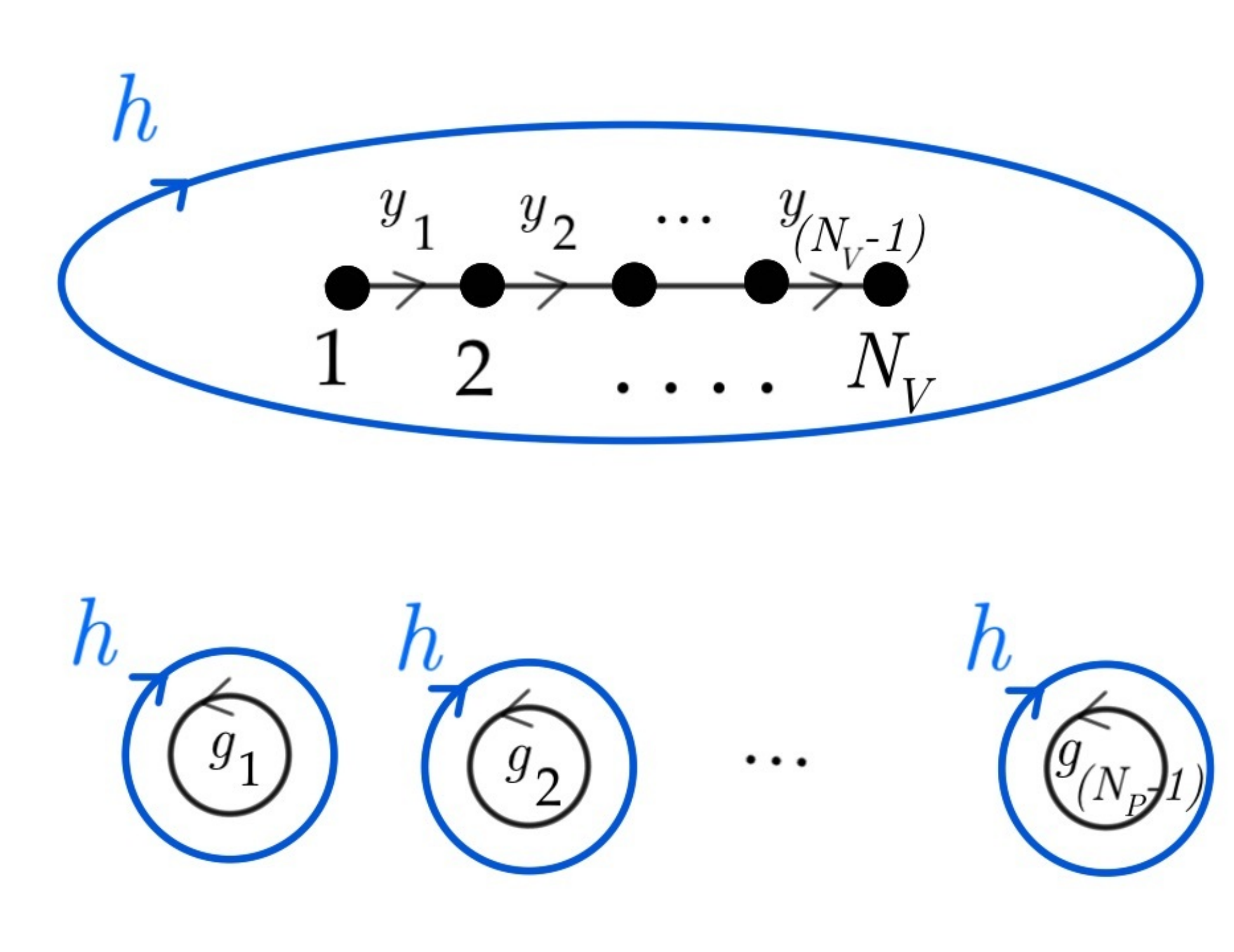}
    \caption{As in Fig.~\ref{fig:canonical-geom-g0h} showing the plaquette projector $\hat P(h)$ for the $N_P^{\rm th}$ plaquette, but now using the same simplification as in Fig.~\ref{fig:canonical-geom-g02}.}
    \label{fig:canonical-geom-g0h2}
\end{figure}

The key to this calculation is handling the final plaquette $i=N_P$.     If we draw the projector $\hat P(h)$ inside a $g$-loop graphically [see Eq.~\eqref{eq:Pheq}] as  
$$
\resizebox{20pt}{!}{$\circlearrowleft$} 
  \!\!\!\!\!\!
  \!\!\!
  \color{blue} 
\raisebox{2pt}{\resizebox{10pt}{!}{$\circlearrowleft$} }
\!\!\!\!\!\! \raisebox{3pt}{\mbox{\tiny $h$} }
$$then the $N_P^{\rm th}$  plaquette (which is the outside region including the point at infinity) has a projection operator $\hat P(h)$ as drawn in Fig.~\ref{fig:canonical-geom-g0h}.  Using the simplification as in Fig.~\ref{fig:canonical-geom-g02}  this can be redrawn as in Fig.~\ref{fig:canonical-geom-g0h2}.

We would now like to find the eigenstate of this last plaquette projector to representation $\R_{N_P}$ [see Eq.~\eqref{eq:PRdef}]
\be
 \hat P_{\R_{N_P}} = \frac{d_{\R_{N_P}}}{|G|} \sum_{h \in G} \chi_{\R_{N_P}}(h) \hat P_{N_P}(h),
\ee
where the subscript $N_P$ on the final projector in this equation indicates that we are applying the projector to the $N_P^{\rm th}$ plaquette. 

Under the action of $\hat P_{N_P}(h)$ we have $y_i \rightarrow h y_i h^{-1}$ but we have $g_i \rightarrow g_i h^{-1}$.   We want to find eigenstates of the $\hat P_{\R_{N_P}}$.

Note that the vertex violations $s_i = y_i y_{i-1}^{-1}$ remain eigenstates of the vertex operator [see Eq.~\eqref{eq:vertex}]
since they remain in the same conjugacy class. 

Now, let us consider what happens to the $g$ loops.  We will work in a basis $|\R n n'\rangle$ for the $g$ loops.    From Eq.~\eqref{eq:Rnneig} we know that these are eigenstates of the plaquette Hamiltonian.    Let us see what happens to the loop under the action of $\hat P_{N_P}(h)$, i.e., under $g \rightarrow g h^{-1}$:
\begin{align}
    |\R n n'\rangle  = & \sqrt{\frac{d_\R}{|G|}} \sum_{g  \in G} \rho_{nn'}^\R(g)|g\rangle, \\
 \rightarrow & \sqrt{\frac{d_\R}{|G|}} \sum_{g  \in G} \rho_{nn'}^\R(g)|g h^{-1}\rangle.
\end{align}
Now, defining $x = g h^{-1}$, this becomes
\begin{align}
|\R n n'\rangle \rightarrow & \sqrt{\frac{d_\R}{|G|}} \sum_{x  \in G} \rho_{nn'}^\R(xh)|x\rangle , \\
= & \sqrt{\frac{d_\R}{|G|}} \sum_{x  \in G} \sum_{m=1}^{d_\R} \rho_{nm}^\R(x) \rho_{mn'}^\R(h)|x\rangle , \\
= & \sum_{m=1}^{d_\R} \rho_{mn'}^\R(h) |\R n m \rangle.
\end{align}
Note that this transformation leaves both $\R$ and the middle index $n$ untouched.   

Consider now a basis for all of the $g_i$ loops for $i=1,\ldots, (N_P-1)$
\begin{align}
& |\R_1 n_1 n_1'\rangle \otimes |\R_2 n_2 n_2'\rangle \otimes \ldots \otimes  |\R_{(N_P -1)} n_{(N_P -1)} n'_{(N_P -1)}\rangle \nonumber\\
& ~~~~~~~~~~~=  \prod_{i=1}^{N_P- 1}  |\R_i n_i n'_i\rangle.    
\end{align}
Under the action of $\hat P_{N_P}(h)$ this transforms as
\begin{align}
&    \hat P_{N_P}(h) \prod_{i=1}^{N_P- 1}  |\R_i n_i n'_i\rangle  = \nonumber \\& \sum_{m_1=1}^{d_{\R_1}} \sum_{m_2=1}^{d_{\R_2}} \ldots \sum_{m_{(N_P-1)}=1}^{d_{\R_{(N_P-1)}}} \prod_{i=1}^{N_P- 1}  \rho_{m_i n_i'}^{\R_i} |\R_i n_i m_i\rangle,
\end{align}
where all of the $\R_i$'s and $n_i$'s are preserved.  Crucially, the energies of these plaquettes are unchanged, as they still correspond to the same irreps $\R_i$ of $G$. 

Under the action of $\hat P_{N_P}(h)$, the $y$'s transform by conjugation, but the conjugacy classes of the defects remain unchanged, hence the vertex energies remain unchanged. 

Now, let us consider the set of $\hat P_{N_P}(h)$ for all possible $h$'s.  The $y$'s transform through orbits and we will consider one orbit at a time.     Let us call a particular orbit $\omega$.
Let us write vectors 
\begin{equation} \label{eq:statesoforbit}
\{y_1, y_2, \ldots, y_{(N_V -1)} \}, 
\end{equation} which will form a basis of the states in orbit $\omega$, and let us call the states of an orbit $|v_\alpha^{(\omega)}\rangle$.    The operator $\hat P_{N_P}(h)$ acts as a group (a representation, but a reducible one) on this basis by conjugation.    In fact, it must permute the elements of the orbit.   Let us call the permutation matrix $\Lambda^{(\omega)}_{\alpha \beta}(h)$.   This set of matrices is a reducible representation.  We want to decompose this representation into irreducible representations
$$
 \Lambda^{(\omega)} = \bigoplus_{q=1}^{\mu^{(\omega)}}   \tilde \R^{(\omega)}_q,
$$
where the $\tilde \R^{(\omega)}_q$ are irreps of $G$ and there are $\mu^{(\omega)}$ irreps in this decomposition (as usual, a particular irrep may occur more than once in the sum).

The number of states $|v_\alpha^{(\omega)}\rangle$ in the orbit $\omega$ (by the orbit-stabilizer theorem) is equal to $|G|/X$ where $X$ is the number of states that are unchanged by the conjugation action of $\hat P_{N_P}$, i.e, $X$ is the number of $g$'s such that $g y_i = y_i g$, for all $y_i$ with $i=1 \ldots (N_V-1)$. Equivalently, again using the definition 
$s_i =  y_{i-1}^{-1} y_i$, 
we could say that $X$ is the number of values of $g$  such that $g s_i = s_i g$, for all $s_i$ with $i=1 \ldots N_V$ given the constraint stemming from Eq.~\eqref{eq:sconstraint}.

We now want to decompose the reducible representation $\Lambda^{(\omega)}$ into irreps $\tilde \R_q$.    How do we generally do such a thing?  The usual strategy is to use group characters and the orthogonality relation for irreps $\R_1$ and $\R_2$
\be
 \frac{1}{|G|} \sum_{g \in G} \chi^*_{\R_1}(g) \chi_{\R_2}(g) = \frac{1}{|G|}  \sum_{g \in G} \chi_{\overline{\R}_1}(g) \chi_{\R_2}(g)  = \delta_{\R_1 \R_2},
\ee
and more generally the fusion of irreps to the identity irrep $\Gamma_0$ has a multiplicity
\be
 N^{\Gamma_0}_{\R_1 \R_2 \ldots \R_z} = \frac{1}{|G|} \sum_{g \in G} \prod_{i=1}^z  \chi_{\R_i}(g).
\ee

So, if we have a reducible representation $\Lambda^{(\omega)}$ with a character $\chi_{\Lambda^{(\omega)}}$, then the decomposition of $\Lambda^{(\omega)}$ into irreps has multiplicity of $\R_i$  given by 
\be
m_i = \frac{1}{|G|} \sum_{g \in G} \chi^*_{\R_i}(g)  \chi_{\Lambda^{(\omega)}}(g).
\ee
Thus, we are interested in the character of the representation $\Lambda^{(\omega)}$
\be
 \chi_{\Lambda^{(\omega)}}(h) = {\rm Tr}[\Lambda^{(\omega)}(h)] = \sum_{\alpha} \Lambda^{(\omega)}_{\alpha \alpha}(h).
\ee
Since $\Lambda^{(\omega)}$ is a permutation matrix, $\chi_{\Lambda^{(\omega)}}$ simply counts the basis states in the orbit that are unchanged by conjugating with $h$, i.e., the diagonal components of $\Lambda^{(\omega)}$. Thus, we have
\begin{align} \label{eq:chiLambda1}
  &  \chi_{\Lambda^{(\omega)}}(h) =\sum_{y_i\mbox{\tiny 's in orbit $\omega$}}  \prod_{i=1}^{N_V -1} \delta_{h y_i, y_i h} ,  \\
&=    \sum_{s_i \mbox{\tiny 's in orbit $\omega$}}  \prod_{i=1}^{N_V} \delta_{h s_i, s_i h} \,\, \delta_{s_1 s_2 \ldots s_{N_V} ,e}.  \label{eq:chiLambda2}
\end{align}

In the end, we will want to fuse these irreps from $\Lambda^{(\omega)}$ with the irreps of the plaquettes $\R_1, \ldots, \R_{N_P}$ in the case of the sphere.  In the higher-genus case (see Sec.~\ref{sub:higher} below)  we will also add representations arising from the handles. 

To see where this is going, note that if we try to fuse $\Lambda^{(\omega)}$ together with the irreps $\R_1 \ldots \R_{N_P}$ to form an identity, we get a multiplicity of
\begin{align} \label{eq:promisedfusion}
 & \frac{1}{|G|} \sum_{h \in G}  \chi_{\Lambda^{(\omega)}}(h)  \prod_{i=1}^{N_P} \chi_{\R_i}(h) = \\
 &\frac{1}{|G|}\sum_{y_i\mbox{\tiny 's in  orbit $\omega$}}  \sum_{h \in G}    \prod_{i=1}^{N_V -1} \delta_{h y_i, y_i h}    \prod_{i=1}^{N_P} \chi_{\R_i}(h).\nonumber
\end{align}
Then, summing over all orbits (hence summing over all $y$'s) we get a multiplicity
\begin{align} \label{eq:promisedfusion2}
 & \frac{1}{|G|} \sum_{h,y_1 \ldots y_{(N_V-1)} \in G}     \prod_{i=1}^{N_V -1} \delta_{h y_i, y_i h}    \prod_{i=1}^{N_P} \chi_{\R_i}(h) \nonumber \\
 =&  \frac{1}{|G|} \sum_{h,s_1 \ldots s_{N_V} \in G}    \delta_{s_1 \ldots s_{N_V} ,e}  \prod_{i=1}^{N_V} \delta_{h s_i, s_i h}    \prod_{i=1}^{N_P} ,\chi_{\R_i}(h) ,
\end{align}
which matches the multiplicity in Eq.~\eqref{eq:manyfus}.   However in addition to this multiplicity, we also want to derive the nontopological degeneracies.

\subsubsection{Nontopological degeneracies and explicit wavefunction construction}
\label{subsub:wavefuction}

So, let us do this more carefully. For a given orbit, let us decompose the reducible representation into its irreducible representations
\begin{equation}
\Lambda^{(\omega)}_{\alpha \beta}(h) = \sum_{q=1}^{\mu^{(\omega)}} \, \sum_{s,t = 1}^{d_{\tilde \R_q^{(\omega)}}}  U^{(\omega)}_{\alpha; (q,s)} \,\, \rho_{st}^{\tilde \R_q^{(\omega)}}\!\!(h)  \,\, U^{(\omega) \dagger}_{(q,t); {\beta}},
\label{eq:splitq}
\end{equation}
where the $U$'s are unitary matrices in their lower indices (these are actually Clebsch-Gordan coefficients), and here $d_{\tilde \R_q^{(\omega)}}$ is the dimension of representation $\tilde \R_q^{(\omega)}$.    

For each orbit $\omega$ we have the $\alpha$ basis $|v_{\alpha}^{(\omega)}\rangle$, but we can switch to a new orthonormal basis
\be
 |\tilde v^{(\omega)}_{(q,s)} \rangle = \sum_{\alpha} U^{(\omega)}_{\alpha; {(q,s)}} |v^{(\omega)}_\alpha\rangle,
\ee
and we can check
\be
\langle \tilde v^{(\omega)}_{(q,s)} | \tilde v^{(\omega')}_{(q',s')} \rangle = \delta_{\omega,\omega'} \delta_{qq'} \delta_{ss'},
\ee
and we have the corresponding reverse transform
\be
|v^{(\omega)}_\alpha\rangle = \sum_{q=1}^{\mu^{(\omega)}} \sum_{s=1}^{d_{\tilde \R_q^{(\omega)}}} U^{{(\omega)} \dagger}_{(q,s); \alpha} |\tilde v^{(\omega)}_{(q,s)}\rangle.
\ee

Note that the set of all $|\tilde v_{(q,s)}^{(\omega)}\rangle$ for all $q$ and all $s$ gives a full basis for all the edge labels in the orbit $\omega$. 

We now have
\be
\hat P_{N_P}(h) \, |v_{\beta}^{(\omega)}\rangle = \sum_{\alpha}  \Lambda_{\alpha \beta}^{(\omega)}(h) \, | v_{\alpha}^{(\omega)}\rangle,
\ee
so that
\begin{align}
&\hat P_{N_P}(h) |\tilde v_{(q,s)}^{(\omega)}\rangle = \sum_{\alpha,\beta}  \Lambda_{\alpha \beta}^{(\omega)}(h) U_{\beta;(q,s)}^{(\omega)} | v_{\alpha}^{(\omega)}\rangle, \\
&=\sum_{\alpha,\beta}  \sum_{q'=1}^{\mu^{(\omega)}} 
\sum_{s', t' = 1}^{d_{\tilde \R^{(\omega)}_{q'}}}
U^{(\omega)}_{\alpha; (q',s')}  \rho_{s't'}^{\tilde \R^{(\omega)}_{q'}}(h) U^{(\omega)\dagger}_{(q',t'), \beta}  
U^{(\omega)}_{\beta; (q,s)} | v^{(\omega)}_{\alpha}\rangle.
\end{align} 
Because of unitarity, the sum over $\beta$ creates $\delta_{qq'} \delta_{t's}$ so that the sum reduces to 
\begin{align}
&\hat P_{N_P}(h) |\tilde v_{(q,s)}^{(\omega)}\rangle = \sum_{\alpha} \sum_{s'= 1}^{d_{\tilde \R^{(\omega)}_{q'}}}
U^{(\omega)}_{\alpha; (q,s')}  \rho_{s's}^{\tilde \R^{(\omega)}_{q}}(h) | v^{(\omega)}_{\alpha}\rangle,
\\ & ~~~~~~~~~~~~~~~~~= \sum_{s'= 1}^{d_{\tilde \R_{q'}}}
  \rho_{s's}^{\tilde \R^{(\omega)}_{q}}(h) | \tilde v^{(\omega)}_{(q,s')}\rangle,
\end{align}
and note that the irreps $\tilde \R^{(\omega)}_q$ are preserved. 

Still focusing on a single orbit, and now also fixing an irrep $\tilde \R^{(\omega)}_q$ we can write a complete set of states as 
\be
 \left[ \prod_{i=1}^{N_P-1} |\R_i n_i n'_i \rangle\right]  \otimes |\tilde v_{(q,s)}^{(\omega)}\rangle.
\ee
All of the terms in the Hamiltonian except the $N_P^{\rm th}$ plaquette term have a fixed eigenvalue.   We finally want to sum all such states to find a basis of states where the $N_P^{\rm th}$ plaquette also has a fixed eigenvalue (i.e., is in an irrep $\R_{N_P}$).   We start with a set of operators on this plaquette which we write as in Eq.~\eqref{eq:setPs},
\be
\hat P_{\R_{N_P}, a b} = \frac{d_{\R_{N_P}}}{|G|} \sum_{h \in G} \rho_{ab}^{\R_{N_P}}(h) \hat P_{N_P}(h).
\ee
Applying these to the above defined kets we have 
\begin{widetext}
\begin{align}
\hat P_{\R_{N_P}, a b}  & \left[ \prod_{i=1}^{N_P-1} |\R_i n_i n'_i \rangle\right]  \otimes |\tilde v_{(q,s)}^{(\omega)}\rangle = \frac{d_{\R_{N_P}}}{ |G|}     \sum_{h \in G} \rho_{ab}^{\R_{N_P}}(h)\hat P_{N_P}(h) \left[ \prod_{i=1}^{N_P-1} |\R_i n_i n'_i \rangle\right]  \otimes |\tilde v_{(q,s)}^{(\omega)}\rangle \\
&=\frac{d_{\R_{N_P}}}{ |G|} \sum_{h \in G} \rho_{ab}^{\R_{N_P}}(h)  \left[ \sum_{m_1 = 1}^{d_{\R_1}} \ldots \sum_{m_{N_P -1} = 1}^{d_{N_P - 1}}  
\prod_{i=1}^{N_P - 1} \rho_{m_i n_i'}^{\R_i}(h) |\R_i n_i m_i\rangle 
\right]  \otimes \sum_{s'=1}^{d_{\tilde \R^{(\omega)}_q}} \rho_{s' s}^{\tilde \R_q^{(\omega)}}(h) |\tilde v^{(\omega)}_{(q,s')}\rangle. \label{eq:Long1}
\end{align}
Here, we have a product of representation matrices.   We will fuse these representations all together [this implements the promised fusion from 
Eq.~\eqref{eq:promisedfusion} above and now justifies it].  The piece that does not vanish when we sum over all $h \in G$ is the part that fuses to the identity representation.   We thus require that
$
 \R_{N_P} \otimes \R_1 \otimes \R_2 \ldots \otimes \R_{N_P - 1} \otimes \tilde \R_q^{(\omega)} \rightarrow \Gamma_0
$
the multiplicity of this fusion to the identity is exactly that shown in Eqs.~\eqref{eq:manyfus} and \eqref{eq:promisedfusion2}.
Now, let us make this decomposition more explicit.  We first decompose into irreps 
\begin{equation}
\left[ \prod_{i=1}^{N_P - 1} \rho_{m_i n_i'}^{\R_i}(h) \right] \rho_{s' s}^{\tilde \R_q^{(\omega)}}(h)  = \sum_{Q}\sum_{\alpha\beta = 1}^{d_{\R_Q}} \tilde U^{(q,\omega)}_{(\{m \}s'); (Q \alpha)}\,\,  \rho^{\R^{(\omega,q)}_Q}_{\alpha,\beta}(h)  \,\, \tilde U^{(q,\omega)^\dagger}_{(Q \beta); (\{n'\} s)}, \label{eq:decom2}
    \end{equation}
where for each orbit $\omega$ and each chosen irrep $\tilde \R_q^{\omega}$, we have  defined $\tilde U^{(q,\omega)}$ a unitary matrix (again Clebsh-Gordan coefficients) and $\{ m \}$ and $\{ n' \}$ are shorthand for the lists of $m_i$'s and $n'_i$'s.    Here, the sum over $Q$ is over all irreps $\R^{(\omega,q)}_Q$ that occur in the product (and the unitary matrices are block diagonal in $Q$).

As mentioned above, in Eq.~\eqref{eq:Long1}, in the sum over $h \in G$, one only obtains a nonzero result if the fusion of all the irreps comes to the identity.  In Eq.~\eqref{eq:decom2}, we have fused all the pieces except the leading $\R_{N_P}$.  Thus, to get a nonzero result the irrep $\R^{(\omega,q)}_Q$ must be the conjugate of the $\R_{N_P}$. 
Using Eq.~\eqref{eq:decom2} in Eq.~\eqref{eq:Long1} then using the grand orthogonality theorem [Eq.~\eqref{eq:grand}], the sum over $h$ applied to the product of the representation matrices now becomes 
\be
\sum_{h \in G} \rho_{ab}^{\R_{N_P}}(h) \left[ \prod_{i=1}^{N_P - 1} \rho_{m_i n_i'}^{\R_i}(h) \right] \rho_{s' s}^{\tilde \R_q}(h) = \frac{|G|}{d_{\R_P}}\sum_{Q} \tilde U^{(q,\omega)}_{(\{m \}s'); (Q b)} \,  \tilde U^{(q,\omega)\dagger}_{(Q a); (\{n'\} s)},
\ee
where the sum over $Q$ is now only over values of $Q$ in Eq.~\eqref{eq:decom2} such that $\R^{(\omega,q)}_Q$ is conjugate to $\R_{N_P}$. Plugging this back into Eq.~\eqref{eq:Long1}, we have 
\begin{align}
\hat P_{\R_{N_P}, a b}  & \left[ \prod_{i=1} |\R_i n_i n'_i \rangle\right]  \otimes |\tilde v_{(q,s)}^{(\omega)}\rangle = \sum_Q    \sum_{m_1 = 1}^{d_{\R_1}} \ldots \sum_{m_{N_P -1} = 1}^{d_{N_P - 1}} 
 \sum_{s'=1}^{d_{\tilde \R_q}} \tilde U^{(q,\omega)}_{(\{m \}s'); (Q b)} \,  \tilde U^{(q,\omega)\dagger}_{(Q a); (\{n'\} s)} 
\prod_{i=1}^{N_P - 1} |\R_i n_i m_i\rangle 
  \otimes |\tilde v^{(\omega)}_{(q,s')}\rangle.  \label{eq:Long2}
\end{align}

Now, fixing the $n$'s, fixing an orbit $\omega$, and fixing a value of $q$ in the decomposition Eq.~\eqref{eq:splitq} (and therefore fixing a representation  $\tilde \R_q^{(\omega)}$), and fixing a value of $Q$ in the decomposition Eq.~\eqref{eq:decom2} such that $\R_Q$ is conjugate to $\R_{N_p}$, we define an orthonormal set
\be
 |\phi^{(\omega,q,Q)}_{b} \rangle = \sum_{m_1 = 1}^{d_{\R_1}} \ldots \sum_{m_{N_P -1} = 1}^{d_{N_P - 1}} \sum_{s'=1}^{d_{\tilde \R_q}} \tilde U^{(q,\omega)}_{(\{m \}s'); (Q b)} \,  
\prod_{i=1}^{N_P - 1} |\R_i n_i m_i\rangle 
  \otimes |\tilde v^{(\omega)}_{(q,s')}\rangle ,
\ee
and since $\tilde U^{(q,\omega)}$ is unitary, and the $|\R n m\rangle$ are orthonormal and the $|\tilde v_{(q,s)}^{(\omega)}\rangle$ are orthonormal we have  
\be
 \langle \phi^{(\omega,q,Q)}_{a}  | \phi^{(\omega',q',Q')}_{ b}\rangle = \delta_{\omega\omega'} \delta_{qq'} \delta_{QQ'} \delta_{ab},
\ee
with $a,b \in 1\ldots d_{\R_Q}$ and $d_{\R_Q} = d_{\R_{N_p}}$. 
We can then rewrite Eq.~\eqref{eq:Long2}
as
\begin{align}
\hat P_{\R_{N_P}, a b} \left[ \prod_{i=1}^{N_P-1} |\R_i n_i n'_i \rangle\right]  \otimes |\tilde v_{(q,s)}^{(\omega)}\rangle = \sum_Q  \tilde U^{(q, \omega) \dagger}_{(a Q); (\{n'\} s)} | \phi^{(\omega,q,Q)}_{b} \rangle,
\end{align}
and so 
\begin{align}
\hat P_{\R_{N_P}, a b} | \phi^{(\omega,q,Q)}_{c} \rangle &=  \sum_{m_1 = 1}^{d_{\R_1}} \ldots \sum_{m_{N_P -1} = 1}^{d_{N_P - 1}} \sum_{s'=1}^{d_{\tilde \R_q}} \tilde U^{(q,\omega)}_{(\{m \}s'); (Q c)} \,  
\hat P_{\R_{N_P}, a b}  \prod_{i=1}^{N_P - 1} |\R_i n_i m_i\rangle 
  \otimes |\tilde v^{(\omega)}_{(q,s')}\rangle,  \\ &=
  \sum_{m_1 = 1}^{d_{\R_1}} \ldots \sum_{m_{N_P -1} = 1}^{d_{N_P - 1}} \sum_{s'=1}^{d_{\tilde \R_q}} \tilde U^{(q,\omega)}_{(\{m \}s'); (Q c)} \,  
\sum_{Q'} \tilde U^{(q,\omega)\dagger}_{(aQ'); (\{ m\} s')} |\phi^{(\omega,q,Q')}_{b}\rangle. 
\end{align}
\end{widetext}
Now, using unitarity of the $\tilde U$, we obtain
\be
\hat P_{\R_{N_P}, a b} | \phi^{(\omega,q,Q)}_{c} \rangle = \delta_{ac} |\phi^{(\omega,q,Q)}_{b}\rangle,
\ee
which is the precise analog of Eq.~\eqref{eq:funnyop}.  Thus, when we construct a projector $P_{\R_{N_P}}$ by tracing over the $a,b$ indices, the $|\phi^{(\omega,q, Q)}_b\rangle$ states become eigenstates, and the index $b$ plays the role of the $n$ index which is preserved.   This completes the proof for the case of the sphere. 

To summarize what we have derived:
\begin{enumerate}
    \item Each plaquette $i=1 \ldots N_P$ is assigned an irrep (of $G$) $\R_i$.
    \item Each vertex $i= 1 \ldots N_V$  is assigned a conjugacy class $C_i$. \end{enumerate}
These two steps fully fix the energy of the system. The irreps  $\Gamma_i$ occur on the plaquettes here (since we are working on the dual lattice) whereas in the main text these were the ``chargeons" which live on the vertices (see Sec.~\ref{subsub:anyons}).  Similarly, the conjugacy classes here occur on the vertices since we are on the dual lattice, but in the main text, they are the plaquette excitations (see Sec.~\ref{subsub:plaquettes}).
\begin{enumerate}
\setcounter{enumi}{2}
\item  Each plaquette $i=1 \ldots N_P$ is assigned an index $n \in {1 \ldots  d_{\R_i}}$, where $d_{\R_i}$ is the dimension of the corresponding irrep $\R_i$.
\end{enumerate}
This is the nontopological index (the energy is independent of this index and it does not carry anyon charge).  For each of the first $N_P-1$ plaquettes, the wavefunction will be in a superposition of $n'$ values of the states $|\R_i n_i n'_i\rangle$.      For the $N_P^{\rm th}$ plaquette the $n$ index is the subscript of $|\phi^{(\omega,q,Q)}_n\rangle$. 
We must also discuss the fusion channel, which is equivalent to fixing $\omega,q,Q$.  Thus, we have the additional steps: 
\begin{enumerate}
\setcounter{enumi}{3}
\item Consistent with the conjugacy classes, we choose a set of $y$'s, which then fixes an orbit $\omega$ under conjugation by $h$. 

\item Choose an irrep $q$ of this orbit $\tilde \R^{(\omega)}_q$ in Eq.~\eqref{eq:splitq}.

\item Once this choice is made, one fixes the irrep $Q$ in the decomposition Eq.~\eqref{eq:decom2} such that $\R^{(\omega,q)}_Q$ is conjugate to $\R_{N_P}$.

\end{enumerate}

When we sum over all of the orbits in step 4 and then count the fusion multiplicity in steps 5 and  6, we obtain all the fusion channels in the fusion multiplicity equation [Eq.~\eqref{eq:promisedfusion2} or \eqref{eq:manyfus}]. 

In principle, these steps constitute a complete construction of all of the eigenstates of the system, although in practice one needs to calculate the Clebsch-Gordon coefficients ($U^{(\omega)}$ and $\tilde U^{(q,\omega)}$), which can be tedious to do explicitly  except for very small systems. Note that once we have the explicit wavefunction on this special geometry, it is easy to reverse the restructuring moves to obtain the wavefunction on any geometry.

\subsection{Higher genus}
\label{sub:higher}

Let us return to Eq.~\eqref{eq:manyfus} and ask how it should generalize to higher genus.  Neglecting the nontopological degeneracies, for a system with $N$ charges in irreps $\R_i$ and $M$ fluxes in conjugacy classes $C_i$ on a manifold of genus $\genus$ we expect a degeneracy of 
\begin{align} \nonumber
& \sum_{\lambda_1, \ldots \lambda_{\genus}} N^{\mathbf{1}}_{\lambda_1 \overline \lambda_1 \ldots \lambda_{\genus} \overline \lambda_{\genus} (e,\R_1)(e,\R_2)...(e,\R_N),(C_1,\Gamma_0)(C_2,\Gamma_0)...(C_M,\Gamma_0)  } \\ &=  \frac{1}{|G|} \sum_{g, \{ \xi \}, \{\xi'\} \{k\} \{s\}}   \left[ \prod_{z=1}^{\genus} \sum_{\lambda_z}  \underline \chi_{\lambda_z} (\xi_z^* g)  \underline \chi_{\overline \lambda_z} (\xi_z'^* g) \right]  \times  \nonumber 
\\ & \nonumber 
~~~~~~~~~\left[ \underline \chi_{(e,\R_1)}(k_1^*g) \ldots \underline \chi_{(e,\R_{N})}(k_{N}^*g)   \right.
\\ & 
\nonumber ~~~~~~~~~~~~~~~ \left.   \underline \chi_{(C_1,\Gamma_0)}(s_1^*g)   \ldots \underline \chi_{(C_{M},\Gamma_0)}(s_{M}^*g)  
\right] \\
& ~~~~~~~~~~~~~\delta_{\xi_1 \xi_1' \ldots \xi_{\genus} \xi_{\genus}' k_1 \ldots k_{N} s_1, \ldots s_{M},e}, \label{eq:longN}
\end{align}
where the sum over $\lambda$'s are over all the anyon types (all irreps of the quantum double).   The first line of Eq.~\eqref{eq:longN}, the $\lambda$ degrees of freedom are the quantum numbers going around the $\genus$ handles (see the discussion in Refs.~\cite{Ritz24_2,Simon_book}).  To go to the second line of Eq.~\eqref{eq:longN}, we use the generalization fo Eq.~\eqref{eq:multifus}.  All sums in this line of the equation are over $G$.  We use shorthand $\{s\}$ in the subscript of the sum to mean sum over all $s$ variables.   Here, the middle two lines are familiar from Eq.~\eqref{eq:manyfus} and simplify to
$$
\left[\prod_{j=1}^N \chi_{\R_j}(g) \delta_{k_j,e} \right] \left[\prod_{k=1}^M \delta_{s_k \in C_k} \delta_{s_k g,g s_k}\right], \rule[15pt]{0pt}{15pt} $$
the only difference being that now the $\delta$ function constraint is
$$
\delta_{\xi_1 \xi_1' \ldots \xi_{\genus} \xi_{\genus}' s_1, \ldots s_{M},e},
$$
where the $\xi$ and $\xi'$ factors did not occur in Eq.~\eqref{eq:manyfus}.

Let us now examine terms in the first line of Eq.~\eqref{eq:longN}.  Let us define
\begin{align}
 f(\xi,x,g) &= \sum_{\xi'} \delta_{\xi\xi',x} \sum_{\lambda} \underline \chi_{\lambda}(\xi^* g) \underline \chi_{\overline \lambda}(\xi'^* g), \\
 &= \sum_{\xi'} \delta_{\xi\xi',x} \sum_{\lambda} \underline \chi_{\lambda}(\xi^* g) \underline \chi^*_{ \lambda}((\xi')^{-1*} g), 
\end{align}
and we note that the sum over $\lambda$ is exactly of the form where we can use the completeness relation Eq.~\eqref{eq:completeness}.  So we obtain
\be
    f(\xi,x,g)  = \sum_{h'} \delta_{\xi\xi',x}  \, \delta_{{\cal C}_{\xi^* g}, \overline{\cal C}_{\xi'^{-1*} g}} \frac{|G|}{|{\cal C}_{\xi^* g}|} \delta_{g\xi,
    \xi g} \delta_{g \xi',\xi'g},
\ee
where  ${\cal C}_{\xi^* g}$ is the double conjugacy class and $|\cal C|$ is the number of elements in this class.   I.e., it is the number of distinct values of the pair $(s \xi s^{-1}, s g s^{-1})$ over all $s \in G$.  Again, by the orbit-stabilizer theorem, the factor $|G|/|\cal C|$ is exactly the number of elements $z \in G$, such that $zg = gz$ and $z\xi = \xi z$.
The $\delta$ function $\delta_{C\bar C}$ is one if and only if $\xi = p \xi'^{-1} p^{-1} $ and $p g = gp$ for some $p \in G$. 

Allowing the $\delta$ function $\delta_{\xi\xi',x}$ to act, we have 
\be
f(\xi,x,g)  = \delta_{{\cal C}_{\xi^* g}, \overline{\cal C}_{(x^{-1} \xi)^* g}} \frac{|G|}{|{\cal C}_{\xi^* g}|} \delta_{g\xi,\xi g} \delta_{g x,x g},
\ee
where we replaced the $\delta_{g \xi',\xi'g}$ at the end with $\delta_{g x,x g}$, which are equivalent once the sum over $\xi'$ is done given the other $\delta$ functions.  The condition for the $\delta_{C\bar C}$ to be unity is now $\xi = p x^{-1} \xi p^{-1}$
(or equivalently $x=\xi p^{-1} \xi^{-1} p$) for some $p$ with $p g=gp$ and recall that $g \xi = \xi g$ as well.   If two different values of $p$ satisfy this condition we can write $p' = \kappa p$ with $\kappa \xi = \xi \kappa$ so we can write instead 
\be
 f(\xi,x,g) = \sum_{p \in G}  \delta_{x,\xi p^{-1} \xi^{-1} p} \delta_{\xi g,g\xi} \delta_{xg,gx} \delta_{pg,gp},
\ee
which now cancels the factor of $|G|/|{\cal C}|$.    

The degeneracy shown  in Eq.~\eqref{eq:longN} can now be written as

\begin{widetext}
\begin{align}
\label{eq:longhigher}
& \frac{1}{|G|} \sum_{g, \{x \}, \{\xi \}, \{s \}} \!\!\!\!\! \delta_{x_1 \ldots x_{\genus} s_1 \ldots s_M,e}  \left\{ \prod_{i=1}^{\genus} f(\xi_i,x_i,g)  \right\}
\left[\prod_{j=1}^N \chi_{\R_j}(g)  \right] \left[\prod_{k=1}^M \delta_{s_k \in C_k} \delta_{s_k g,g s_k}
\right] \\
& =  \frac{1}{|G|} \sum_{g, \{x \}, \{\xi \}, \{s \}, \{ p \}}  \!\!\!\!\! \delta_{x_1 \ldots x_{\genus} s_1 \ldots s_M,e}  \left\{ \prod_{i=1}^{\genus}  \delta_{x_i,\xi_ip_i^{-1} \xi_i^{-1} p_i} \delta_{\xi_ig,g\xi_i} \delta_{x_ig,gx_i} \delta_{p_ig,gp_i}\right\}
\left[\prod_{j=1}^N \chi_{\R_j}(g) \right] \left[\prod_{k=1}^M \delta_{s_k \in C_k} \delta_{s_k g,g s_k}
\right]. \nonumber
\end{align}
\end{widetext}
This looks a bit messy, but compared to our result for the sphere it is the same except for the $f$-terms inserted (the curly bracket terms) and now the overall $\delta$ function out front contains the $x$'s as well as the $s$'s. 

\subsubsection{Explicit construction}

\begin{figure}[h]
\vspace*{10pt}
~~~\includegraphics[width=3.2in]{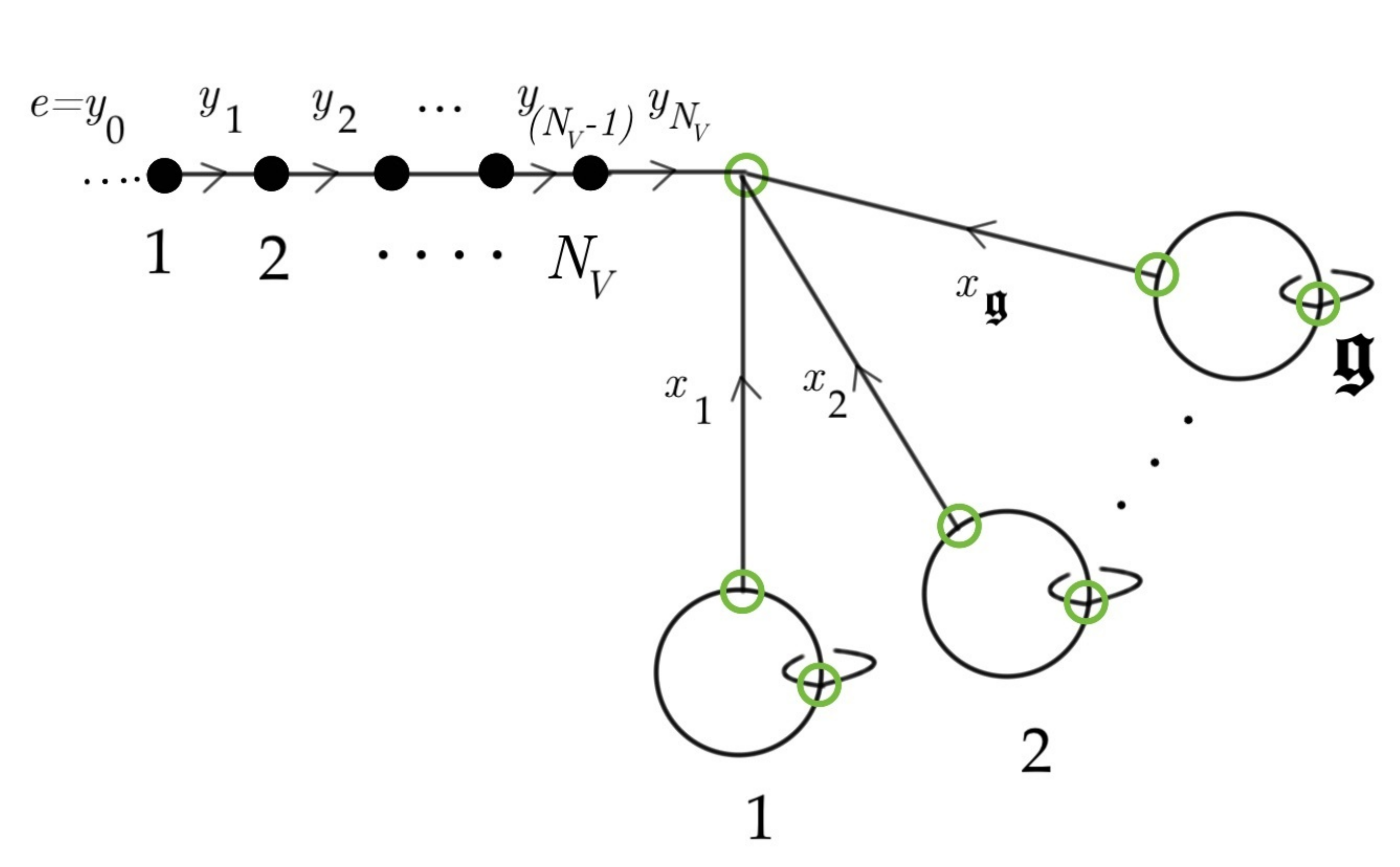}

\vspace*{-10pt}
    \caption{The top line of  Fig.~\ref{fig:canonical-geom-g02} but modified to include the handles of Fig.~\ref{fig:canonical-geom}.}
    \label{fig:canonical-geom-g02gg}
\end{figure}

The full construction of the wavefunction follows closely that of Sec.~\ref{sub:sphere} on the sphere.    Here, however, we need to include the handles, so our geometry is now that of Fig.~\ref{fig:canonical-geom}.   When we 
label the edges and enforce the constraints from the green circles to obtain the analog of Fig.~\ref{fig:canonical-geom-g02}, we again get $N_P-1$ decoupled loops along the bottom row, but now the top row is connected to the handle and looks like Fig.~\ref{fig:canonical-geom-g02gg}.

Note that the green circle at the top vertex (with ${\genus}+1$ lines incoming) enforces
\begin{equation}
\label{eq:consyform}   x_1 \ldots x_{\genus}  y_{N_V} = e ,
\end{equation}
or
\begin{equation}
 x_1 \ldots x_{\genus} s_1 \ldots s_N  = e ,\label{eq:conssform} \end{equation}
which is exactly the constraint we want in  Eq.~\eqref{eq:longhigher}.

\begin{figure}[h]
\vspace*{10pt}
~~~~~~\includegraphics[width=1.8in]{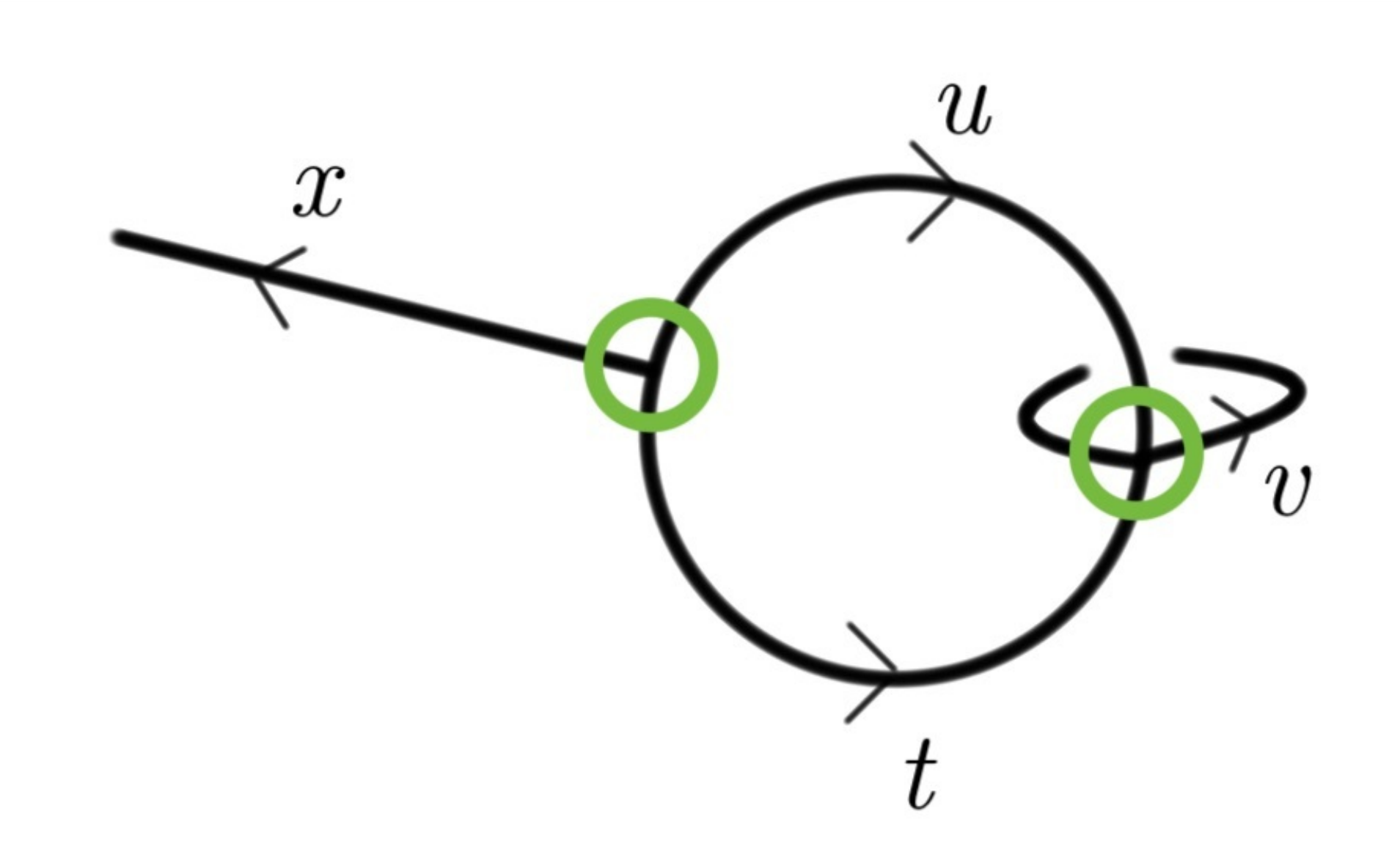}
    \caption{Closeup of a handle}
    \label{fig:close}
\end{figure}

\medskip
Let us now examine the handles more closely as shown in  Fig.~\ref{fig:close}.
Here, the green circles impose the constraints $xut=e$ and $tv^{-1}uv= e$, which we can rewrite as
$
 t = v^{-1} u^{-1} v$, 
 and we completely eliminate $t$ from the discussion.   Thus, we now have $x = t^{-1} u^{-1} =  v^{-1} u v u^{-1}$.

Under the operator $P_{N_P}(h)$, again all of these variables transform by conjugation
$
 x \rightarrow h x h^{-1},  t \rightarrow hth^{-1}, u \rightarrow h u h^{-1}, v \rightarrow h v h^{-1}$.
Now, when we list out the states of an orbit $\omega$, our vector is not just that shown in Eq.~\eqref{eq:statesoforbit} but now includes these additional variables as well
$$
\{y_1, y_2, \ldots, y_{N_V}, x_1 \ldots x_{\genus}, u_1 \ldots u_{\genus}, v_1 \ldots v_{\genus} \}.
$$
Again, the elements of an orbit permute under the group action and we are interested in the permutation matrices $\Lambda^{(\omega)}(h)$, which form a representation. To find the characters of this representation, we want to count the elements which are unchanged under the action of $P_{N_P}(h)$.   For an element of the orbit to be unchanged, we must have $h$ commuting with all of the $y's$, and all the $x$'s, and all of the $u$'s and $v$'s and further we need to restrict that $ x= v^{-1} u v u^{-1}$.

The character is thus
\begin{align}
 & \chi_{\Lambda^{(\omega)}}(h) = \sum_{ \{y, x, u, v\}\mbox{\tiny's in orbit $\omega$}}   \delta_{x_1 \ldots x_{\genus} y_{N_V},e} \\
 &  \prod_{j=1}^{N_V} \delta_{y_j h,hy_j} \prod_{i=1}^{\genus} \delta_{x_i h, h x_i} \delta_{u_i h, h u_i} \delta_{v_i h, h v_i} \delta_{x_i, v^{-1}_i u_i v_i u^{-1}_i}, \nonumber
   \end{align}
analogous to Eq.~\eqref{eq:chiLambda1} but now accounting for the handles.   The $\delta$ function on the first line is implementing the constraint Eq.~\eqref{eq:consyform}.

As in going to Eq.~\eqref{eq:chiLambda2} we can make the change of variables $s_i  = y_{i-1}^{-1} y_i$ with $y_0=e$ and use a sum over $s_i$ instead with the constraint now being that in Eq.~\eqref{eq:conssform} and the first $\delta$ function on the second line becoming $\delta_{s_i h,h s_i}$. 

As in the going to Eq.~\eqref{eq:promisedfusion2}, we then multiply this character by the result of fusing the representations $\R_1, \ldots, \R_{N_P}$ together and then sum over orbits (having fixed conjugacy classes $C_i$) to obtain the result
\begin{widetext}
$$
   \frac{1}{|G|} \sum_{h, \{x \}, \{s \}, \{u \}, \{ v \}}  \!\!\!\!\! \delta_{x_1 \ldots x_{\genus} s_1 \ldots s_M,e}  \left\{ \prod_{i=1}^{\genus}  \delta_{x_i,v_i^{-1} u_i v_i u_i^{-1}} \delta_{x_i g,g x_i} \delta_{v_i h,h v_i} \delta_{h u_i,u_ih}\right\}
\left[\prod_{j=1}^{N_P} \chi_{\R_j}(h) \delta_{k_j,e} \right] \left[\prod_{k=1}^{N_V} \delta_{s_k \in C_k} \delta_{s_k h,h s_k}
\right] , 
$$
\end{widetext}
which (up to redefinition of dummy variables) matches Eq.~\eqref{eq:longN}.

Now, when we continue on to the analog of Sec.~\ref{subsub:wavefuction}, everything follows almost exactly the same as above.  The only difference is that the orbits will decompose into a different set of irreps $\tilde \R_q^{(\omega)}$ and then into a different set of irreps $\R_Q^{(\omega,q)}$, but the remaining calculation remains unchanged.   Thus, the calculation holds in this case as well, which completes the proof in general.

{\bf Grand orthogonality theorem for discrete groups:}

For reference it is worth presenting the grand orthogonality theorem for discrete groups.   Given a discrete group $G$ and two irreps $\R$ and $\R'$ we have 
\beqn\label{eq:grand}
    \sum_{h \in G} \rho_{mn}^{\R}(h^{-1}) \rho_{pq}^{\R'}(h) &=& \\ \sum_{h \in G} \rho_{nm}^{\R*}(h) \rho_{pq}^{\R'}(h) &=& \frac{|G|}{d_{\R}} \delta_{\R \R'} \delta_{np} \delta_{mq}, \nonumber
\eeqn
where $d_\R$ is the dimension of irrep $\R$.

\section{Comparison with string-net models}
\label{app:sn}
The string-net model~\cite{Levin05} is another exactly-solvable lattice model realizing topological orders. This model is built from a unitary fusion category $\mathcal{C}$ and produces a topological order corresponding to the Drinfeld center $\mathcal{Z(C)}$. When the input category $\mathcal{C}$ is built either from the irreducible representations of a group $G$ (category called $\repg$) or from the group elements (category called $\vecg$), this model realizes the same topologically-ordered phase $\mathcal{Z}(\vecg)$ as the KQD model with group $G$. In particular, the ground-state degeneracy is then the same for both models. Degeneracies of the excited states of string-net models can also be obtained using the Moore-Seiberg-Banks formula~\cite{Vidal22,Ritz24_1}. In the following, we discuss how the excited-states degeneracies (and the partition function) of both KQD and string-net models, are related to each other.  

Contrary to the KQD model, the string-net model is generally studied in a limit where only plaquette excitations are allowed (vertex excitations are not allowed: branching rules are satisfied at every vertex). In the following, when we refer to the string-net model, we always consider this  restricted Hilbert space.

Given a unitary fusion category $\mathcal{C}$, the corresponding Drinfeld center $\mathcal{Z(C)}$ can be constructed via the tube algebra (for an introduction on the subject, see Ref.~\cite{Simon_book}).
In particular, the tube algebra provides a rectangular matrix of nonnegative integers $n_{A,s}^\mathcal{C}$, where $A\in \mathcal{Z(C)}$ and $s\in \mathcal{C}$ (see, e.g., Ref.~\cite{Ritz24_1}). The integer $n_{A,s}^\mathcal{C}$ is the number of ``input strings'' of type $s$ (i.e., simple objects of $\mathcal{C}$) that are contained in the anyon (or ``output string'', i.e., simple object of $\mathcal{Z(C)}$) of type $A$. For example, the quantum dimension $d_A$ of an anyon $A$ is given by
\be
d_A = \sum_{s\in \mathcal{C}} n_{A,s}^\mathcal{C} d_s,
\ee
where $d_s$ is the quantum dimension of the input string $s$. More precisely, an anyon type $A$ is defined by $n_{A,s}^\mathcal{C}$ and by a half-braiding $\Omega_A$ and usually written:
\be
A=(\underset{s\in\mathcal{C}}{\oplus} n^{\mathcal{C}}_{A,s}s,\, \Omega_A).
\ee

\subsubsection{$\repg$ string-net model}
The $\mathcal{C}=\repg$ string-net model is equivalent to the KQD model on the same lattice (say, the honeycomb lattice) but only with plaquette excitations~\cite{Buerschaper09}. Indeed, setting $n=0$ in Eq.~\eqref{eq:kqddeg2}, one recovers the degeneracies for a $\repg$ string-net model with $m$ excited plaquettes~\cite{Ritz24_1}. Plaquette excitations $A\in\mathcal{Z(C)}$ in the string-net model are identified by $n_{A,1}^{\repg} \neq 0$ and correspond to fluxons Fl in the KQD model. It turns out that $n_{A,1}^{\repg}=n^\text{Pl}_A$, which we introduced in Eq.~\eqref{eq:nJPl}  to identify the internal multiplicity of plaquette excitations of the KQD model.

However, the number of subtypes of plaquette excitations for the $\repg$ string-net model  is not the same as the number of subtypes of fluxons in the KQD model $G$. In the first case, the number of subtypes of a plaquette excitation $A$ is given by 
\be 
n_A = \sum_{s\in \repg}n_{A, s}^{\repg}. 
\label{eq:dimrel}
\ee
In the second case, the number of subtypes of a fluxon $A$ is rather given by 
\be 
n_A = \sum_{s\in \repg}n_{A, s}^{\repg} d_s = d_A, 
\ee
where $s$ labels the irreducible representations of $G$ and $d_s$ is the dimension of these representations.

In the limit where there are no vertex excitations (i.e., $J_\text{v}\to \infty$), apart from a global shift in the total energy by $J_\text{v} N_\text{v}$, the KQD partition function~\eqref{eq:kqdPF} on the honeycomb lattice is the same as  the one found for the $\repg$ string-net model [see Eqs.(16) and (32) in Ref.~\cite{Ritz24_2}]:
\be
Z\simeq {\rm e}^{\beta J_\text{v} N_\text{v}} \sum_{A\in \mathcal{Z}(\vecg)} \!\!\!\!S_{\mathbf{1},A}^{2-2\genus} \left( q_A^\text{Pl}-1+{\rm e}^{\beta J_{\rm p}} \right)^{N_{\rm p}},
\label{eq:zsnrep}
\ee
with $q_A^\text{Pl} =n^\text{$\repg$}_{A,1} / S_{\mathbf{1}, A}$.

\subsubsection{$\vecg$ string-net model}
The $\mathcal{C}=\vecg$ string-net model is equivalent to a KQD model on the dual lattice and with only vertex excitations (see, e.g., Ref.~\cite{Simon_book}). Here, we have in mind the KQD model on the triangular lattice and the string-net model on the dual honeycomb lattice. The degeneracies of the $\vecg$ string-net model are recovered by setting $m =0$ in Eq.~\eqref{eq:kqddeg2}; see Ref.~\cite{Ritz24_1}. Plaquette excitations in the string-net model  are identified by $n_{A,1}^{\vecg} \neq 0$ and correspond to chargeons Ch (or vertex excitations Ve) in the KQD model.  It turns out that $n_{A,1}^{\vecg}=n^\text{Ve}_A$, which we introduced in Eq.~\eqref{eq:nJVe}  to identify the internal multiplicity of vertex excitations of the KQD model.

In the limit where there are no plaquette excitations (i.e., $J_\text{p}\to \infty$), apart from a global energy shift $J_\text{p} N_\text{p}$, the partition function of the KQD model~\eqref{eq:kqdPF} on the triangular lattice recovers that for the $\vecg$ string-net model on the dual honeycomb lattice~\cite{Ritz24_2},
\be
Z \simeq {\rm e}^{\beta J_\text{p} N_\text{p}} \sum_{A\in \mathcal{Z}(\vecg)} \!\!\!\!S_{\mathbf{1},A}^{2-2\genus} \left( q_A^\text{Ve}-1+{\rm e}^{\beta J_{\rm v}} \right)^{N_{\rm v}},
\ee
with $q_A^\text{Ve} =n^{\vecg}_{A,1} / S_{\mathbf{1}, A}$. Of course, on the dual lattice, plaquettes and vertices are exchanged so that vertex excitations of the triangular lattice KQD model become plaquette excitations of the honeycomb string-net model.

\subsubsection{Extended string-net models}
Extended string-net models are a variation of the string-net models that incorporate extra degrees of freedom in the form of added tails to the lattice. In these models, and contrary to the original string-net model, even when remaining in the sector where vertex excitations are forbidden, it is possible to obtain all anyons of the corresponding Drinfeld center as single plaquette excitations~\cite{Hu18,Soares25}. As a consequence, the partition function still has the form (\ref{eq:zsnrep}) but with a different $q_A^\text{Pl}$ [see Eq.~(26) in Ref.~\cite{Soares25}].

\end{document}